\documentclass[10pt]{article}

\usepackage[superscript,sort]{cite}  

\usepackage{ifthen,ifpdf}
\ifpdf
	\usepackage[pdftex]{hyperref}	 \hypersetup{colorlinks=true,linkcolor=black,citecolor=black,urlcolor=blue,backref=page,bookmarks=true,breaklinks=true,plainpages=false }
	\usepackage[pdftex]{graphicx}
	\usepackage[usenames,pdftex]{color} 
\else
	\usepackage[colorlinks=true,breaklinks=true,plainpages=false]{hyperref}
	\usepackage{graphicx}
	\usepackage[usenames]{color}
\fi

\usepackage{amsthm,amscd,amsxtra,amsfonts,amsmath,amssymb,multirow}
\usepackage{wrapfig}
\usepackage[footnotesize]{caption}
\usepackage[tiny,compact]{titlesec}
\usepackage[textwidth=0.8in,textsize=footnotesize]{todonotes}

\usepackage{helvet}

\usepackage[letterpaper,top=0.5in,bottom=0.5in,left=0.5in,right=0.5in]{geometry}

\titlespacing*{\section}{0pt}{*0}{*0}
\titlespacing*{\subsection}{0pt}{*0}{*0}
\titlespacing*{\subsubsection}{0pt}{*0}{*0}
\titlespacing{\paragraph}{0pt}{*0}{*1}
\definecolor{MyPurple}{rgb}{1,0,1}

\newtheorem{theorem}{Theorem}[section]
\newtheorem{lemma}{Lemma}[section]
\newtheorem{definition}{Definition}[section]

\newtheorem{remark}{Remark}

\begin{document}


\title{ Objective-oriented  Persistent Homology  }

\author{Bao Wang$^{1}$, and Guo-Wei Wei$^{1,2,3}$
\footnote{
Address correspondences  to Guo-Wei Wei. E-mail: wei@math.msu.edu}  \\
$^1$Department of Mathematics \\
Michigan State University, MI 48824, USA\\
$^2$Department of Electrical and Computer Engineering \\
Michigan State University, MI 48824, USA \\
$^3$Department of Biochemistry and Molecular Biology\\
Michigan State University, MI 48824, USA
}

\date{\today}

\maketitle

\begin{abstract}
Persistent homology provides a new approach for the topological simplification of big data via measuring the life time of intrinsic topological features in a filtration process and has found  its success  in scientific and engineering applications. However, such a success is essentially limited to   qualitative data characterization, identification and analysis (CIA). Indeed, persistent homology has rarely been employed for quantitative modeling and prediction. Additionally, the present persistent homology is  a passive tool, rather than a proactive  technique, for CIA.  In this work, we outline a general protocol to construct objective-oriented  persistent homology methods. By means of differential geometry theory of surfaces, we construct an objective functional, namely, a surface free energy defined on the data of interest. The minimization of the objective functional  leads to a Laplace-Beltrami operator  which generates a multiscale representation of the initial data and  offers an objective oriented filtration process. The resulting differential geometry based objective-oriented  persistent homology  is able to preserve desirable  geometric features in the evolutionary filtration and enhances the corresponding topological persistence. The cubical complex based homology algorithm is employed in the present work to be compatible with  the   Cartesian representation of the Laplace-Beltrami flow.  The proposed Laplace-Beltrami flow based persistent homology method is extensively validated. The consistence between Laplace-Beltrami flow based filtration and Euclidean distance based filtration  is confirmed on the Vietoris-Rips complex for a large amount of numerical tests. The convergence and reliability of the present  Laplace-Beltrami flow based cubical complex filtration approach are analyzed over various spatial and temporal mesh sizes. The Laplace-Beltrami flow based persistent homology approach is utilized to study  the intrinsic topology of  proteins and fullerene molecules. Based on a quantitative model which correlates the topological persistence of fullerene central cavity with the total curvature energy of  the fullerene structure, the proposed method is used for the prediction of fullerene isomer stability. The efficiency and robustness of the present method are verified by more than 500 fullerene molecules. It is shown that the proposed persistent homology based quantitative model offers good predictions of   total curvature energies for ten types of fullerene isomers. The present work offers the first example to design  objective-oriented  persistent homology  to enhance or preserve desirable  features in the original data during the filtration process and then automatically detect or extract the corresponding topological traits  from the data.   
\end{abstract}

\vskip 1cm
{\it Keywords:}~
Objective-oriented  persistent homology,
Differential geometry based persistent homology, 
Laplace-Beltrami flow based filtration, 
Objective-oriented partial differential equation,
Protein,
Fullerene,
Total curvature energy.


\newpage
{\setcounter{tocdepth}{4} \tableofcontents}

\newpage

\section{Introduction}
In mathematical science, homology is a general procedure to associate a sequence of abelian groups or modules to a given  topological space and/or manifold  \cite{Hatcher:2001,Edelsbrunner:2010}. The idea of homology dates back to Euler and Riemann, although homology class was first rigorously defined by Henri Poincar\'e,  who built the foundation of the modern algebraic topology. The  topological  structure of a given manifold can be studied via defining the different dimensional homology groups on the manifold such that the bases of the homology groups are  isomorphic to the bases of the corresponding topological spaces. In computational perspective, a given manifold can be approximated by the triangulated simplicial complex, on which  homology groups can be further defined. The triangulation of a manifold or a topological space can be realized through a number of methods, such as Delaunay triangulation. There are many triangulation software packages, such as \href{http://wias-berlin.de/software/tetgen/}{TetGen} and \href{https://www.cgal.org/}{CGAL}. In scientific computing, the Cartesian representation is one of most  important approaches in numerical analysis. Consequently,  cubical complex based homology analysis has also become a popular research topic in the past decade. A systematic description of  homology analysis in the cubical complex setting has been  given by Kaczynski et al \cite{Kaczynski:2004}.

Persistent homology creates a multiscale representation of topological structures via  a scale parameter relevant  to topological events. The basic concept was introduced by Robins~\cite{Robins:1999} and  Frosini and Landi~\cite{Frosini:1999}, independently.   Edelsbrunner et al.~\cite{Edelsbrunner:2002} introduced first efficient algorithm for persistent homology analysis. The generalization of persistent homology was given by Zomorodian and Carlsson \cite{Zomorodian:2005}. In the past decade, persistent homology has been developed as an efficient computational tool for the characterization, identification and analysis (CIA) of topological features in large data sets  \cite{Edelsbrunner:2002,Zomorodian:2005,Zomorodian:2008}.  Topological persistence over the  filtration process  can be captured  continuously over a range of spatial  scales in  persistent homology analysis. Unlike commonly used computational homology which results in  truly metric free or coordinate free representations, persistent homology is able to embed geometric information to topological  invariants so that  the ``birth"  and ``death" of  isolated components, circles, rings, loops, pockets, voids or cavities at all geometric scales  can be  monitored by topological measurements. Compared with traditional computational topology \cite{Krishnamoorthy:2007,YaoY:2009,ChangHW:2013}  and/or computational homology, persistent homology  inherently has an additional dimension, namely, the filtration parameter, which can be utilized to embed some crucial geometry or quantitative information into the topological invariants.  Barcode representation has been proposed for the visualization  of topological  persistence  \cite{Ghrist:2008}, in which various horizontal line segments or bars are  utilized to represent the persistence of the topological features. Efficient computational algorithms  such as, pairing algorithm \cite{Edelsbrunner:2002,Dey:2008}, smith normal form \cite{Edelsbrunner:2010,Zomorodian:2005} and  Morse reduction \cite{Harker:2010,Harker:2013,Wagner:2012},  have been proposed to track topological variations during the  filtration process \cite{Bubenik:2007, Edelsbrunner:2010,Dey:2008,Dey:2013,Mischaikow:2013}.   Some of these persistent homology  algorithms have been implemented in  many software packages, namely     Perseus \cite{Perseus,Mischaikow:2013}, JavaPlex \cite{Javaplex} and \href{http://www.mrzv.org/software/dionysus/}{Dionysus}.  In the past few years, persistent homology has been applied to  image analysis \cite{Carlsson:2008,Pachauri:2011,Singh:2008,Bendich:2010}, image retrieval \cite{Frosini:2013}, chaotic dynamics verification \cite{Mischaikow:1999,Kaczynski:2004}, sensor network \cite{Silva:2005}, complex network \cite{LeeH:2012,Horak:2009}, data analysis \cite{Carlsson:2009,Niyogi:2011,BeiWang:2011,Rieck:2012,XuLiu:2012}, computer vision \cite{Singh:2008}, shape recognition \cite{DiFabio:2011} and computational biology \cite{Kasson:2007,Dabaghian:2012,Gameiro:2013, KLXia:2014c}.

Nevertheless, the applications of persistent homology have been essentially limited to qualitative  CIA. Indeed, there is little literature about the use of persistent homology as a quantitative tool, i.e., for  mathematical modeling and physical prediction, to our best knowledge. Recently, we have introduced molecular topological fingerprints (MTFs) as an efficient approach for protein CIA \cite{KLXia:2014c}. We have also utilized MTFs to reveal quantitative structure-function relationships in  protein folding, protein stability and protein flexibility \cite{KLXia:2014c}.

In the past few decades, geometric analysis, which combines  differential equations and differential geometry, has become a popular approach for data analysis, signal and image processing,  surface generation and computer visualization \cite{Feng:2004a,Gomes:2001,Mikula:2004,Osher:2001,Sarti:2002,Sethian:2001,Sochen:1998,Zhang:2006c}. Geometric partial differential equations (PDEs) \cite{Willmore:1997}, i.e., the Laplace-Beltrami flows, are efficient apparatuses in applied  mathematics and computer science \cite{Cecil:2005,Chopp:1993,Smereka:2003}. Osher and Sethian \cite{SOsher:1988,Sethian:2001} have devised level set as a computational tool for solving geometric PDEs. An alternative approach is to make use  of the Euler-Lagrange variation to derive a desirable set of geometric PDEs from  a functional, such as a Mumford-Shah functional \cite{Mumford:1989}, for image or surface analysis  \cite{Blomgren:1998,Carstensen:1997,Li:1996a,Osher:1990,Rudin:1992,Sapiro:1996}. Wei introduced  some of the first families of high-order geometric PDEs for image analysis in 1998 \cite{Wei:1999}. Mathematical analysis of high-order geometric PDEs was reported in the literature \cite{Greer:2004,Greer:2004b,MXu:2007,ZMJin:2010}. Geometric PDE based high-pass filters was pioneered by Wei and Jia by  coupling two nonlinear geometric PDEs \cite{Wei:2002a}. Recently, this approach has been extended  to a more general formalism, the PDE transform, for image and surface analysis \cite{YWang:2011c,YWang:2012b,YWang:2012d,QZheng:2012}.

In 2005,    the use of curvature-controlled PDEs for the construction of biomolecular surfaces was introduced for the first time in computational biophysics \cite{Wei:2005}. The geometric PDE was shown to provide a multiscale representation of biomolecular surfaces \cite{Wei:2005}. Based on  differential geometry,  Wei and coworkers introduced  the first variational solvent-solute interface: the minimal molecular surface (MMS)  for molecular surface representation  in 2006 \cite{Bates:2006,Bates:2006f,Bates:2008}.  Since the surface free energy is the product of surface tension and  surface area, 
the minimization of the surface free energy leads to the Laplace-Beltrami operator. One then obtains  the Laplace-Beltrami flow   by adopting an artificial time.   The  Laplace-Beltrami flow approach has been used to calculate both solvation energies and electrostatics of proteins \cite{ZhanChen:2012, Bates:2008,Bates:2009}. We have proposed potential-driven geometric flows, which admit non-curvature-driven terms, for biomolecular surface construction subject to potential interactions \cite{Bates:2009}. While our approaches were employed by many others \cite{Cheng:2007e,Yu:2008g,SZhao:2011a,SZhao:2014a} for molecular surface analysis, our curvature-controlled PDEs and the  differential geometry based   Laplace-Beltrami   models  \cite{Wei:2005,Bates:2006,Bates:2008,Bates:2009} are, to our knowledge, the first of their kind for biomolecular surface and electrostatics/solvation modeling.

Using the variational principle, Wei introduced  differential geometry based multiscale models in 2009 \cite{Wei:2009}.  The essential idea is to use the differential geometry theory of surfaces as a natural means to geometrically separate the macroscopic domain of the biomolecule from the microscopic domain of the solvent, and to dynamically couple the continuum treatment  of the solvent from the discrete description of the biomolecule.  In the past few years, differential geometry based multiscale models have been implemented for nonpolar solvation analysis \cite{ZhanChen:2012},  full solvation analysis \cite{ZhanChen:2010a, ZhanChen:2010b, ZhanChen:2011a},  proton transport \cite{DuanChen:2012a,DuanChen:2012b}, ion permeation across membrane channel proteins \cite{QZheng:2011a,QZheng:2011b,Wei:2012,Wei:2013,JKPark:2013a}.  The performance of our methods has been extensively validated with experimental data, including solvation energies and  current-voltage (I-V) curves.

In this work, we propose objective-oriented persistent homology methods   to proactively extract desirable topological traits from biomolecular data.  As a general procedure, we  construct an objective functional to optimize desirable  features in data. In our specific example, such an optimization is realized through a geometry-embedded  filtration process and leads to an objective-oriented persistent homology method.   As a proof of principle, we utilize  differential geometry theory of surfaces  to minimize the surface free energy, which results in an objective-oriented partial differential equation, i.e., the  Laplace-Beltrami flow. The evolution of  the  Laplace-Beltrami flow creates a multiscale representation of a nano-bio object, which  naturally constitutes a  filtration and  gives rise to a differential geometry   based persistent homology method. The proposed  differential geometry based persistent homology is utilized to analyze nano-bio data.    The topological invariants of a given nano-bio object are extracted from  the evolutionary profiles of the Laplace-Beltrami flow.  Then topological persistence  is analyzed to identify the intrinsic topological signature of a given data.  Such information is further utilized to unveil quantitative topology-function relationships.  It is well known that geometric PDEs can be designed to preserve certain geometric features in the time evolution \cite{Wei:1999}. Specifically, Laplace-Beltrami flow minimizes the mean curvature or surface area \cite{Bates:2008}.  As a result, topological invariants computed from the geometric PDE based filtration enhance the corresponding features. This idea is potentially useful and powerful for automatic feature detection and extraction  from big data.

The rest of this paper is organized as follows. In Section \ref{geo_flow_theory}, we give a brief introduction to the theory of Laplace-Beltrami flows for nano-bio systems, such as   proteins and carbon fullerene molecules. A computational protocol,  including  numerical implementation, for integrating the evolution Laplace-Beltrami flow is described in detail. In Section \ref{cub_hom_theory}, the brief review of  homology and persistent homology theories are given in the cubical complex setting. The construction of objective-orieted persistent homology is discussed in Section \ref{comb}. As a specific example of this new method,  we propose the Laplace-Beltrami flow based persistent homology.  The validity of the proposed method  is carefully carried out in Section \ref{Validation} using carbon fullerene data.  The consistence with radius based filtration and the numerical convergence are verified.  The proposed method is applied to the CIA of proteins and fullerene molecules in Section \ref{Application}. We consider the topological persistence  of a beta barrel, which has an intrinsic ring structure.  We demonstrate that the specific intrinsic feature of the beta barrel, namely the inner ring structure, is enhanced during the time evolution. Whereas, some undesirable topological feature due to the Vietoris-Rips complex can be effectively suppressed in the present approach. We further apply this differential geometry based persistent homology to the quantitative prediction of fullerene isomer total curvature energies. This paper ends with a conclusion.


\section{Laplace-Beltrami flows  for  nano-bio systems }\label{geo_flow_theory}

In this section, we provide a brief summary of differential geometry based Laplace-Beltrami flows. To this end, we discuss differentiable manifold and curvature, and followed by the construction of Laplace-Beltrami  operator  using an objective functional. The implementation of the Laplace-Beltrami flow for biomolecular data is described in detail. 

\subsection{Differentiable manifolds and curvatures}

Consider an immersion of an open set $U\subset {\mathbb R}^3$ to ${\mathbb R}^4 $  via a differentiable hypersurface element ${\bf f}: U\rightarrow {\mathbb R}^4 $. Here the
hypersurface element is a  vector-valued $C^2$  function: ${\bf f}({\bf u}) =(f_1(u),f_2(u),f_3(u),f_4(u))$ and ${\bf u}=(u_1,u_2,u_3)\in U$.

Tangent vectors (or directional vectors) of ${\bf f}$ are ${\bf X}_i=\frac{\partial {\bf f}} {\partial u_i}$.
The Jacobi  matrix of the mapping ${\bf f}$ is given by $D{\bf f}=({\bf X}_1,{\bf X}_2,{\bf X}_3)$.

As a symmetric and  positive definite metric tensor of ${\bf f}$, the first fundamental form is
${\bf I}:=(g_{ij})=(D{\bf f})^T\cdot(D{\bf f})$, where matrix elements are $g_{ij}=<{\bf X}_i,{\bf X}_j>$. Here
$<,>$ is the Euclidean inner product in ${\mathbb R}^4$, $i,j =1,2,3$. 

The Gauss map  $\nu: U\rightarrow S^3$ is defined by  the unit normal vector $\nu({\bf u})$
\begin{equation}\label{normal}
\nu(u_1,u_2,u_3):=\frac{{\bf X}_1 \times {\bf X}_2 \times {\bf X}_3 }{ \|{\bf X}_1 \times {\bf X}_2 \times {\bf X}_3\|} \in \bot_u{\bf f},
\end{equation}
where the cross product in ${\mathbb R}^4$ is a generalization of that in ${\mathbb R}^3$.
Here $\bot_u{\bf f}$ is the normal space of ${\bf f}$ at point ${\bf p}={\bf f}({\bf u})$. It is easy to verify that $$<\nu,\nu>=1.$$
Locally at ${\bf p}$, the normal vector $\nu$ is perpendicular to the tangent hyperplane $T_u{\bf f}$:
$$<\nu,{\bf X}> =0.$$
Note that  $T_u{\bf f}\oplus \bot_u{\bf f} =T_{{\bf f}({\bf u})}{\mathbb R}^3$, which is the tangent space at point ${\bf p}$.
The second  fundamental form is of crucial importance and can be defined by  means of the normal vector $\nu$ and tangent vector ${\bf X}_i$,
\begin{equation}
{\bf II}({\bf X}_i,{\bf X}_j)=(h_{ij}) \equiv \left(\left<-\frac{\partial\nu}{\partial u_i}, {\bf X}_j\right>\right).
\end{equation}
The definition of the second fundamental form can be systematically generalized by using the Weingarten map, a shape operator of ${\bf f}$:
$${\cal L}:=-D\nu \circ (D{\bf f})^{-1}.$$
 Since ${\cal L}$ is a self-adjoint operator, we have
\begin{equation}
{\bf II}({\bf X}_i,{\bf X}_j)={\bf I}({\cal L}{\bf X}_i,{\bf X}_j)=(h_{ij})=\left(\left<-\frac{\partial\nu }{\partial u_i}, {\bf X}_j \right>\right)
=\left(\left< \frac{\partial^2 {\bf f} }{\partial u_i\partial u_j}, \nu \right>\right).
\end{equation}
The third and fourth  fundamental forms are conveniently given in terms of the shape operator
\begin{eqnarray}
{\bf III}({\bf X}_i, {\bf X}_j)&=&{\bf I}({\cal L}^2{\bf X}_i,{\bf X}_j)=(e_{ij})=\left(\left<\frac{\partial \nu }{ \partial u_i},\frac{\partial \nu }{\partial u_j}\right>\right)\\
{\bf IV}({\bf X}_i, {\bf X}_j)&=&{\bf I}({\cal L}^3{\bf X}_i,{\bf X}_j).
\end{eqnarray}

The Laplace-Beltrami can be calculated by
\begin{equation}\label{meanc}
H=\frac{1}{3}h_{ij}g^{ji},
\end{equation}
where we use the Einstein summation convention, and $(g^{ij})$ denotes the inverse matrix $(g^{ij})= (g_{ij})^{-1}$.

Principal curvatures $\kappa_i ~(i=1,2,3)$ are defined as the eigenvalues of Weingarten map ${\cal L}$ with eigenvectors being unit tangent vectors.
Appropriate organization of the principal curvatures gives rise to the first three Laplace-Beltramis
\begin{eqnarray}
K_1&=&\frac{1}{3}(\kappa_1+\kappa_2+\kappa_3)\\
K_2&=&\frac{1}{3}(\kappa_1\kappa_2+\kappa_1\kappa_3+\kappa_2\kappa_3)\\
K_3&=&\kappa_1\kappa_2\kappa_3
\end{eqnarray}
where $K_1=H=\frac{1}{3}{\rm Tr}({\cal L})$ is the Laplace-Beltrami and $K_3=K={\rm Det}({\cal L})$ is the Gauss-Kronecker curvature or Gauss curvature.
The local property of the Gauss curvature is used to classify the point as elliptic, hyperbolic, parabolic, etc. The combination of Gauss and Laplace-Beltramis has been used to characterize protein surfaces and predict protein-ligand binding sites \cite{XFeng:2013b, KLXia:2014a}.
It follows from the Cayley-Hamilton theorem  that  the first four fundamental  forms satisfy: ${\bf IV}-3H{\bf III}+3K_2{\bf II}-K{\bf I}=0$.

We discuss an iterative procedure to generate a family of hypersurfaces that have vanishing Laplace-Beltrami except at the boundary.
Let  $U \subset {\mathbb R}^3$ be an open set  with a compact closure $\overline{U}$ and boundary $\partial U$.
Consider a family of hypersurface elements  ${\bf f}_\varepsilon:\overline{U}\rightarrow {\mathbb R}^4$   ($\varepsilon>0$)    generated by  deforming ${\bf f}$ in the normal direction  with speed of  the Laplace-Beltrami:
\begin{equation}\label{evol}
{\bf f}_\varepsilon (x,y,z):={\bf f}(x,y,z)+\varepsilon H \nu(x,y,z).
\end{equation}
Equation (\ref{evol}) is iterated until  $H=0$ in all of $U$, except at  boundary $\partial U$, which can be a set of atomic surface constraints.   This procedure leads to a minimal hypersurface \cite{Bates:2008}.

As discussed above, the hypersurface element is a vector-valued function which is cumbersome in biophysical application. We therefore construct a scalar hypersurface function by setting  ${\bf f}({\bf u})=(x,y,z,S)$, where $S(x,y,z)$ is a hypersurface function of interest. The first fundamental form can be explicitly computed
\begin{equation}
(g_{ij})=\left(
       \begin{array}{lll}
   1+S_x^2  & S_xS_y  & S_xS_z \\
   S_xS_y   & 1+S_y^2 & S_yS_z \\
   S_xS_z   & S_yS_z  & 1+S_z^2
       \end{array}\right).
\end{equation}
Matrix tensor $(g_{ij})$ has the inverse
\begin{equation}
(g^{ij})=\frac{1}{g}\left(
       \begin{array}{lll}
   1+S_y^2+S_z^2  & -S_xS_y  &- S_xS_z \\
   -S_xS_y   & 1+S_x^2+S_z^2 &- S_yS_z \\
   -S_xS_z   & -S_yS_z  & 1+S_x^2+S_y^2
       \end{array}\right),
\end{equation}
where $g={\rm Det}(g_{ij})=1+S_x^2+S_y^2+S_x^2$ is the Gram determinant.
From Eq. (\ref{normal}), the normal vector is given by
\begin{equation}
\nu=(-S_x,-S_y,-S_z,1)/\sqrt{g}.
\end{equation}
The second fundamental form, the Hessian matrix of $S$, is obtained as
\begin{equation}
(h_{ij})=\left(\frac{1}{\sqrt{g}}S_{x_ix_j}\right).
\end{equation}
Using Eq. (\ref{meanc}),  one can obtain the Laplace-Beltrami
\begin{equation}
H=\frac{1}{3} \nabla \cdot \left(\frac{\nabla S} {\sqrt{g}}\right).
\end{equation}

\subsection{Laplace-Beltrami flow} \label{Laplace-Beltrami}
\subsubsection{ Laplace-Beltrami equation}
 According to differential geometry theory of surfaces,  a surface area is minimized if and only if the Laplace-Beltrami is zero 
everywhere on the surface except for a set of boundary points. Following Eq. (\ref{evol}), we construct  a family of hypersurfaces $S_\varepsilon$ as
\begin{equation}\label{evolut}
S_\varepsilon(x,y,z)=S(x,y,z)+ \frac{\varepsilon}{3\sqrt{g}}
\nabla \cdot \left(\frac{\nabla S} {\sqrt{g}}\right).
\end{equation}
The iteration of the hypersurface function so that $S_\varepsilon(x,y,z)\rightarrow S(x,y,z)$, i.e., $\nabla \cdot \left(\frac{\nabla S} {\sqrt{g}}\right)=0$, leads to the desired minimal hypersurface function $S$.  

A more general procedure  is to construct an objective functional, i.e., a surface free energy functional,  for the molecular data of interest
\begin{equation}
E=\int_{\partial \Omega} \gamma  d\Omega,
\end{equation}
where $\partial \Omega$ is the boundary of the molecule,  $\gamma$ is the surface tension and $d\Omega=\sqrt{g}dxdydz$.
 Using  the Euler Lagrange equation, we minimize the surface free energy density $e=\gamma\sqrt{g}$ with respect to $S$
\begin{equation}\label{EL}
\frac{\partial e}{\partial S} -
\frac{\partial}{\partial x} \frac{\partial e}{\partial S_x} -
\frac{\partial}{\partial y}\frac{\partial e}{\partial S_y} -
\frac{\partial}{\partial z}\frac{\partial e}{\partial S_z}
=0.
\end{equation}
Since $\gamma\neq 0$ in general,  we arrive at the vanishing of the mean curvature operator $\nabla \cdot \left(\frac{\nabla S} {\sqrt{g}}\right)=3H=0$ again.

From the computational point of view, the iteration process can be efficiently achieved by  introducing an artificial time variable $t$ so as to change the elliptic PDE into a parabolic one.  Specifically, instead of iterating Eq. (\ref{evolut}),  we set the hypersurface function $S$ to be $S(x, y, z, t)$ in the computational perspective and construct  the following  Laplace-Beltrami equation
\begin{equation}
\label{Geo}
\frac{\partial S}{\partial t}=\sqrt{g}\nabla\cdot\left(\frac{\nabla S}{\sqrt{g}}\right).
\end{equation}
A similar approach  is to set $\sqrt{g}$ as $|\nabla S|$, leading to another popular form of the Laplace-Beltrami equation \cite{Bates:2008}
\begin{equation}
\label{Geo2}
\frac{\partial S}{\partial t}=|\nabla S|\nabla\cdot\left(\frac{\nabla S}{|\nabla S|}\right).
\end{equation}
These equations were employed to construct minimal molecular surfaces of proteins and other biomolecules \cite{Bates:2008,Bates:2009,ZhanChen:2010a,KLXia:2014a}.

\subsubsection{ Initial value and boundary condition for nano-bio Laplace-Beltrami flows}

In the present  work, we generate a family of hypersurface functions indexed by the artificial time $t$ by using  Laplace-Beltrami equataion (\ref{Geo}). We call this family of hypersurface functions the profiles of Laplace-Beltrami flows. Note that we do not seek the minimal molecular surfaces described in our earlier work \cite{Bates:2008,Bates:2009,ZhanChen:2010a,KLXia:2014a}. Instead, we look for a geometric PDE or Laplace-Beltrami flow representation of nano-bio molecules. To apply this approach to  proteins and nano-molecules, we start with a given set of $N$ atomic coordinates $\{{\bf r}_i\}, ~ (i=1,2,\cdots, N)$, which can be obtained from Protein Data Bank (PDB) available on web or from the literature.  We define a set by ${\bf R}_N= \cup_{i=1}^N B_{\epsilon}(r_i)$, where $B_{\epsilon}({\bf r}_i,r_i)$ is the ball centered at ${\bf r}_i$ of radius $r_i=\varepsilon r_{vdW}$. Here $\epsilon>0$ is a parameter and  $r_{vdW}$ is the van der Waals radius of the $i$th atom.

The  initial value of the hypersurface $S$ can be chosen in a number of ways. One choice is
\begin{equation}\label{InitiaV1}
S({\bf r}, 0)=
\begin{cases}
   1 &  \mbox{if ${\bf r} \in  {\bf R}_N $},\\
   0 &  \mbox{otherwise}.
\end{cases}
\end{equation}
\begin{remark}
The initial radius of an atom $\epsilon r_i$ in a molecule can  be  adjusted by parameter $\varepsilon $.  For different applications, one can choose different initial radius. In our earlier work, $\epsilon>1$ was used \cite{Bates:2008,Bates:2009,ZhanChen:2010a,KLXia:2014a}. In the present work, we set $\epsilon=\frac{1}{2}$.
\end{remark}

Alternatively, another choice is a heaviside function $\theta$
\begin{equation}\label{InitiaV2}
S({\bf r}, 0)=\theta (\mu({\bf r})-\mu_0),
\end{equation}
where $\mu_0$ is a cutoff value and $\mu({\bf r})$ is a rigidity function \cite{Opron:2014}
\begin{equation}
\mu({\bf r}) =\sum_i^Nw_i \Phi(|{\bf r}-{\bf r}_i|;\eta_i).
\end{equation}
Here $w_i$ is a weight associated with the atomic type of the $i$th atom and is set to 1 in the present work. Additionally, correlation function  $\Phi(|{\bf r}-{\bf r}_i|; \eta_i)$ is monotonically decreasing radial basis functions, such as generalized exponential functions or generalized Lorentz functions \cite{Opron:2014}.  The scaling function $\eta_i$ can be set to $\eta_i\propto r_{\rm vdW}$ and should be systematically adjusted for different choices of $\Phi$.

Obviously, the other choice of the initial value is to directly use the rigidity function
\begin{equation}\label{InitiaV3}
S({\bf r}, 0)= \mu({\bf r}).
\end{equation}

The initial  values given by Eq. (\ref{InitiaV2}) are smoother than those given by Eq. (\ref{InitiaV1}). However, Eq. (\ref{InitiaV3}) provides the smoothest initial values. The results reported in this work are based on  Eq. (\ref{InitiaV1}). However, our tests indicate that other two types of initial values work well.

Both the Dirichlet boundary  ($S({\bf r},t)=0~\forall {\bf r}\in \partial U)$ or the Neumann boundary ($\frac{\partial S}{\partial {\bf r}}=0~ \forall {\bf r}\in \partial U$) can be employed.  The  solution of Eq. (\ref{Geo}) gives a family of  hypersurface functions $S(x,y,z,t)$. We extract desirable nano-bio information from $S$ by using two different procedures. One is to take an iso-surface for a given iso-value, i.e., $S=c$, which can be extracted by the level set method. For our applications, the iso-value of the hypersurface for carbon fullerene molecules is set to be $c=0.1$, and that for   protein molecules is set  $c=0.01$.  The other approach is to evaluate the structural information   contained in $S(x,y,z,T)$ at a given time $T>>0$. We typically set $T$ to be a quite large value so the hypersurface profile is well developed. However, to avoid boundary effect, $T$ should not be too large.


\section{Cubical complex based homology and persistent homology}\label{cub_hom_theory}
In this section, a brief review of the homology and persistent homology in the cubical complex setting is provided. The reader is referred to the literature   \cite{Kaczynski:2004,Strombom:2007} for more comprehensive  discussion  and treatment.

\subsection{Geometric building blocks}

The cubes are the basic geometric building blocks of the homology and persistent homology theory in the cubical complex setting.  First of all, we need to introduce a few basic concepts about cubes.

\begin{itemize}
\item An elementary non-degenerate interval is a closed interval $I\subset \mathbb{R}$ of the form $I=[m, m+1]$ (or $I=[m]$ for simplicity) for some integer $m$.  An elementary degenerate interval is a point $I=[m, m]$.

\item An elementary cube $Q$ or $d$ cube is a $d$-product of elementary intervals, i.e.,
$$
Q=I_1\times I_2\times ...\times I_d\subset \mathbb{R}^d,
$$
where each $I_i, i=1, 2,..., d$ is an elementary interval of non-degenerated or degenerated type, and $d$ is called the embedding number of $Q$, denoted as ${\rm emb}~ Q=d$.
The dimension of $Q$, denoted by ${\rm dim}~ Q$, is defined to be the number of non-degenerated components in $Q$, and $\mathcal{K}_k$ denotes the set of all $k$ dimensional elementary cubes. Let $\mathcal{K}:=\bigcup_{d=1}^\infty \mathcal{K}^d$  be the set of all elementary cubes, and $\mathcal{K}^d$ be the set of all elementary cubes in $\mathbb{R}^d$.

\item The set of $k$-dim cubes with embedding number $d$ is $\mathcal{K}_k^d:=\mathcal{K}_k\bigcap \mathcal{K}^d$. Obviously, if $Q\in \mathcal{K}_k^d$ and $P\in \mathcal{K}_{k'}^{d'}$, then $Q\times P\in \mathcal{K}_{k+k'}^{d+d'}$.
\end{itemize}

With the above building blocks, we say that  set $X\subset \mathbb{R}^d$ is cubical if $X$ can be written as a finite union of elementary cubes.

For a given cubical set $X\subset \mathbb{R}^d$, we  define the following cubical  set $\mathcal{K}(X)$ and $k$-cube set $\mathcal{K}_k(X)$ of $X$:
$$
\mathcal{K}(X)=\{Q\in \mathcal{K}|Q\subset X\},
$$

$$
\mathcal{K}_k(X):=\{Q\in \mathcal{K}(X) | {\rm dim}~ Q=k\}.
$$
The elements of $\mathcal{K}_k(X)$ are called the $k$-cube  of $X$.

\subsection{Algebraic building blocks}

With the above geometric building blocks, we   define the algebraic operations on the building blocks, following the line of Kaczynski et al. \cite{Kaczynski:2004}

First, each elementary $k$-cube $Q\in \mathcal{K}_k^d$ is associated with an algebraic object $\hat{Q}$ which is
called an elementary $k$-chain of $\mathbb{R}^d$. The set of all elementary $k$-chains of $\mathbb{R}^d$ is
$$
\hat{\mathcal{K}}_k^d:=\{\hat{Q}| Q\in \mathcal{K}_k^d\},
$$
and the set of all elementary chains of $\mathbb{R}^d$ is
$$
\hat{\mathcal{K}}^d:=\bigcup_{k=0}^\infty \hat{\mathcal{K}}_k^d.
$$

Second,  addition operation and boundary operator are defined for the further algebraic treatment of the cubical complex.
\subsubsection{Addition operation}
To define the addition operation on  elementary chains, first, the following $k$-chains, i.e., a linear combination of $k$-chain,
$$
c=a_1\widehat{Q_1}+a_2\widehat{Q_2}+\cdots a_m\widehat{Q_m},\ \  a_i\in \mathbb{Z}, i=1, 2, ..., m,
$$
is allowed for any given finite collection $\{\widehat{Q_1}, \widehat{Q_2}, ..., \widehat{Q_m}\}$,
and, if all the $a_i=0$, then we set $c=0$.

The set of all the above  $k$-chains is denoted by $C_k^d$.
The addition of two $k$-chains is defined by:
$$
\sum a_i\widehat{Q_i}+\sum b_i\widehat{Q_i}=\sum (a_i+b_i)\widehat{Q_i}.
$$

It is easy to check for $\forall$ $k$-chains $c=\sum_{i=0}^m a_i \widehat{Q_i}$, there is an inverse element $-c=\sum_{i=0}^m -a_i \widehat{Q_i}$
with the property $c+(-c)=0$, note the addition operation is commutable, thus $C_k^d$  is an abelian group.

\subsubsection{Boundary operator}
Before we define the boundary operator, the scalar product and cubical product operation on the $k$-chain group $C_k^d$ need to be defined.

\begin{definition}
Let $c_1, c_2\in C_k^d$, where $c_1=\sum_{i=1}^m a_i\widehat{Q_i}$ and $c_2=\sum_{i=1}^m b_i \widehat{Q_i}$. The scalar product of chains
$c_1$ and $c_2$ is defined as \cite{Kaczynski:2004}:
$$
<c_1, c_2>:=\sum_{i=1}^m a_ib_i.
$$
\end{definition}

\begin{definition}
For all elementary cubes $P\in \mathcal{K}_k^d$ and $Q\in \mathcal{K}_{k'}^{d'}$, the cubical product between $P, Q$ is defined to be \cite{Kaczynski:2004}:
$$
\widehat{P}*\widehat{Q}:=\widehat{P\times Q}.
$$
And for all chains $c_1\in C_k^d$ and $c_2\in C_{k'}^{d'}$, the cubical product is:
$$
c_1*c_2=\sum_{P\in \mathcal{K}_k, Q\in \mathcal{K}_{k'}} <c_1, \widehat{P}><c_2, \widehat{Q}>\widehat{P\times Q},
$$
and $c_1*c_2\in C_{k+k'}^{d+d'}$.
\end{definition}

For the cubical product, the following important factorization property holds \cite{Kaczynski:2004}:
\begin{lemma}
For $\forall \widehat{Q}\in \widehat{\mathcal{K}}^d$ with $d>1$. There exists unique elementary cubical chains $\widehat{I}$ and $\widehat{P}$ with
${\rm emb}~ I=1$ and ${\rm emb} ~P=d-1$, such that $\widehat{Q}=\widehat{I}*\widehat{P}$.
\end{lemma}

With the above preparation, the boundary operation can be defined inductively in the following way \cite{Kaczynski:2004}.
\begin{definition}
For $k\in \mathbb{Z}$, the cubical boundary operator
$$
\partial_k: \mathcal{C}_k^n\rightarrow \mathcal{C}_{k-1}^n
$$
is a homomorphism of abelian groups, defined for an elementary chain $\widehat{Q}\in \widehat{\mathcal{K}}_k^n$ by induction on the embedding number $n$ as follows:
\begin{itemize}
\item For $n=1$ $Q$ is an elementary interval, i.e., $Q=[m]$ or $Q=[m, m+1]$ for some $m\in \mathbb{Z}$, and one defines:
$$
\partial_k\widehat{Q}=
\begin{cases}
   0 &\mbox{if $Q=[m]$}\\
   \widehat{[m+1]}-\widehat{[m]} &\mbox{if $Q=[m, m+1]$}.
\end{cases}
$$

\item For $n>1$, let $I=I_1(Q)$ and $P=I_2(Q)\times \cdots\times I_n(Q)$ so that $\widehat{Q}=\widehat{I}*\widehat{P}$, then one defines:
$$
\partial_k\widehat{Q}=\partial_{{\rm dim}~ I}\widehat{I}*\widehat{P}+(-1)^{{\rm dim}~ I}\widehat{I}*\partial_{{\rm dim} ~P}\widehat{P}.
$$
\end{itemize}
\end{definition}

By linearity this can be extended to chains, i.e., if $c=\sum_{i=1}^p a_i \widehat{Q}_i$, then:
$$
\partial_k c=\sum_{i=1}^p a_i\partial_k \widehat{Q}_i.
$$

\begin{theorem}
The boundary operator operator satisfies:
$$
\partial_k\circ \partial_{k-1}=0,\ \  \forall k>1,
$$
which is consistent with the simplicial complex setting.
\end{theorem}

Now, for a given cubical set $X\subset \mathbb{R}^d$, let $\{\widehat{K}_k(X):={\widehat{Q}|Q\in \mathcal{K}_k(X)}\}$ and let
$C_k(X)$ be the subgroup of $C_k^d$ generated by the elements of $\widehat{K}_k(X)$, which is called the set of $k$-chains of $X$.
The boundary operator maps $C_k(X)$ to a subset of $C_{k-1}(X)$, thus one can restrict the boundary operator to the cubical set $X$.

\begin{definition}
The boundary operator for the cubical set $X$ is defined to be:
$$
\partial_k^X: C_k(X)\rightarrow C_{k-1}(X),
$$
obtained by restricting $\partial_k: C_k^d\rightarrow C_{k-1}^d$ to $C_k(X)$.
\end{definition}

\begin{definition}
The cubical chain complex for the cubical set $X\subset \mathbb{R}^d$ is
$$
\mathcal{C}(X):=\{C_k(X), \partial_k^X\}_{k\in \mathbb{Z}},
$$
where $C_k(X)$ are the groups of cubical $k$-chains generated by $\mathcal{K}_k(X)$ and $\partial_k^X$ is the cubical boundary operator restricted to $X$.
\end{definition}

\subsection{Homology of cubical sets}
As discussed above,  one has  the corresponding  $k$-chains group $C_k(X)$,  for a given cubical set $X$, now one can define two subgroups of $C_k(X)$.
\begin{itemize}
\item $k$-cycle group $Z_k(X):={\rm ker} ~\partial_k^X=C_k(X)\bigcap {\rm ker}~ \partial_k\subset C_k(X)$.
\item $k$-boundary group $B_k(X):={\rm im} ~\partial_{k+1}^X=\partial_{k+1}(C_{k+1}(X))\subset C_k(X)$.
\end{itemize}

Following from $\partial_k\circ\partial_{k-1}=0\ \ \forall k>1$, one has $B_k(X)\subset Z_k(X)$.   Therefore, one has the following homology group \cite{Kaczynski:2004}.
\begin{definition}
The $k$th homology group of the cubical set $X$ is the quotient group:
$$
H_k(X):=Z_k(X)/B_k(X).
$$
\end{definition}

The $k$th Betti number is defined as the rank of the $k$th homology group,
$$
\beta_k={\rm rank}~ H_k.
$$

From the topological point of view, $H_k(X)$ describes $k$-dimensional holes of $X$, e.g., $H_0(X)$ measures  connected components, $H_1(X)$ measures  loops and  $H_2(X)$ measures  voids. In other words, $\beta_0$ is the number of connected components, $\beta_1$ is the number of loops, $\beta_2$ is the number of voids, and so on. We are particularly interested in behavior of $\beta_0$, $\beta_1$ and  $\beta_2$ for proteins and fullerenes.

\subsection{Persistent homology of cubical complex}
Homology gives a characterization of a manifold, while it does not distinguish different holes in the same dimension. To measure these topological
features,  the concept of persistent homology was proposed based on the simplicial complex.  Persistence measures the birth, death and the lifetime of the topological attributes during the filtration process.

To define the persistent homology, first we need a filtration, i.e., a complex $K$ together with a nested sequence of sub-complexes $\{K^i\}_{0\leq i\leq n}$, such that
$$
\emptyset =K^0\subset K^1\subset\cdots \subset K^n=K.
$$
Each sub-complex $K^i$ in the filtration has an associated chain group $C_k^i$, cycle group  $Z_k^i$ and boundary group $B_k^i$ $\forall i\geq 1$, and thus one  has the following definition \cite{Strombom:2007}.
\begin{definition}
The \textbf{$p$-persistent $k$th homology group} of $K^i$ is:
$$
H_k^{i,p}=Z_k^i/\left(B_k^{i+p}\bigcap Z_k^i\right).
$$
\end{definition}
Here $H_k^{i,p}$ captures the topological features of the filtrated complex that persists for at least $p$ steps in the filtration.


\begin{figure}
\begin{center}
\includegraphics[width=0.4\textwidth]{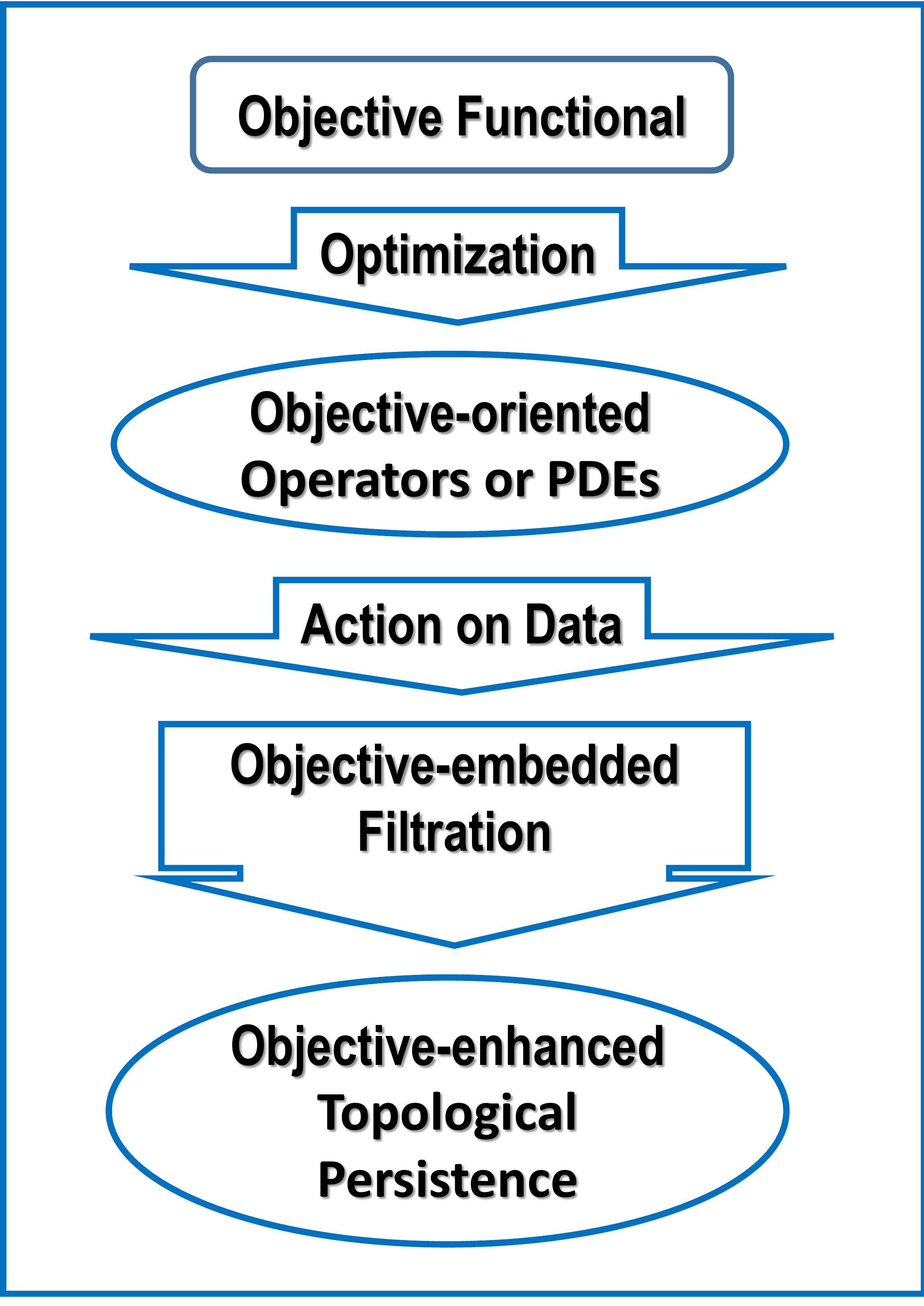}
\end{center}
\caption{ A flow chart for the construction of objective-oriented persistent homology. 
}
\label{Flowchart}
\end{figure}

\section{Objective-oriented persistent homology} \label{comb}

In this section, we propose  a general procedure for constructing objective-oriented persistent homology. We start with an objective functional for the data of interest. By the optimization of the objective  functional, we arrive at  one or a set of objective-oriented  operators, or objective-oriented PDEs. The number of operators  depends on how the objective functional is parametrized.  The action of the objective operators leads to a series of objective-embedded representations of   the original data.   We then utilize such objective-embedded  representations for the filtration of original data to construct objective-oriented persistent homology.  We illustrate this procedure by  a flow chart in Fig. \ref{Flowchart}

As discussed in Section \ref{Laplace-Beltrami}, the minimization of the surface free energy functional gives rise to the mean curvature operator for the biomolecular data. 
We formulate  the Laplace-Beltrami flow to computationally minimize the surface free energy. The integration of the  Laplace-Beltrami flow leads to a family of minimal surface representations of the original data.  In this part, we construct a filtration $\{K_T\}_{T\geq 1}$ of the data of interest based on the Laplace-Beltrami  flow. Here, $T=0,1,2,\cdots,$ is the time steps. For a given initial structure, we embed it in an enlarged bounding box, which defines the whole computational domain. Then  uniform Cartesian mesh is employed for our computation:
$$
\{(i, j, k)|1\leq i\leq n_x, 1\leq j\leq n_y, 1\leq k\leq n_z\}.
$$

The initial values of the grid points that is inside the initial geometric object is set to be $1$, and $0$ for grid points outside the object.

Under the geometric flow action, the following vertex set can be constructed at each evolution time:
$$
V_0:=\emptyset,
$$
$$
V_t:=\{(i, j, k)|S(i, j, k,t)\geq S_0, 1\leq i\leq n_x, 1\leq j\leq n_y, 1\leq k\leq n_z\},\ \  \forall t > 0,
$$
where $S_0$ is the threshold value for extracting the iso-surface.

Furthermore, let $\tilde{V}_T:=\bigcup_{0\leq t\leq T}V_t$, which is the set of vertices that have value greater than the threshold value at time $T$.

The $T$th component of the filtration is set to be:
$$
K_T:=\{\mbox{cubes whose vertices set is a subset of}\  \tilde{V}_T\}.
$$

Based on the above construction, it is obvious that $K_T\subset K_{T+1}, \forall~ T\geq 0$.

\begin{remark}
Neumann boundary condition is utilized to make the Laplace-Beltrami flow computationally well posed. Since the Laplace-Beltrami flow is dispersive, when the evolution time is large enough, the value of $S$ will be less than a given $S_0$ for all the grid points.  Therefore the evolutionary flow based filtration is upper bounded.
\end{remark}

\begin{figure}
\begin{center}
\begin{tabular}{ccc}
\includegraphics[width=0.25\textwidth]{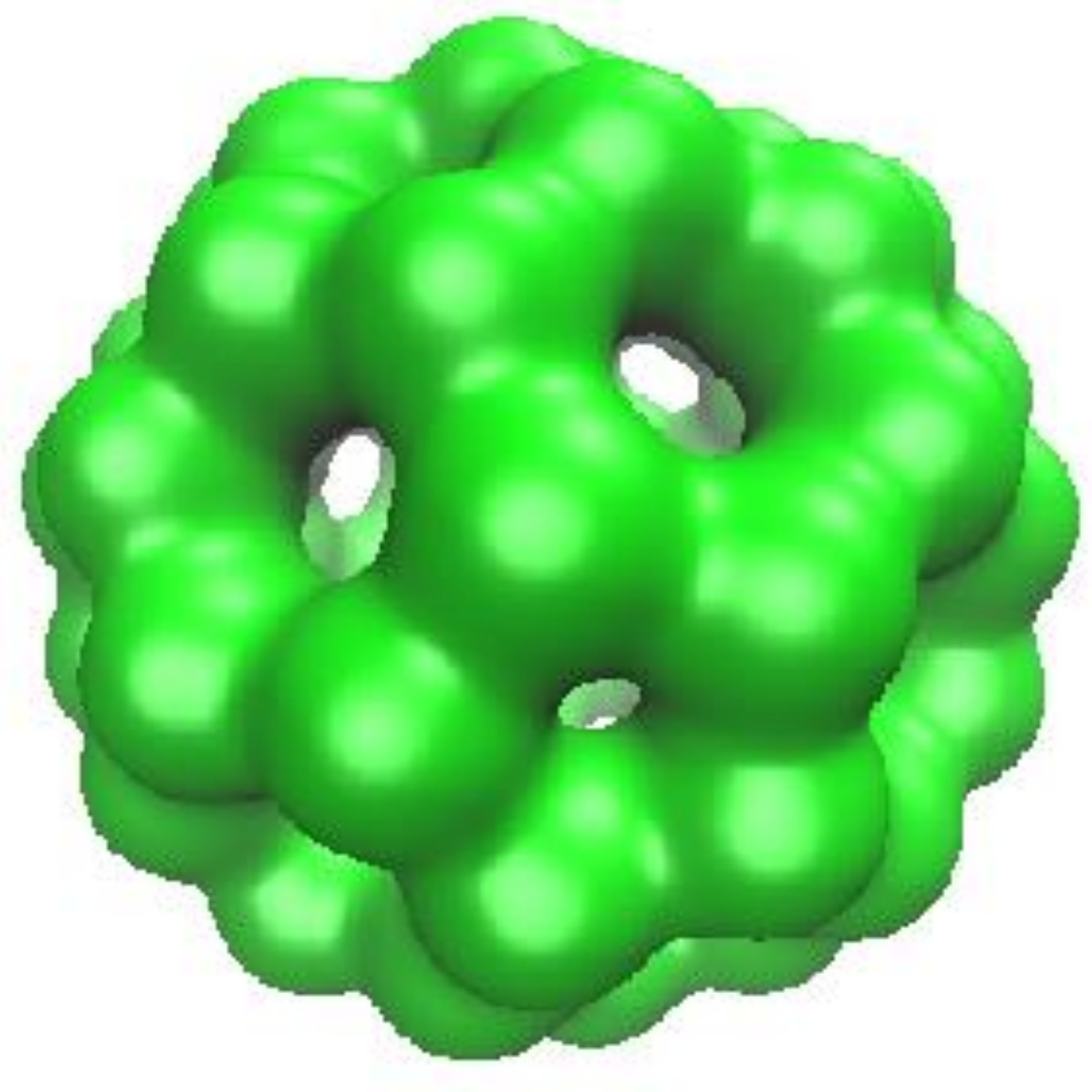}
\includegraphics[width=0.25\textwidth]{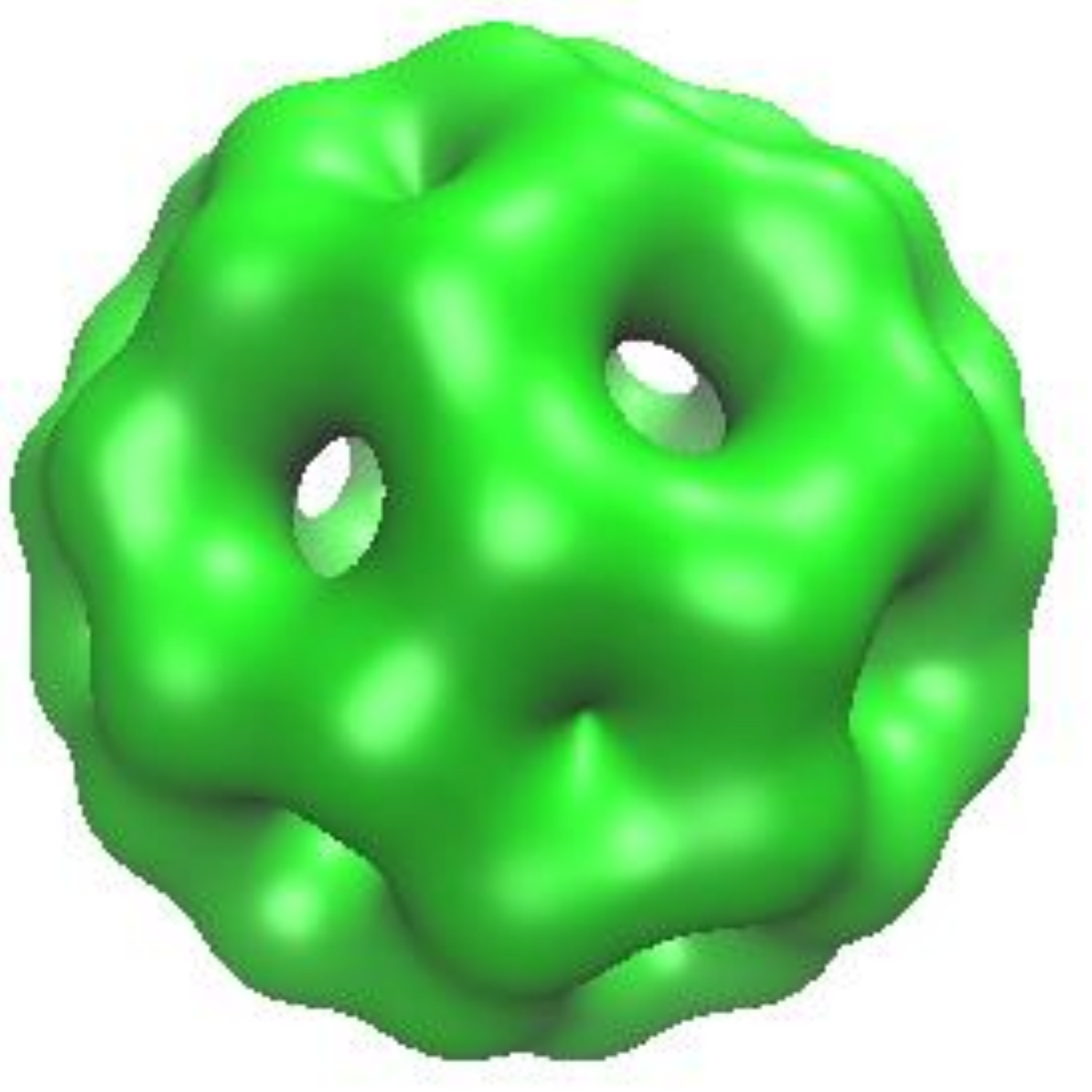}
\includegraphics[width=0.25\textwidth]{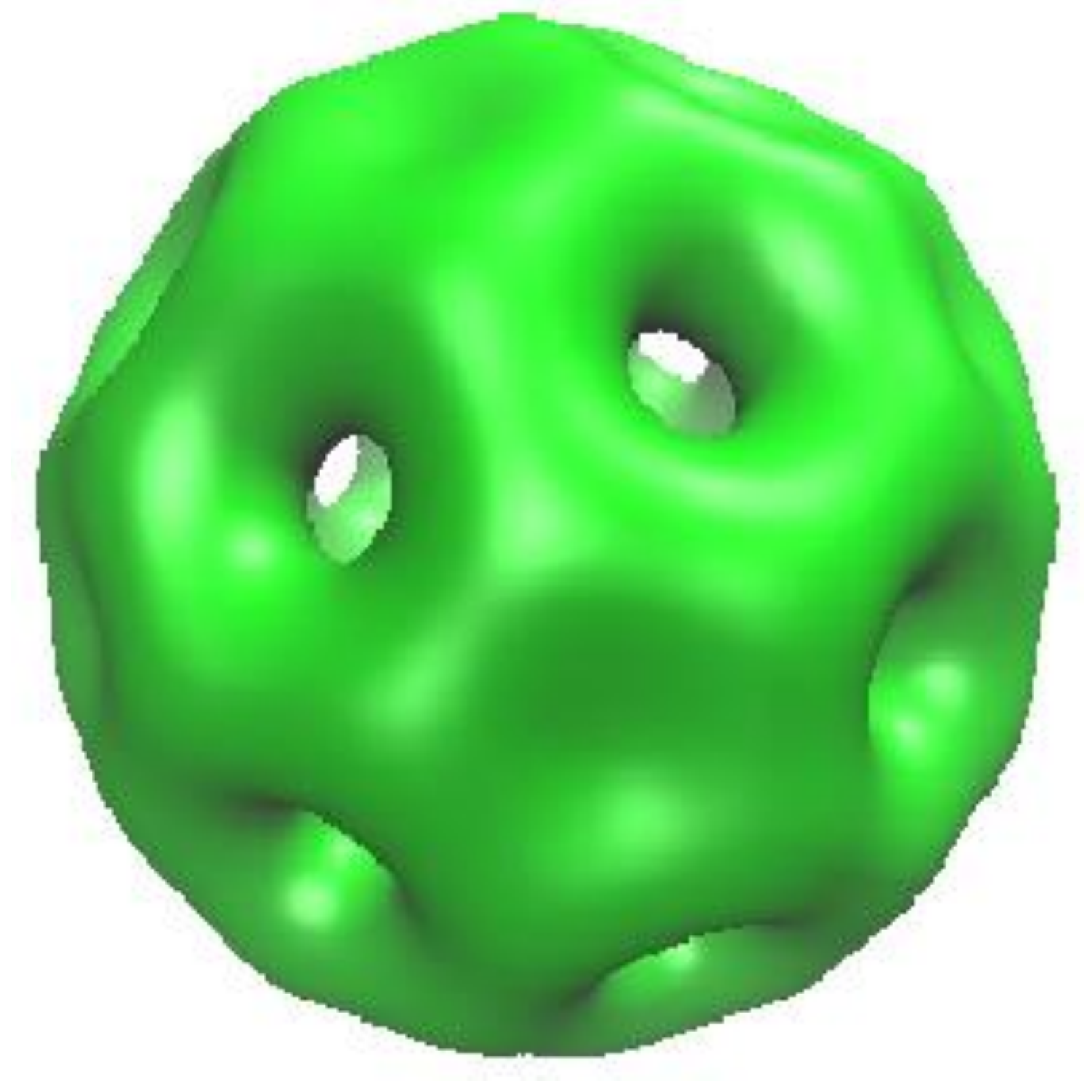}\\
\includegraphics[width=0.25\textwidth]{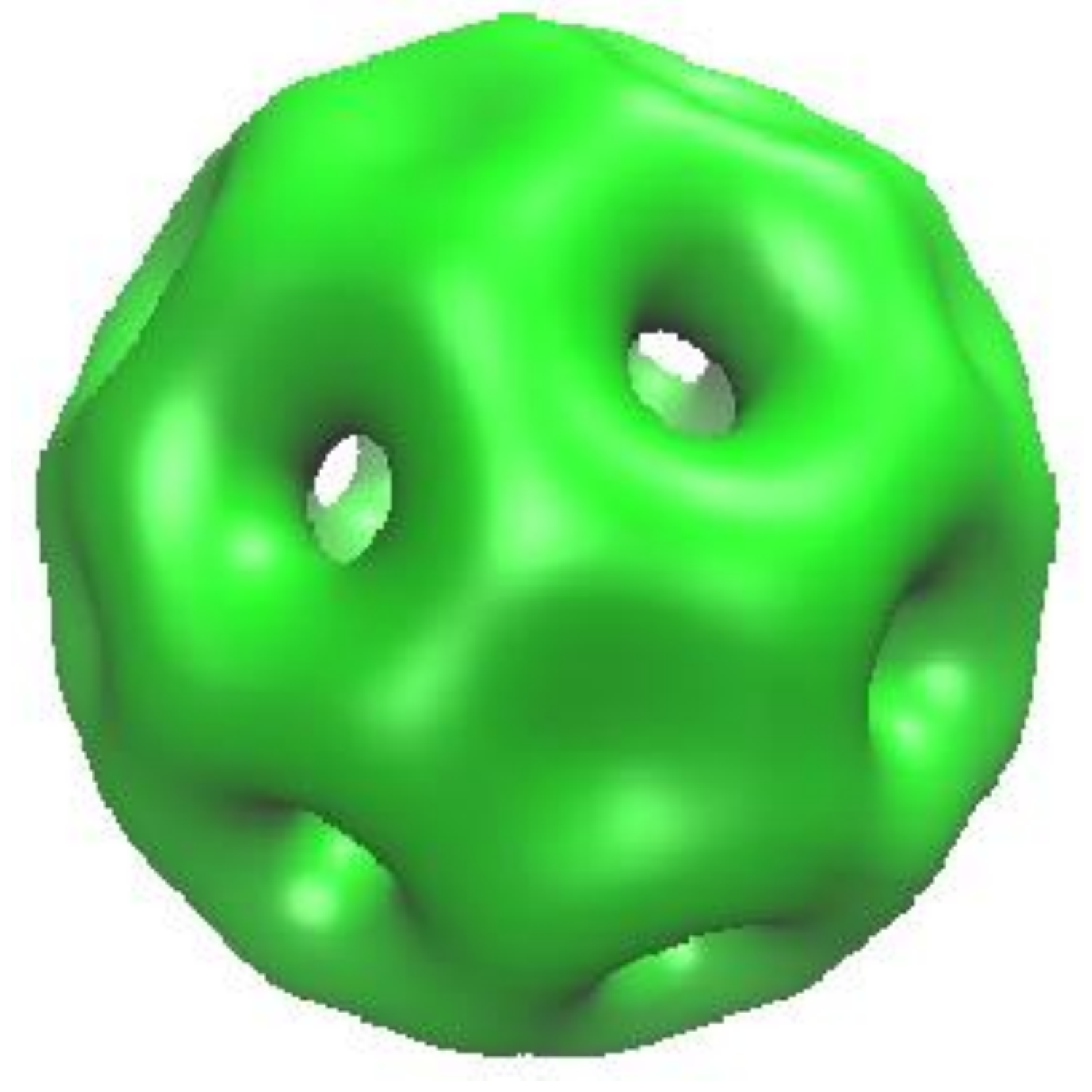}
\includegraphics[width=0.25\textwidth]{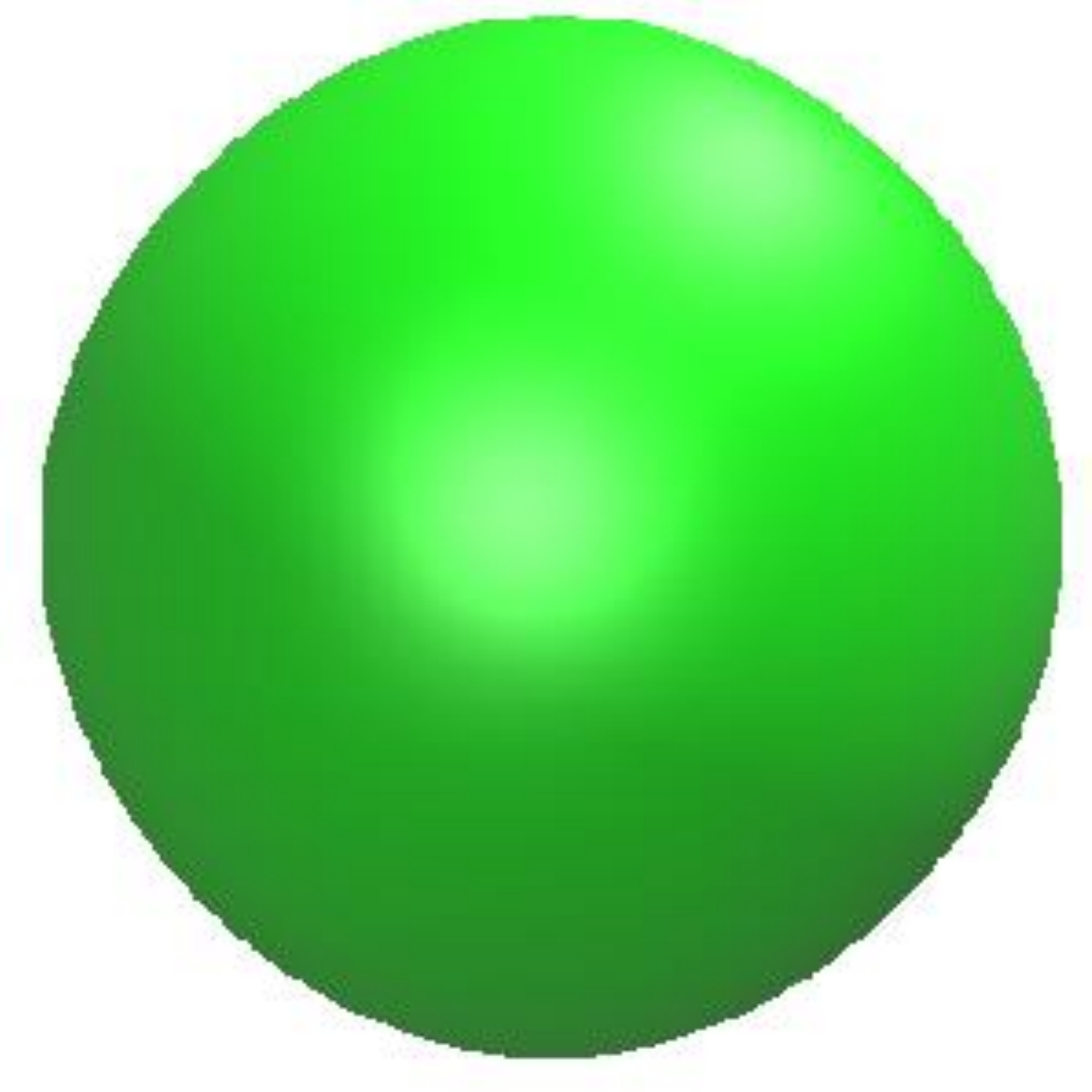}
\includegraphics[width=0.25\textwidth]{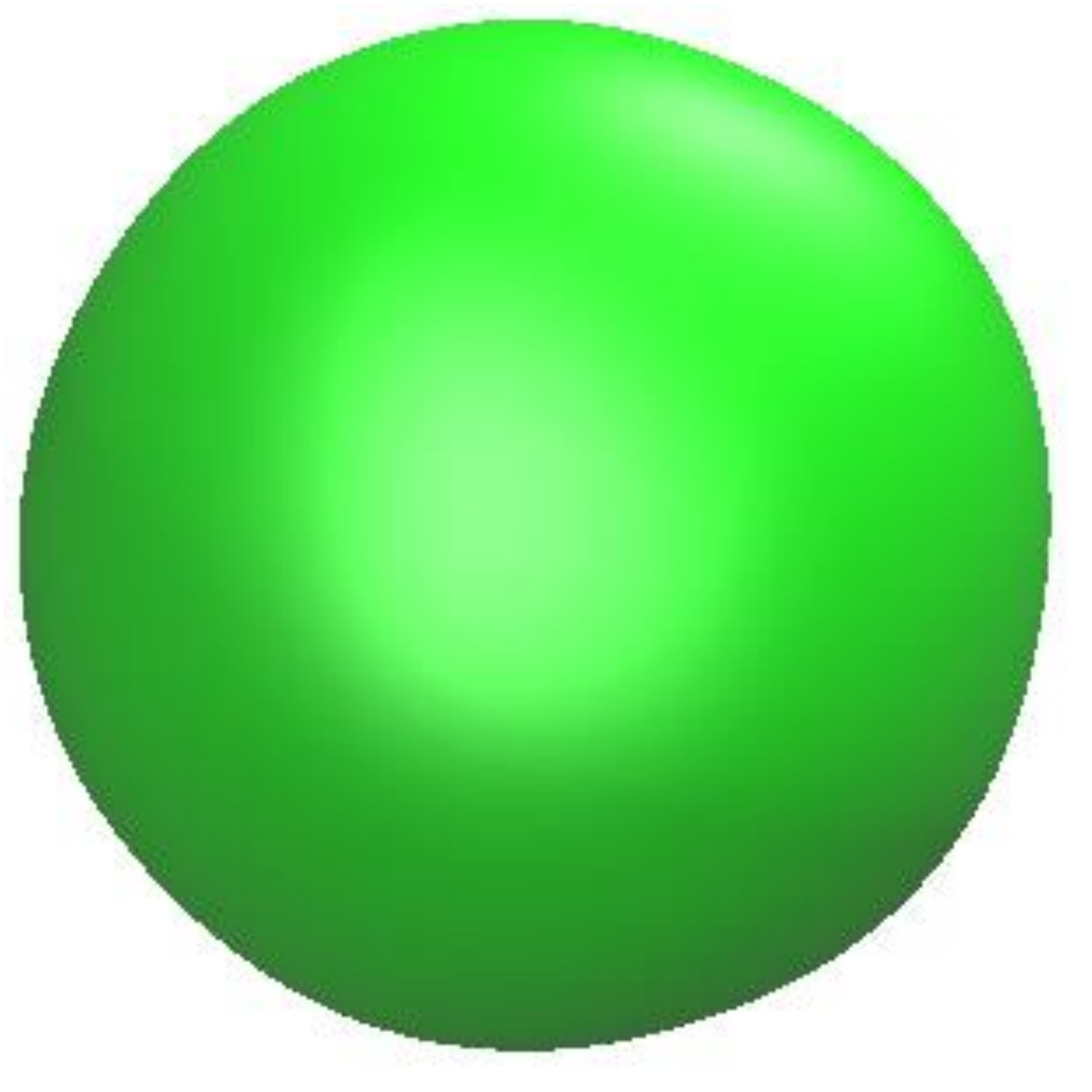}
\end{tabular}
\end{center}
\caption{Selected frames of fullerene C$_{60}$ generated from the  time evolution of the Laplace-Beltrami flow.  Charts from left to right and from top to bottom are frames 1 to 6, respectively.  According to Table  \ref{C60_evo_topo}, the second last frame has a central cavity while the last frame has no void.  
}
\label{evo_c60}
\end{figure}
\subsection{Laplace-Beltrami flow based persistent homology}

\subsection{Computing Laplace-Beltrami flow based persistent homology}

The objective-oriented persistent homology   on the cubical complex can be computed by existing software packages. In the present work, we utilize \href{http://www.sas.upenn.edu/~vnanda/perseus/index.html}{Perseus} \cite{Mischaikow:2013} for persistent homology calculation. The sparse grid data structure is utilized as the input data format for the Perseus software in the present work.

\begin{remark}
Since the Laplace-Beltrami flow minimizes the surface area of the surface defined on the initial data,  
the persistence of   topological features associated with minimal surfaces is enhanced in the Laplace-Beltrami flow based persistent homology approach.
\end{remark}

\section{Validation}\label{Validation}

\subsection{Topological invariant analysis}

In this subsection, we examine accuracy and reliability of  the proposed geometric flow based persistent homology method. To this end, 
we consider a fullerene molecule, C$_{60}$, which has distinct topological  loops, namely pentagon and hexagon loops.
The structural data of fullerene molecules and isomer total curvature energies \cite{Guan:2014} used in our tests are downloaded from the webpage: \href{http://www.nanotube.msu.edu/fullerene/fullerene-isomers.html}{fullerene-isomers}. In these structural data, coordinates of fullerene carbon atoms  and isomer total curvature energies are given.   The atoms of all these molecules form only two types of polygons, namely, pentagons and hexagons. For the fullerene cage composed only pentagons and hexagons, according to Euler Characteristics, the number of pentagons must be $12$ and that for hexagons  is $\frac{N}{2}-10$, where $N$ is the number of atoms of the fullerene molecule.

\begin{table}[!ht]
\centering
\caption{The evolution  of  the topological features of C$_{60}$ molecule under the time evolution of  the Laplace-Beltrami  flow. }
\begin{tabular}{lllll}
\cline{1-5}
Frame &Time   &$\beta_0$  &$\beta_1$  &$\beta_2$ \\
\hline
1 &0.01  &1  &31 &0  \\
2 &0.07  &1  &30 &0  \\
3 &0.15  &1  &19 &0  \\
4 &0.57  &1  &18 &0  \\
5 &0.67  &1  &0  &1  \\
6 &2.31  &1  &0  &0  \\
\hline
\end{tabular}
\label{C60_evo_topo}
\end{table}

Figure \ref{evo_c60} depicts six   frames extracted from the solution of the Laplace-Beltrami equation for   C$_{60}$ fullerene molecule. Note that the initial setting is a set of balls with half van der Waals radii as described in Eq. (\ref{InitiaV1}).   It is seen that during the time evolution, many  pentagonal  rings disappear and followed by the disappearance of hexagonal rings.  Table \ref{C60_evo_topo} gives a summary of topological invariants in these six frames. From this table we notice that pentagons persist in the time interval $[0, 0.15]$ and the hexagon persist in the time interval $[0, 0.67]$.  The difference of the last two frames is that the second last frame has a cavity, whereas the last frame has no cavity.

\begin{figure}
\begin{center}
\begin{tabular}{ccc}
\includegraphics[width=0.25\textwidth]{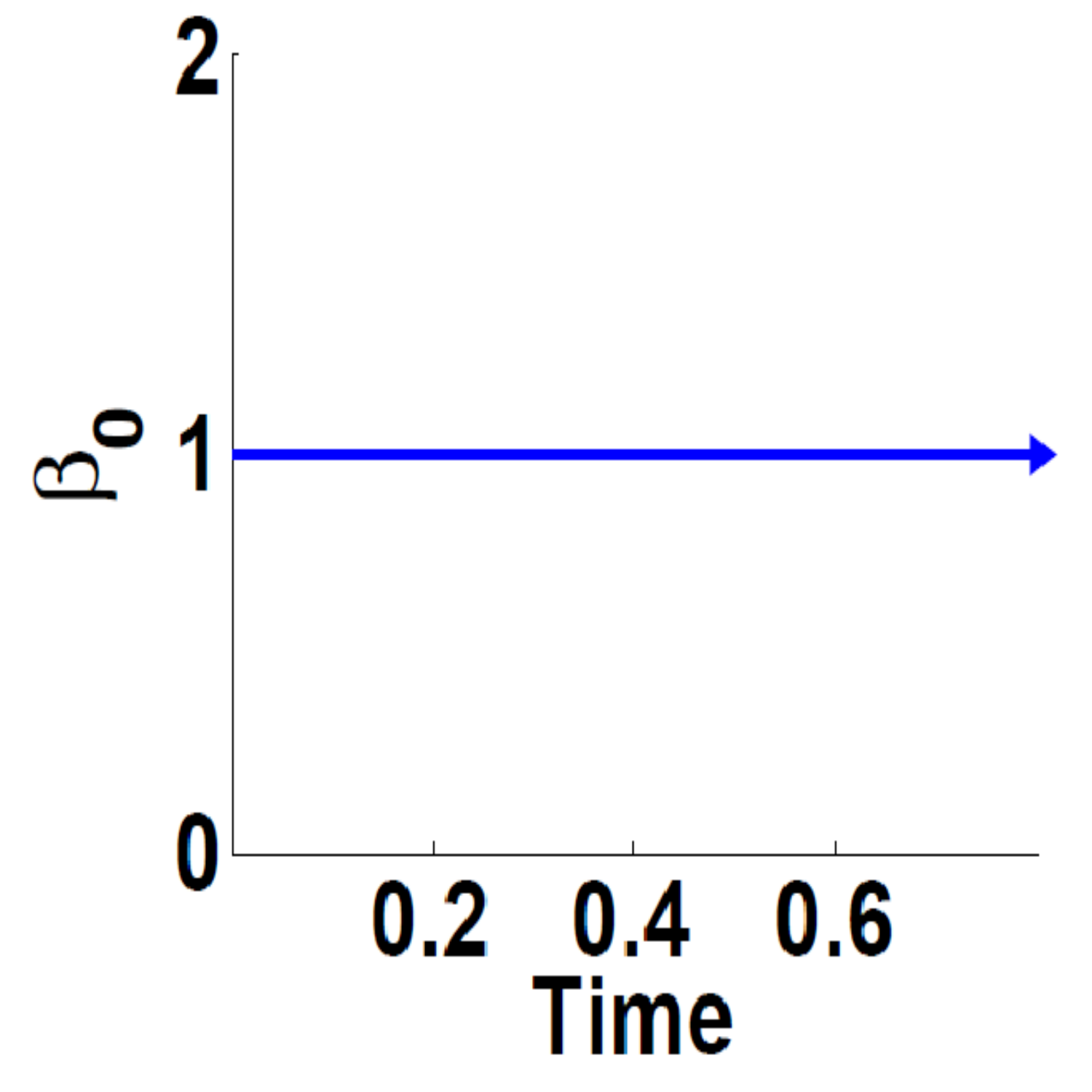}
\includegraphics[width=0.25\textwidth]{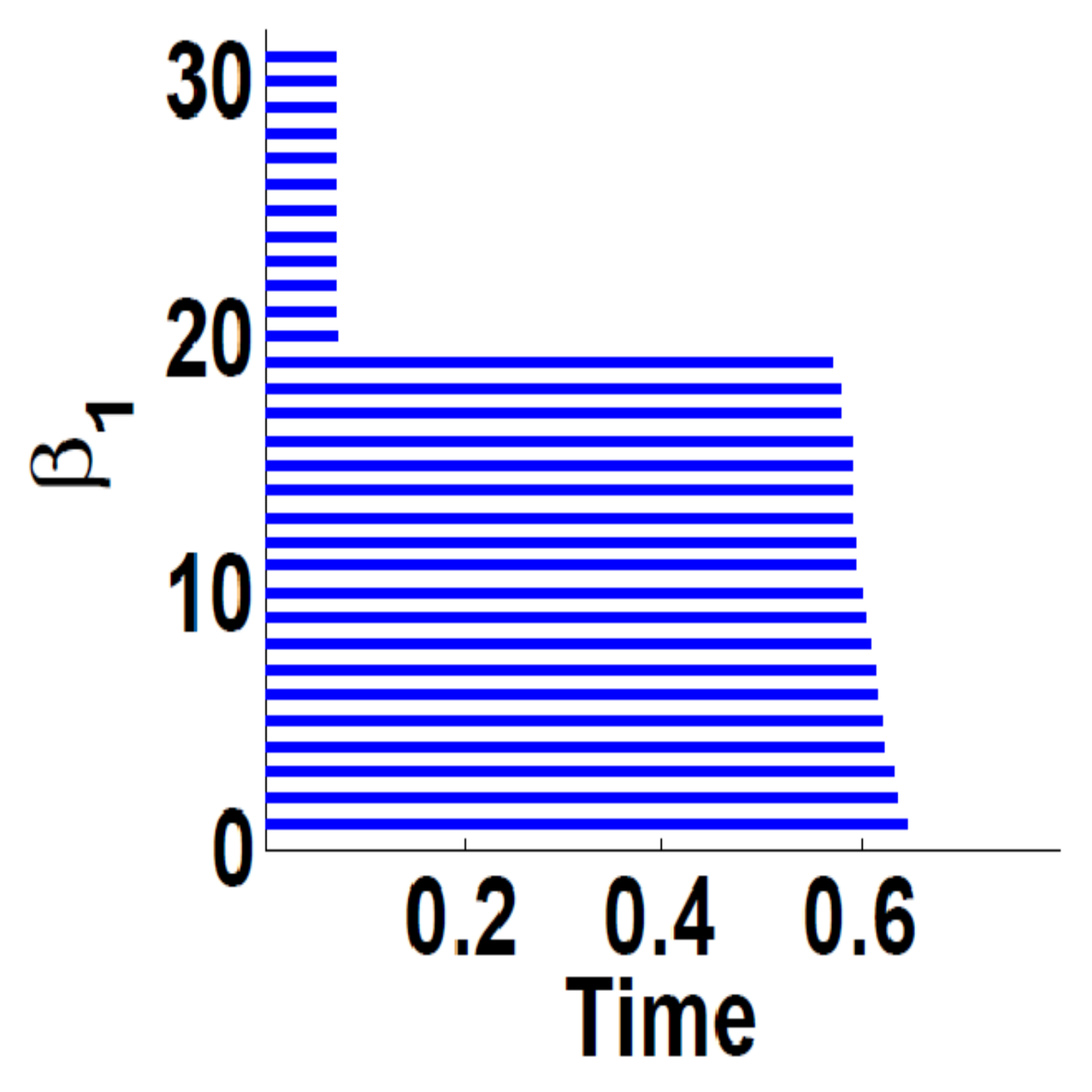}
\includegraphics[width=0.25\textwidth]{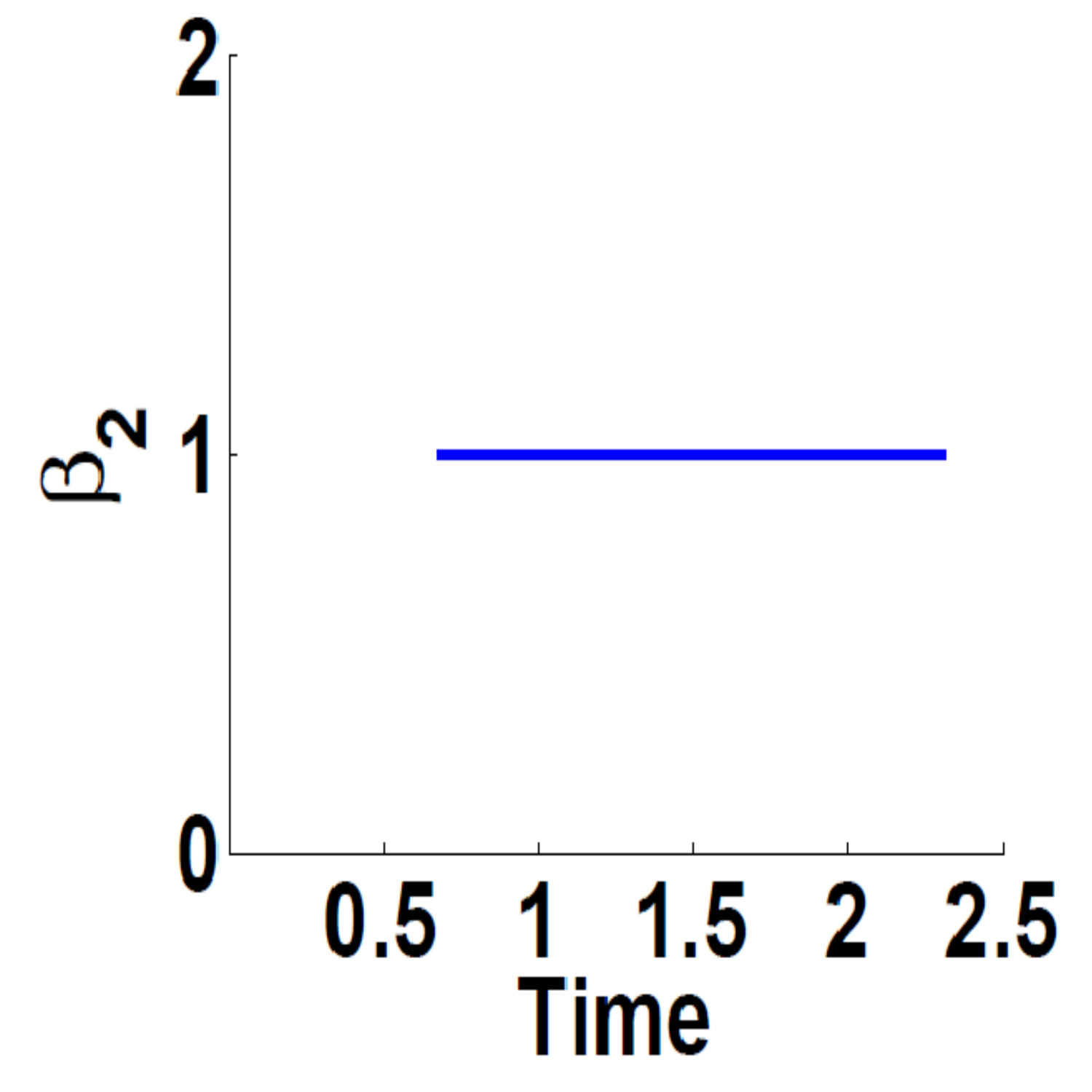}\\
\includegraphics[width=0.25\textwidth]{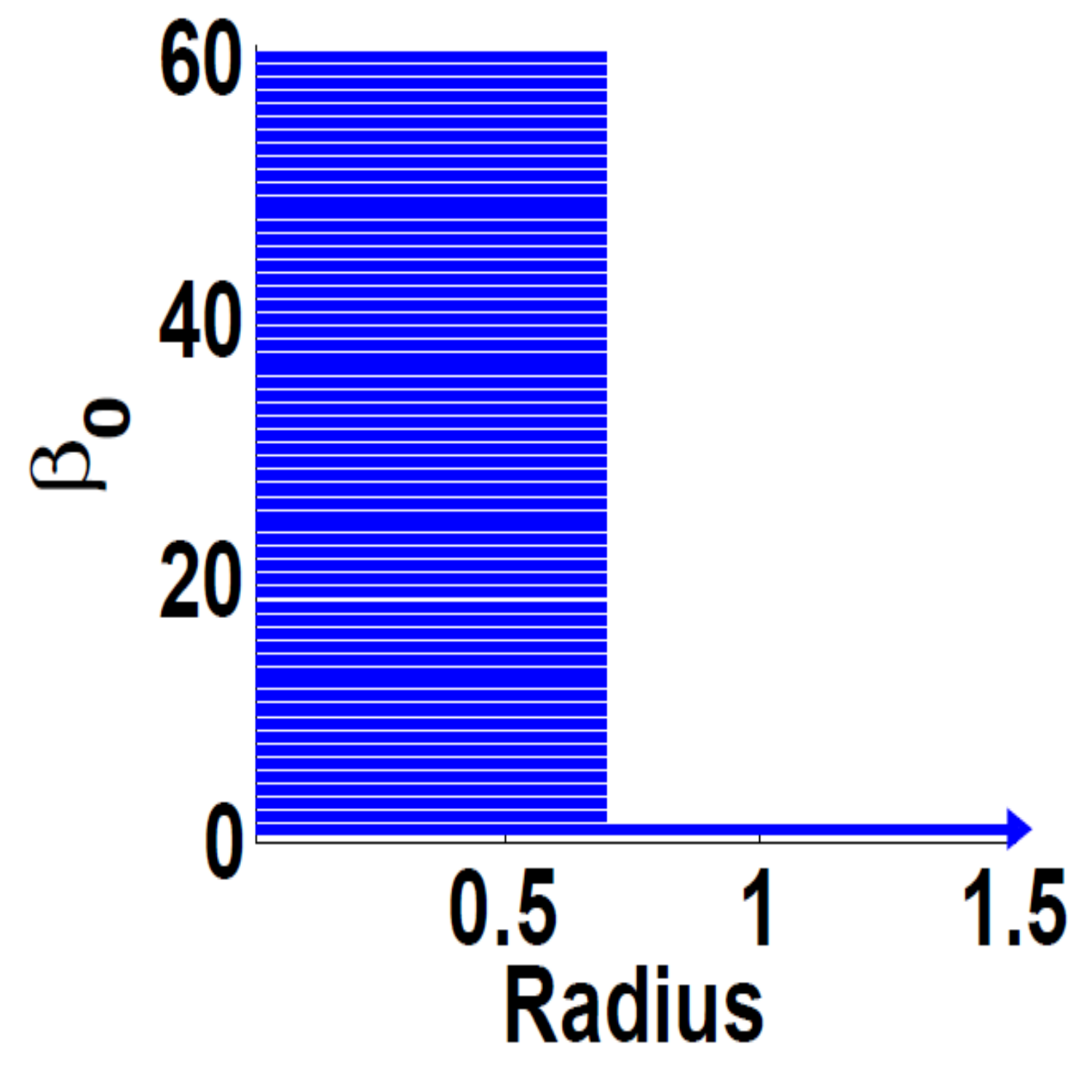}
\includegraphics[width=0.25\textwidth]{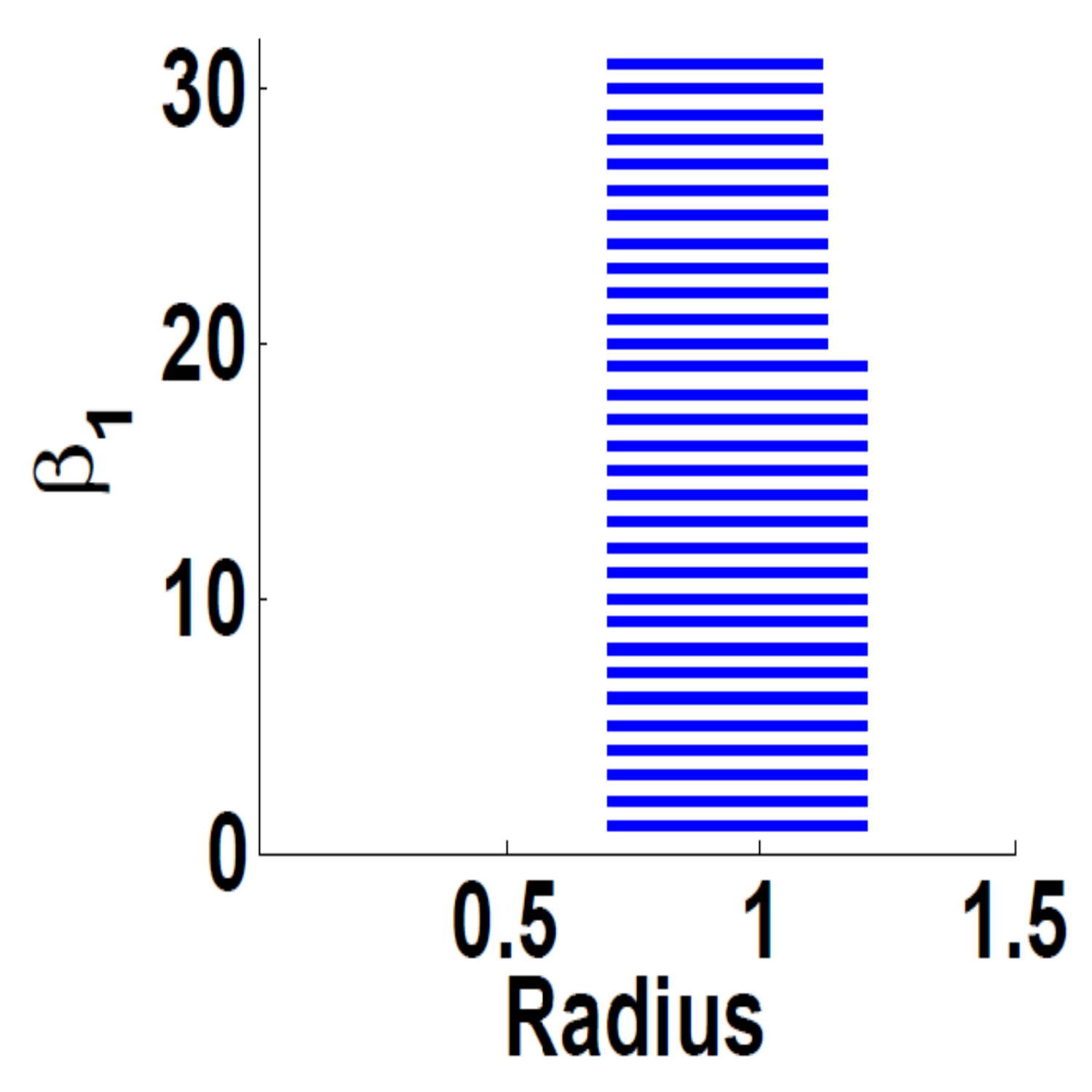}
\includegraphics[width=0.25\textwidth]{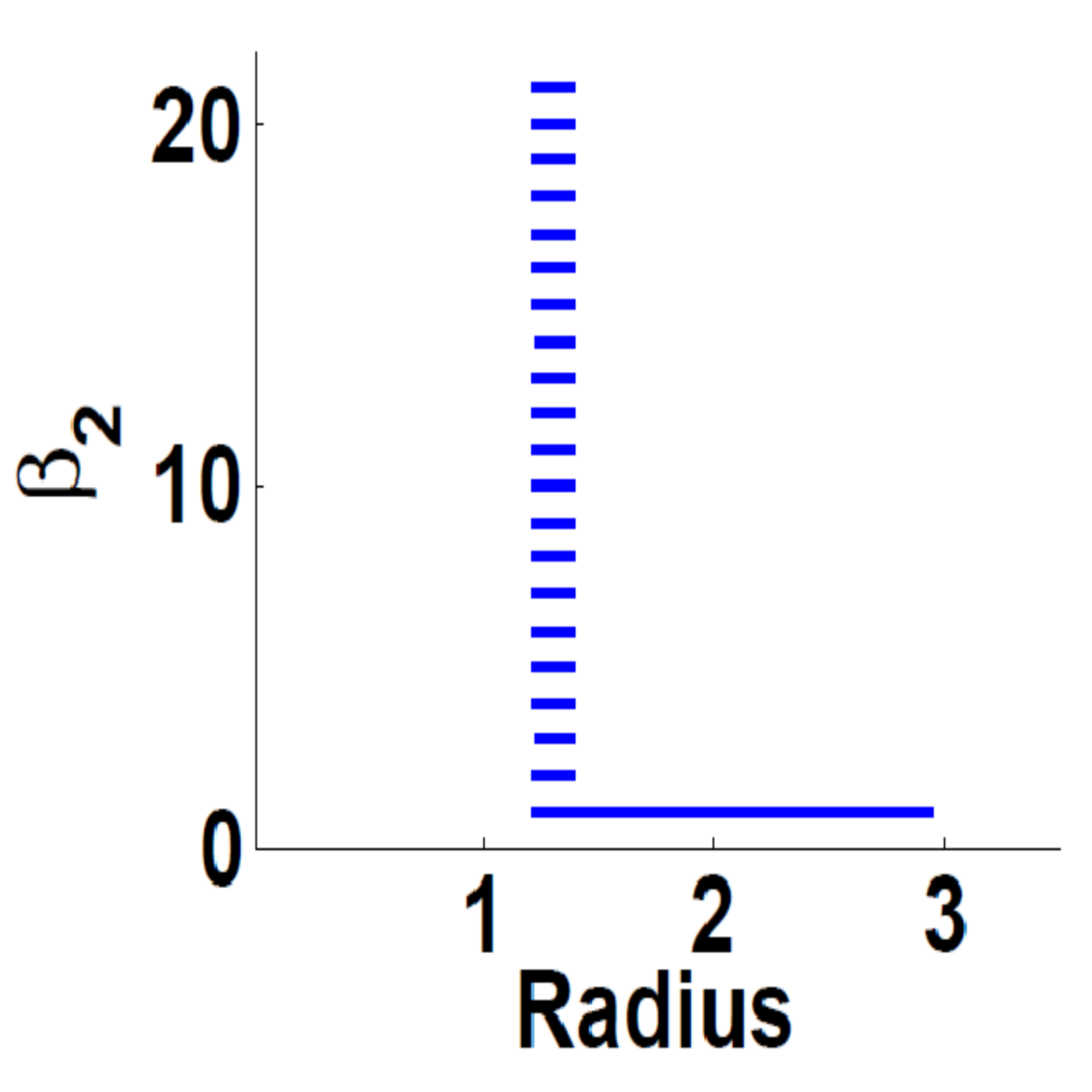}
\end{tabular}
\end{center}
\caption{Comparison of the topological evolution  and persistence  of  the  C$_{60}$ molecule.
Top row: Barcodes obtained from the proposed Laplace-Beltrami flow based filtration;
Bottom row: Barcodes obtained from the Rips complex filtration.
}
\label{c60_be}
\end{figure}

The  evolution of the topological features of  carbon fullerene C$_{60}$ under the geometric flow  is demonstrated in Fig. \ref{c60_be}. As a comparison, we also plot the result generated by using the Rips complex. In  $\beta_0$ panels, one sees a long-lasting bar from the present method, while a reduction from 60 bars to one bar in the Rips complex representation. This behavior is expected because the starting point of the present method is a set of connected balls as described above, while Rips complex filtration starts from the zero radius. In the $\beta_1$ panels, there is a good consistence between two approaches. One sees 12 short-lived bars, which correspond to 12     pentagonal  rings. However, there are only 19 relatively long bars for 20 hexagonal rings  because one of $H_1$ element can be expressed as the combination of other $H_1$ basis elements. Finally, in  $\beta_2$ panels, the present method provides a single  relatively long-lived bar for the inner cavity, while the Rips complex filtration gives rise to additional 20 short-lived bars for 20 hexagons. The disappearance of the short-lived $\beta_2$ bars in the present approach is due to the cubical complex  used in our calculation. Short-lived bars are often regarded as topological noise in  the literature, while used in our models for physical modeling \cite{KLXia:2014c}.  However, in the present work, we only need the long-lasting $\beta_2$ bar for our quantitative modeling as discussed in Section \ref{FullereneTotalCurvature}.

\begin{remark}
The persistent homology derived from the  Laplace-Beltrami flow   results in   nonlinear modification of certain topological features. Because the geometric PDE is able to preserve certain geometric features \cite{Wei:1999}, the persistence of the corresponding intrinsic topology can be amplified.  This feature is a fundamental property of the  objective-oriented persistent homology  constructed in this work. It is possible to design objective-oriented PDEs to selectively  enhance and/or extract other desirable topological features from big data.
\end{remark}

\subsection{Convergence analysis}

\begin{figure}
\begin{center}
\begin{tabular}{ccc}
\includegraphics[width=0.4\textwidth]{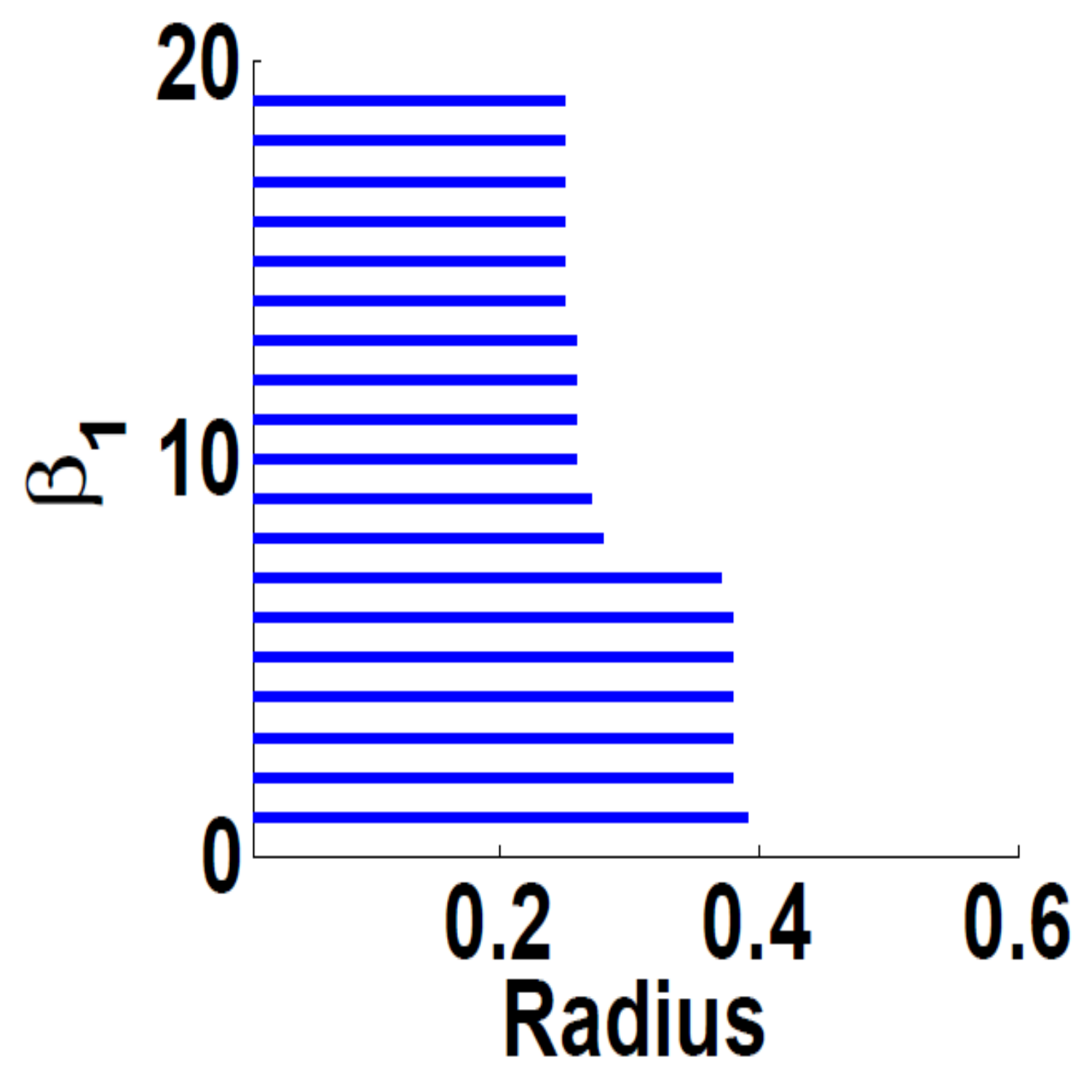}
\includegraphics[width=0.4\textwidth]{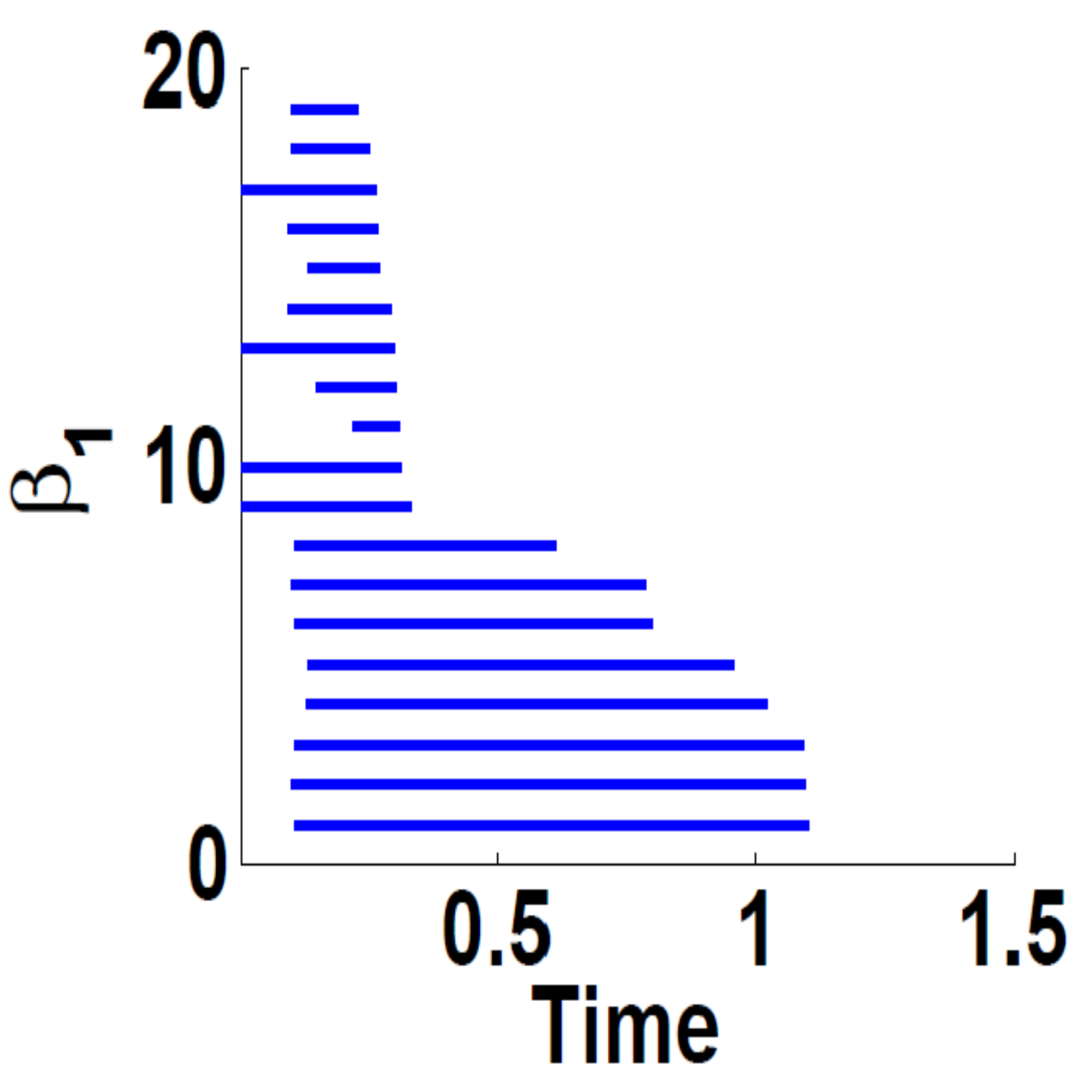}\\
\includegraphics[width=0.4\textwidth]{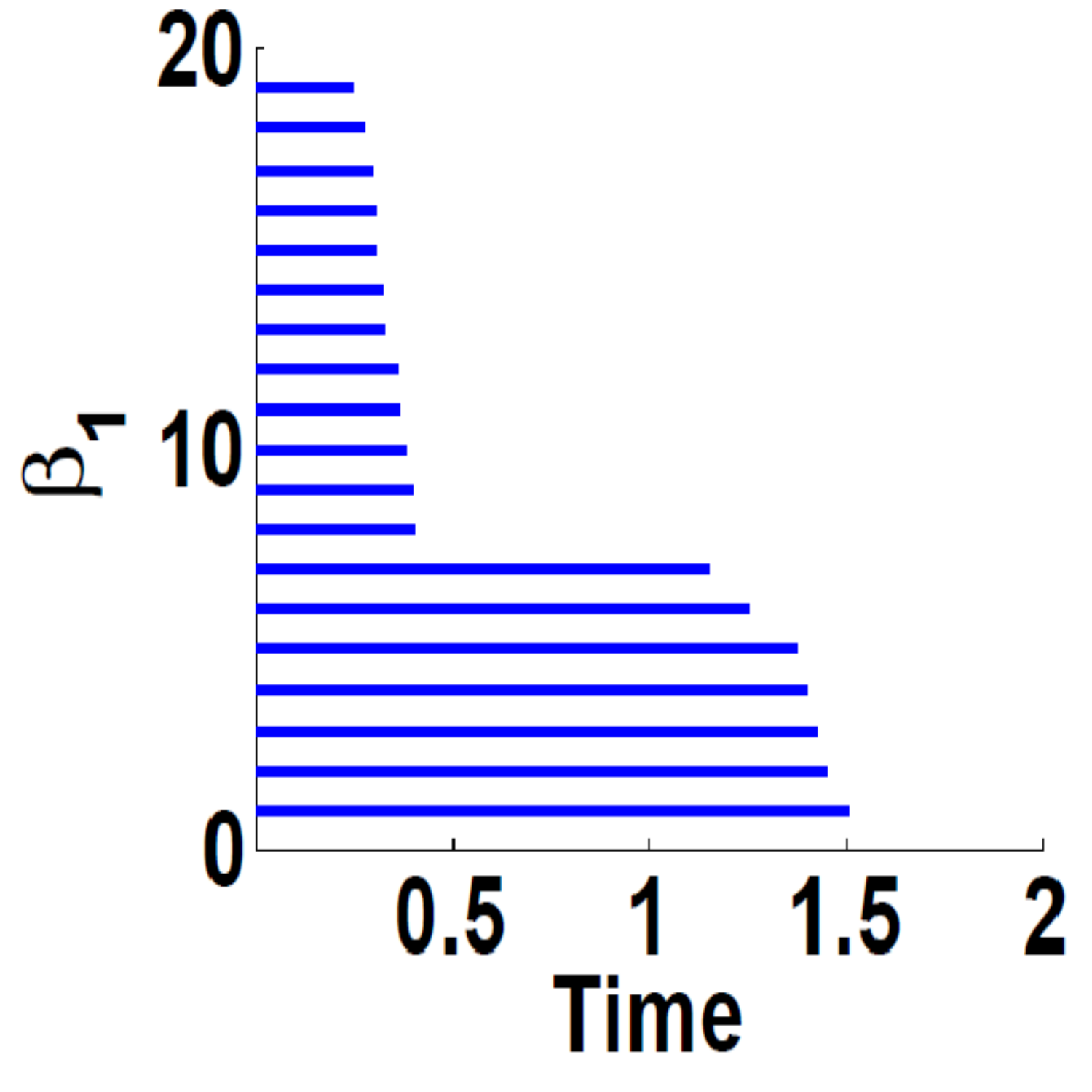}
\includegraphics[width=0.4\textwidth]{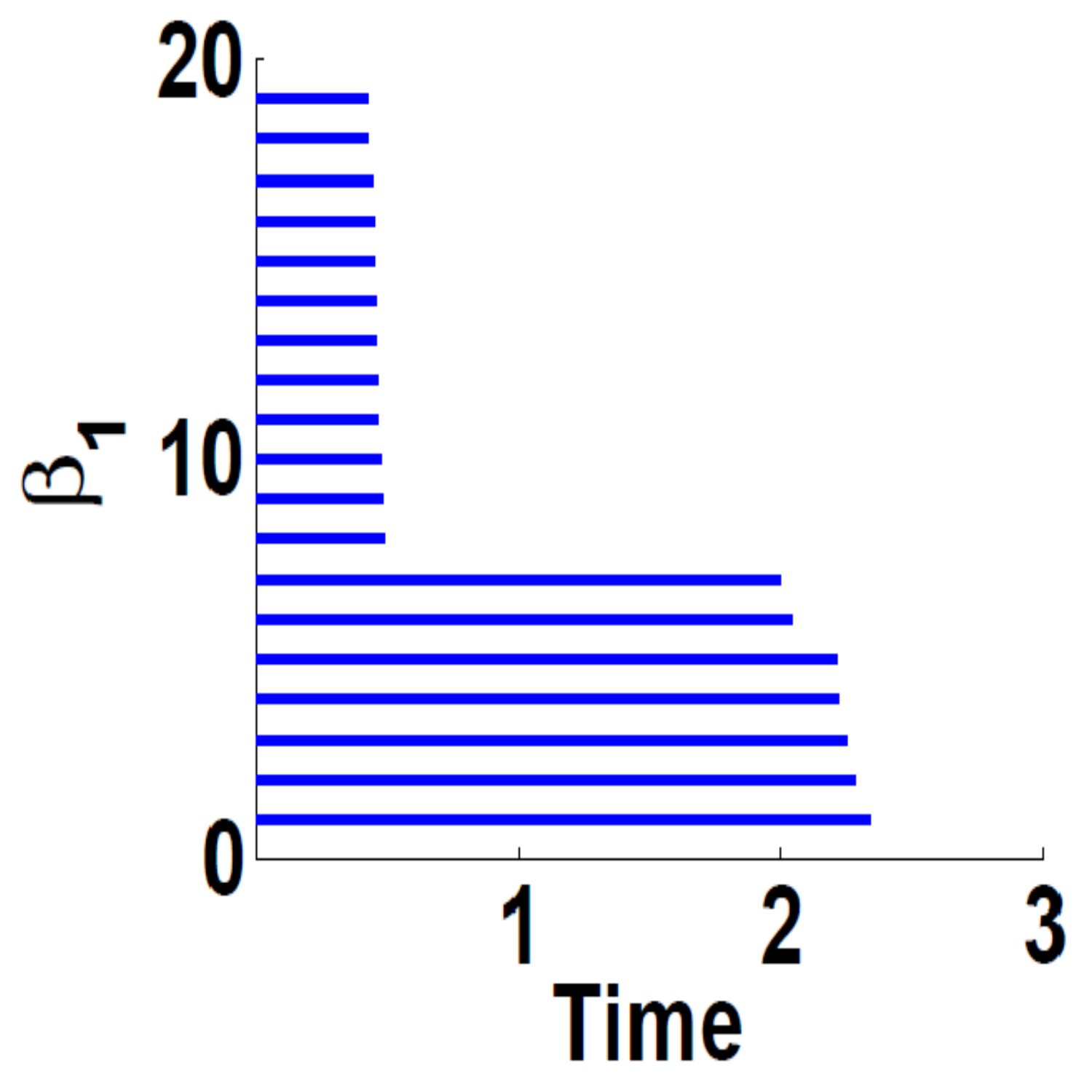}
\end{tabular}
\end{center}
\caption{Comparison of the persistence of $\beta_1$ barcodes obtained from the growth of atomic radius filtration and from the geometric flow based filtration  for  fullerene  C$_{36}$.
Top left: Atomic radius filtration;
Top right: Geometric flow filtration, $h=0.5$\AA;
Bottom left:  Geometric flow filtration, $h=0.25$\AA;
Bottom right:  Geometric flow filtration, $h=0.125$\AA.
}
\label{refine_compare1}
\end{figure}
\begin{figure}
\begin{center}
\begin{tabular}{ccc}
\includegraphics[width=0.4\textwidth]{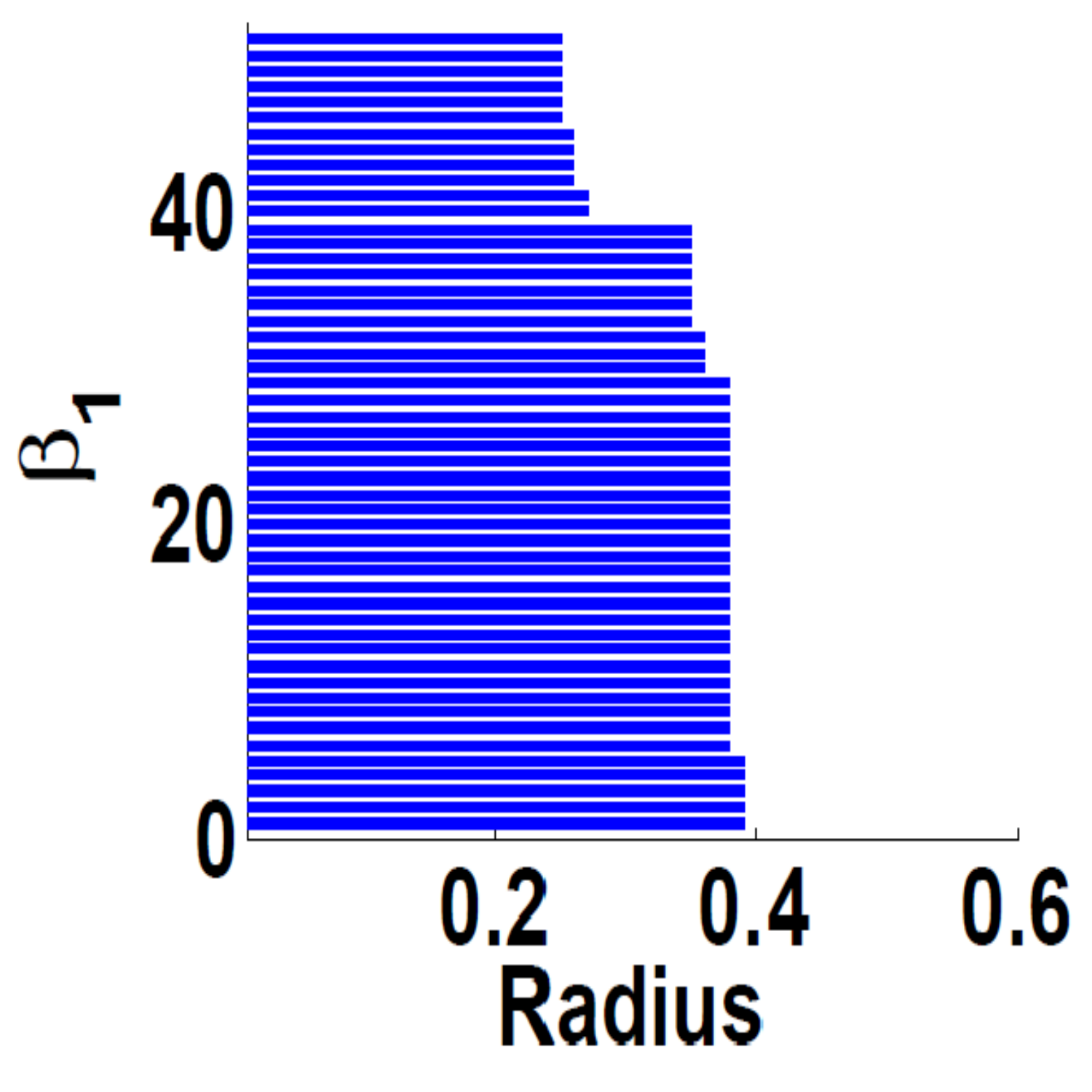}
\includegraphics[width=0.4\textwidth]{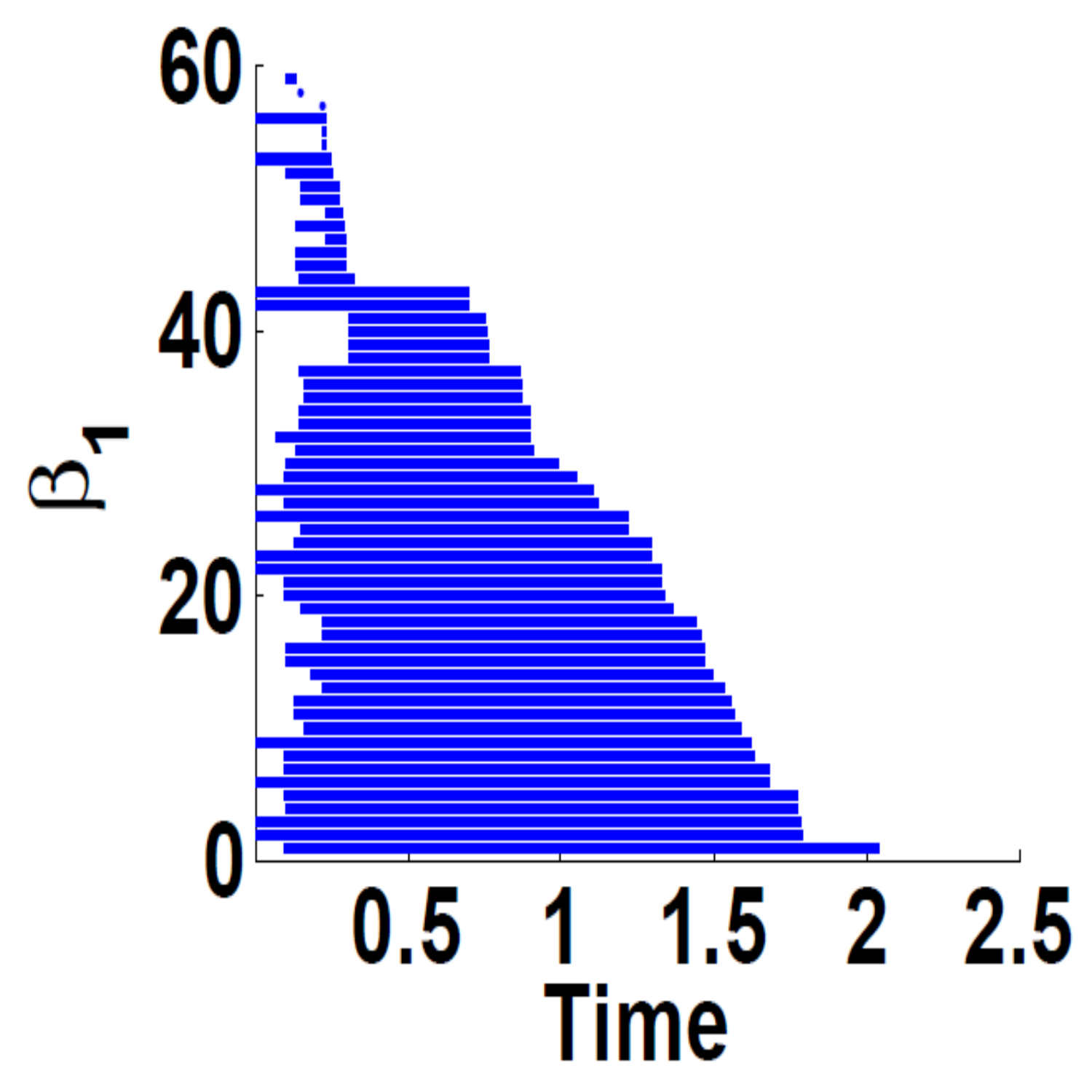}\\
\includegraphics[width=0.4\textwidth]{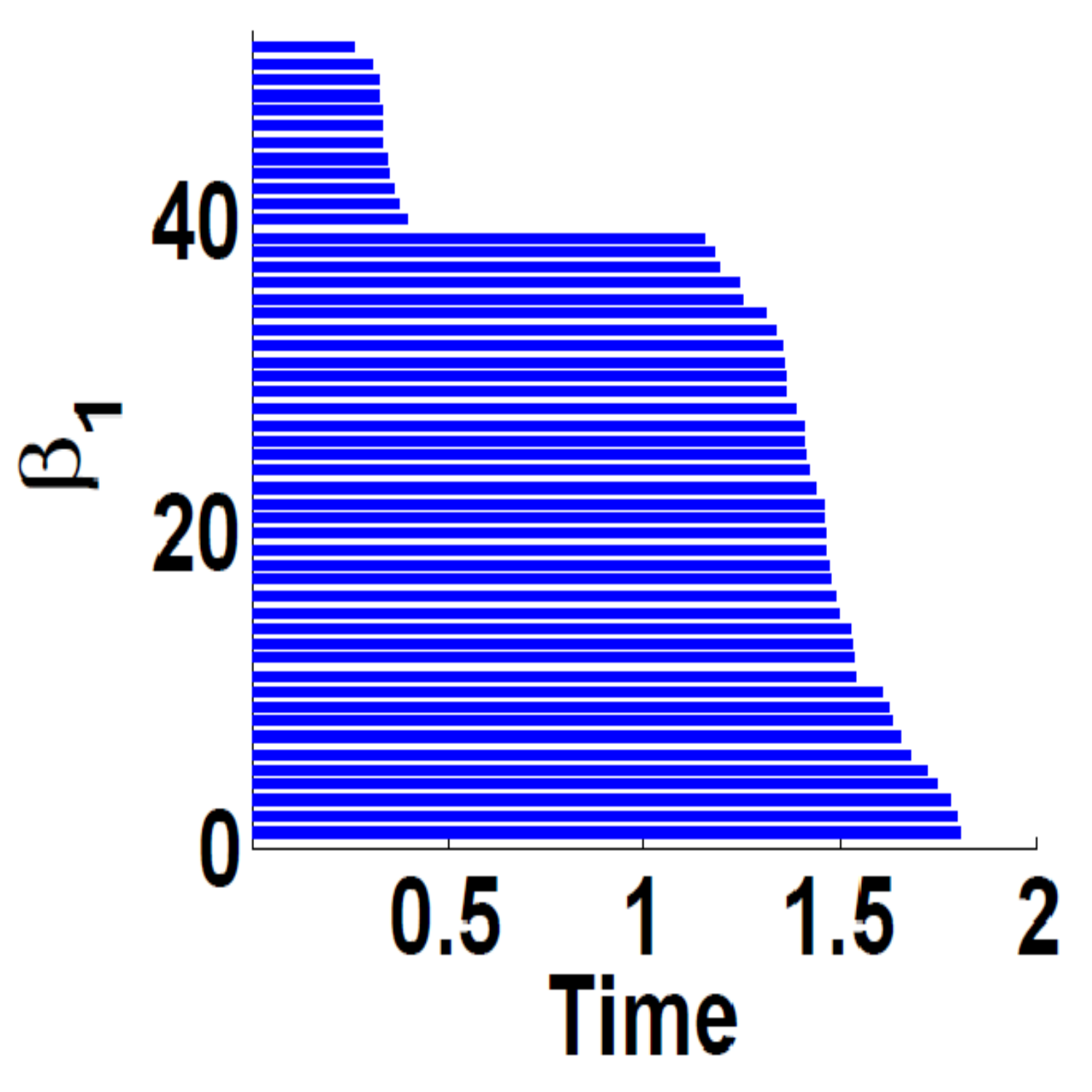}
\includegraphics[width=0.4\textwidth]{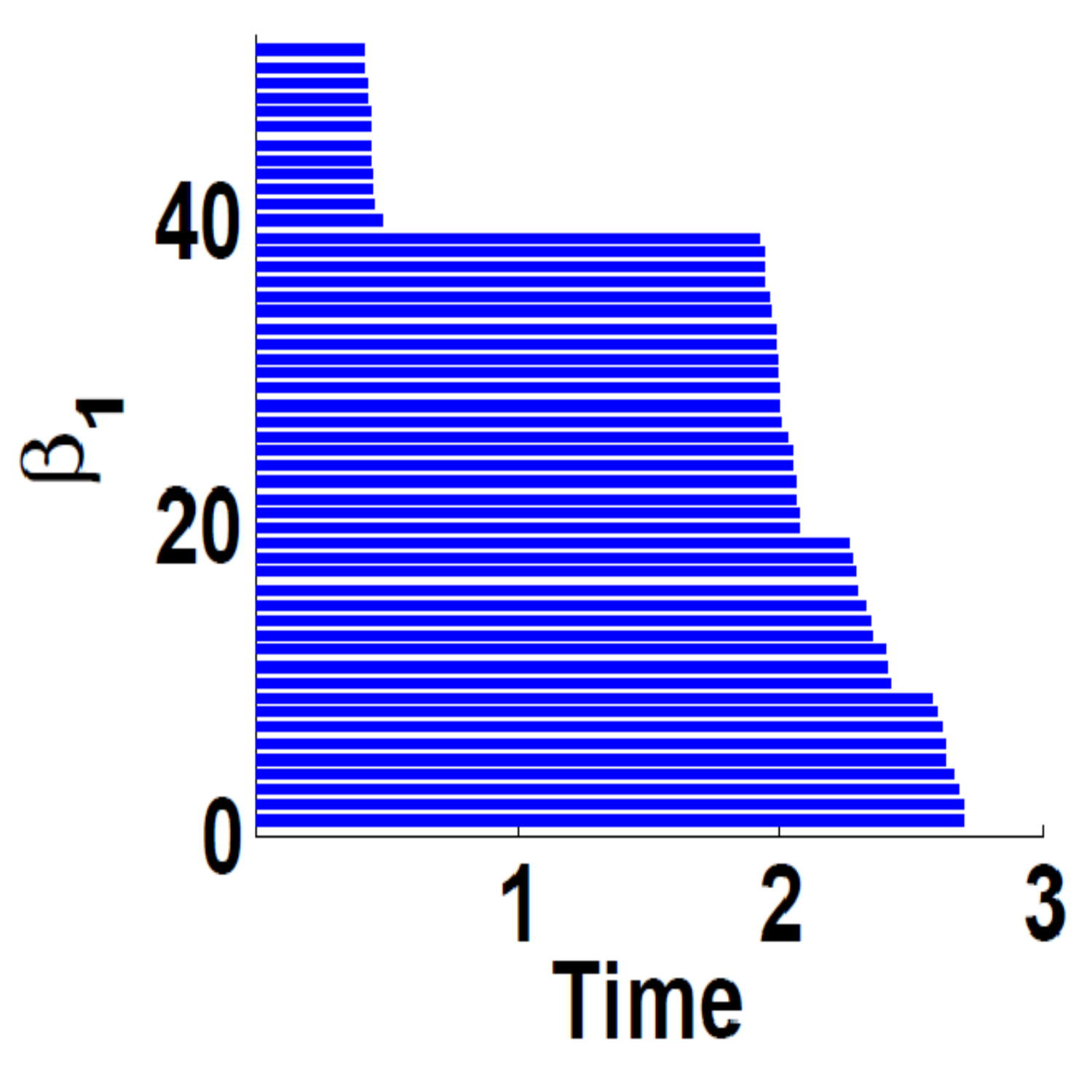}
\end{tabular}
\end{center}
\caption{Comparison of the persistence of $\beta_1$ barcodes obtained from the growth of atomic radius filtration and from the geometric flow based filtration  for  fullerene  C$_{100}$.
Top left: Atomic radius filtration;
Top right: Geometric flow filtration, $h=0.5$\AA;
Bottom left:  Geometric flow filtration, $h=0.25$\AA;
Bottom right:  Geometric flow filtration, $h=0.125$\AA.
}
\label{refine_compare10}
\end{figure}

Figures \ref{refine_compare1} and \ref{refine_compare10} demonstrate the numerical convergence of of proposed Laplace-Beltrami flow approach for computing the persistence of $\beta_1$ invariants.  We present the time evolution of the persistence of $\beta_1$ invariants collected over a sufficiently long period at different grid sizes.  It can be seen that the persistent pattern at grid size $0.25$\AA~ is essentially the same as that at grid size $0.125$\AA, which shows the convergence with respect to grid spacing variations.

As another validation of the proposed Laplace-Beltrami flow based  persistent homology method, we examine the numerical convergence of the proposed method.  Additionally, we demonstrate that topological invariants computed from our Laplace-Beltrami flow method 
converge to the right ones, where we regard the $\beta_1$ barcodes obtained via the conventional  Rips complex  filtration based on the growth of the radius of the point cloud data as the benchmark. To this end, we consider the persistent homology of the two approaches for two fullerene structures, namely, C$_{36}$ 
and  C$_{100}$.
The coordinates of these fullerene structures are downloaded from Web
\href{http://www.nanotube.msu.edu/fullerene/fullerene-isomers.html}{fullerene-isomers} and are saved. For isomers, the first structure in the isomer family is used. These fullerene molecules contain pentagon and hexagon loops, which give rise to appropriate $\beta_1$ bars.

It remains to show that our persistent homology results converge to the right ones. As shown in Figures \ref{refine_compare1} and \ref{refine_compare10}, there are a total  of 12 pentagon $\beta_1$ bars. The numbers of hexagon bars are 7 and 39, respectively for C$_{36}$ and C$_{100}$, as expected. Therefore, the proposed geometric flow based filtration  captures the intrinsic topological features of fullerenes. Additionally,  the Rips complex based filtration is employed as a reference with a fine atomic radius growth rate of 0.001\AA~ per step.  The comparison of topological invariants computed from the proposed method and that obtained from the Rips complex is given in Figures \ref{refine_compare1}-\ref{refine_compare10}. Clearly,  persistent patterns obtained by Laplace-Beltrami flow based method capture all topological features   generated from the Rips complex, which indicates the reliability of the proposed method.

In fact, we have carried out similar tests for many other fullerenes, including C$_{38}$,  C$_{40}$,  C$_{44}$, C$_{52}$,  C$_{84}$, C$_{86}$,  C$_{90}$ and  C$_{92}$. Although these results are omitted for simplicity, our findings are the same.

The above validations verify that the Laplace-Beltrami flow based filtration in conjugation with the cubical complex setting is convergent and accurate. The resulting  topological invariants are consistent with those obtained with the  Rips complex using  radius based filtration.  On the other hand, our results also indicate that the Laplace-Beltrami flow based method is very sensitive to grid resolution. Some topological features barely show up at the grid size of 0.5\AA. Therefore, the grid resolution better than 0.25\AA~ is recommended for nano-bio data.


\section{Application}\label{Application}
Having verified the reliability, accuracy  and efficiency of the present Laplace-Beltrami flow based persistent homology analysis, we utilize it for the study of proteins and nano-material in this section.

\subsection{Protein structure analysis}
\subsubsection{Protein 2GR8}
In this subsection, we explore the topological structures and their persistence of the protein molecules using the Laplace-Beltrami flow based  persistent homology.
 We consider a  beta-barrel protein (PDB ID: 2GR8).

\begin{figure}
\begin{center}
\begin{tabular}{cc}
\includegraphics[width=0.34\textwidth]{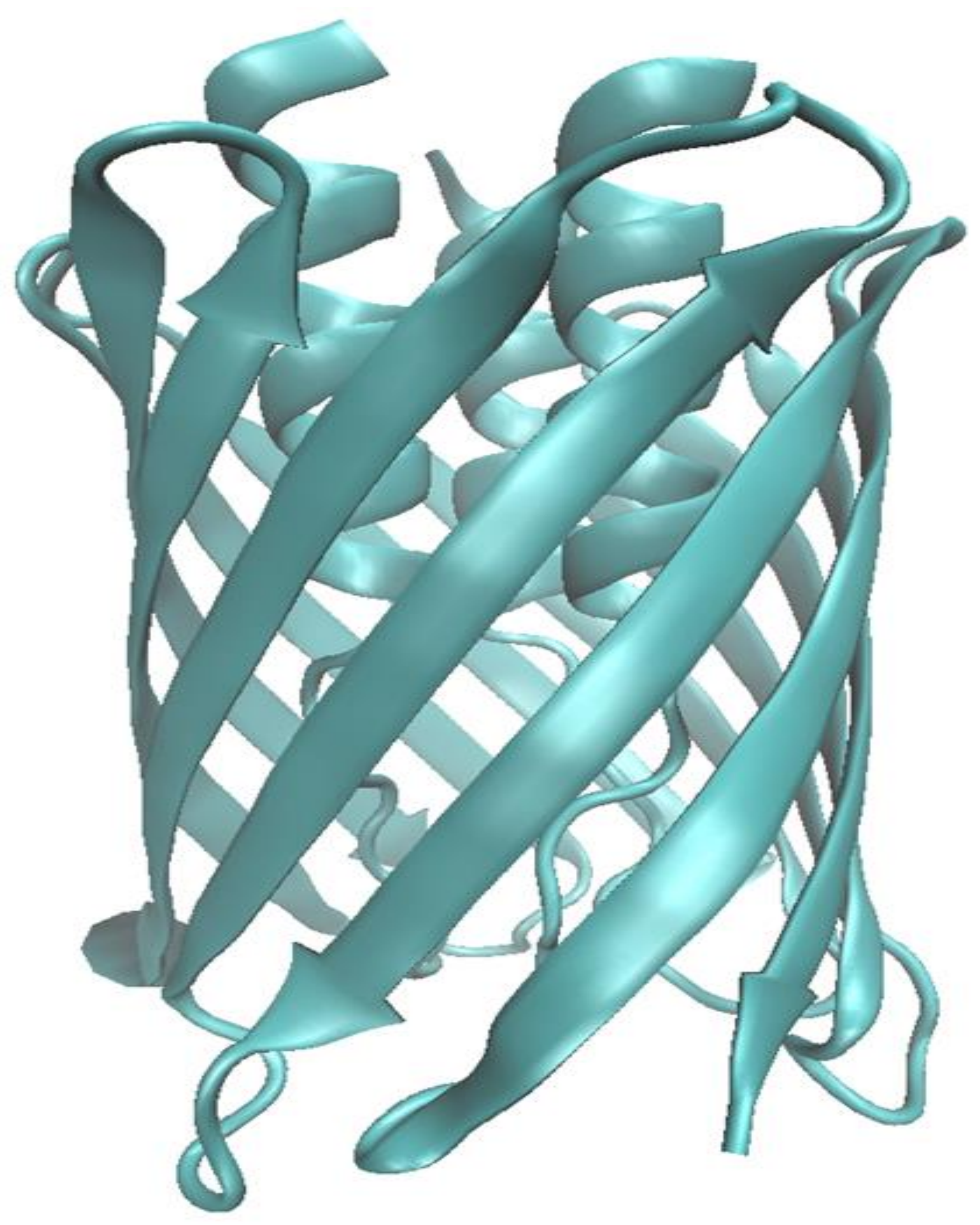}
\includegraphics[width=0.34\textwidth]{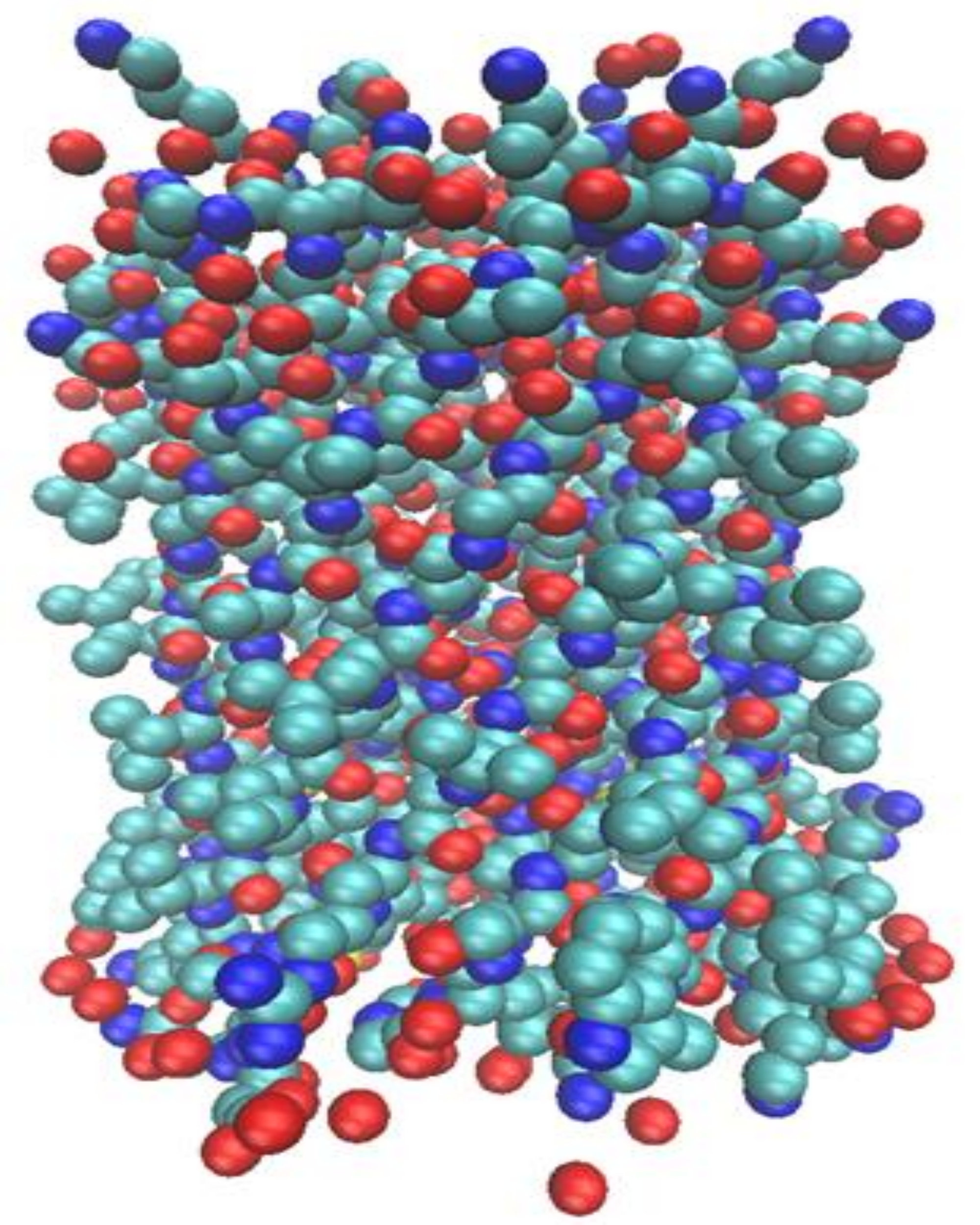}\\
\end{tabular}
\end{center}
\caption{The initial structure of  protein 2GR8. Left chart: Secondary structure representation; Right chart: atomic representation. Colors indicate different types of  atoms.
}
\label{beta_ori}
\end{figure}

Figure \ref{beta_ori} shows the initial structure of  protein 2GR8 in both secondary structure and atomic representations.  Clearly, it is a beta barrel   with 12 twisted  beta strands coiled together in an antiparallel fashion to form a  cylindrical structure in which the first strand is hydrogen bonded to the last.  However, inside the beta barrel, there are  also three alpha  helices as shown in the left chart of  Fig.  \ref{beta_ori}. The topological structure is complicated due to the presence of these alpha helices.

\begin{figure}
\begin{center}
\begin{tabular}{ccc}
\includegraphics[width=0.25\textwidth]{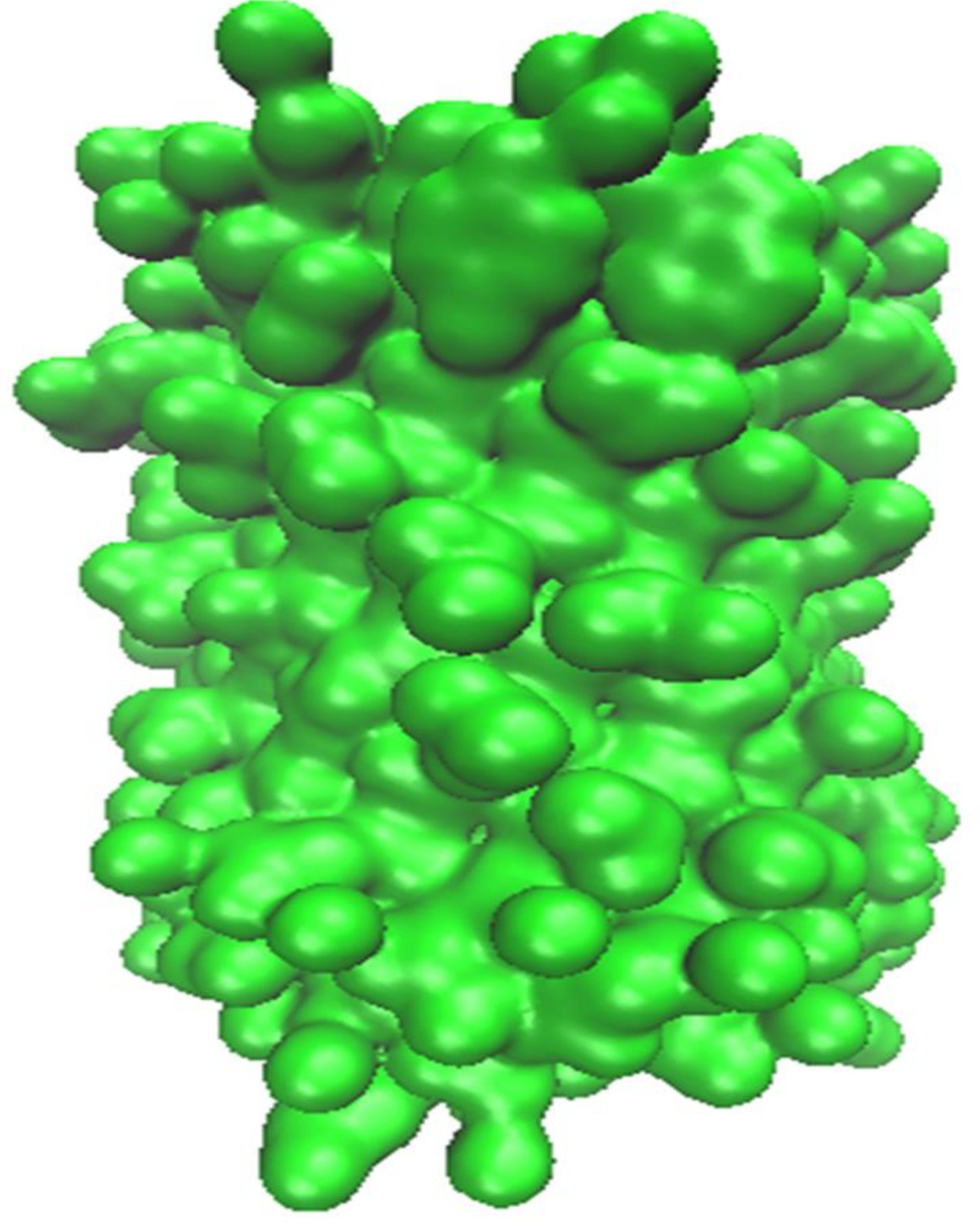}
\includegraphics[width=0.25\textwidth]{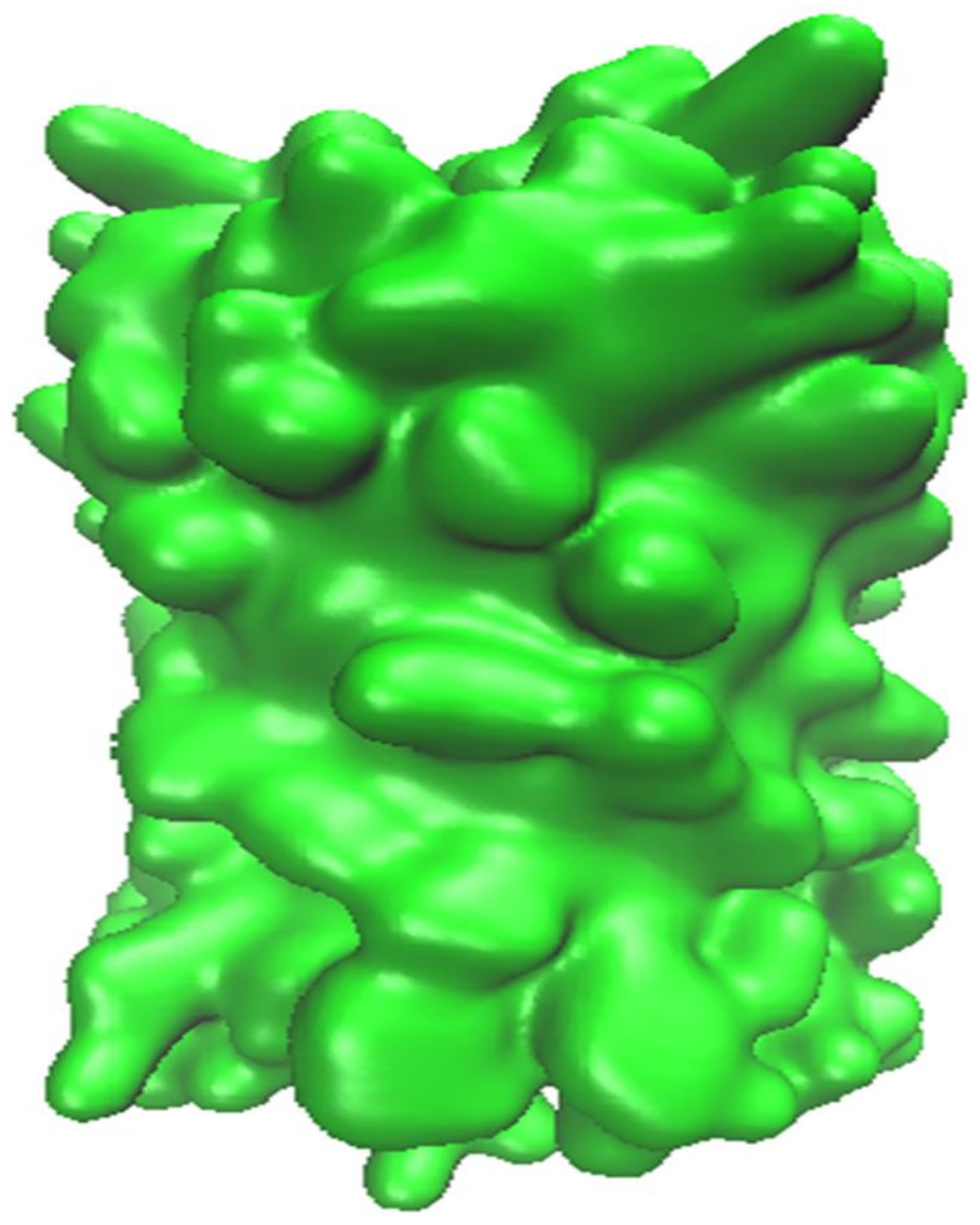}
\includegraphics[width=0.25\textwidth]{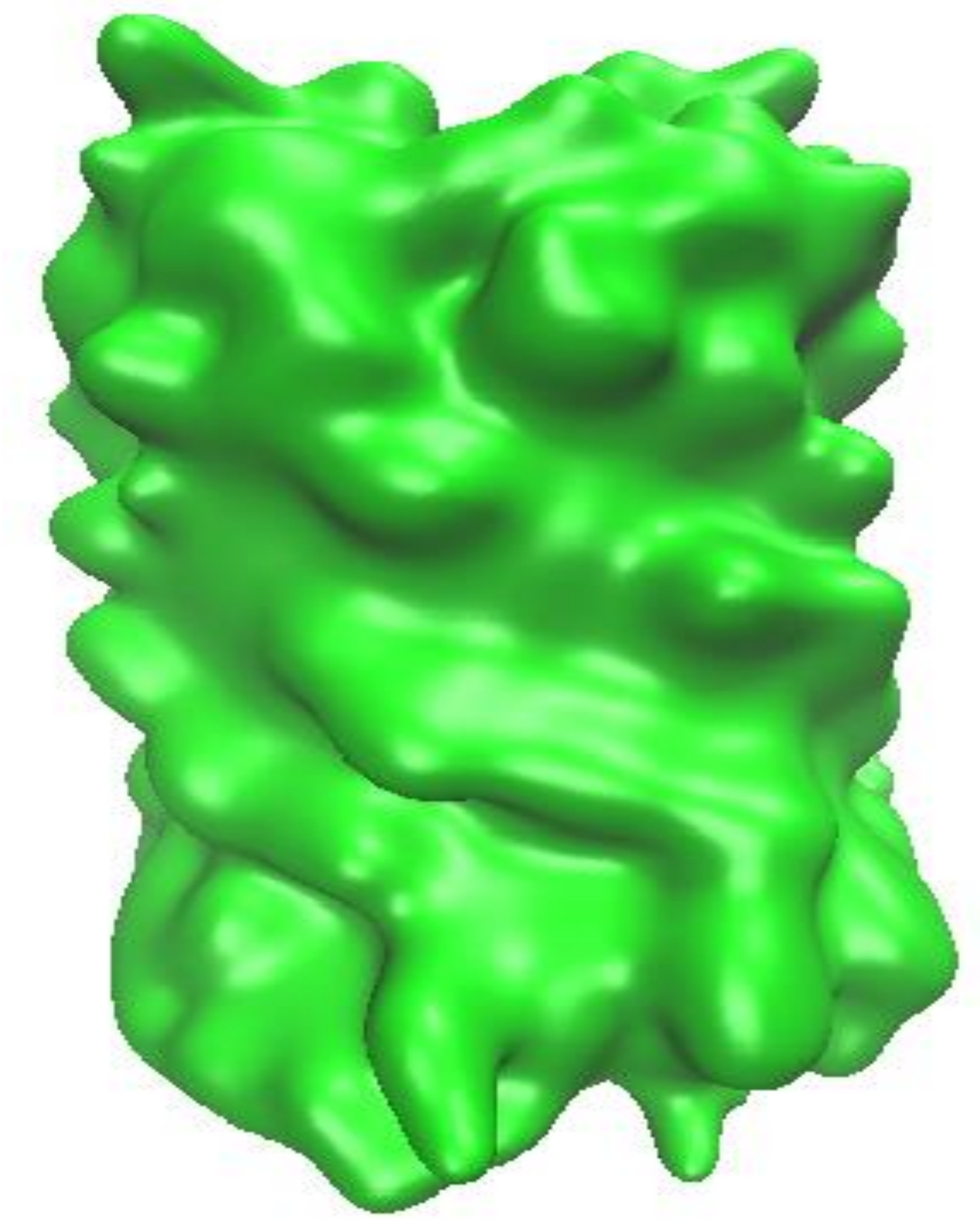}\\
\includegraphics[width=0.24\textwidth]{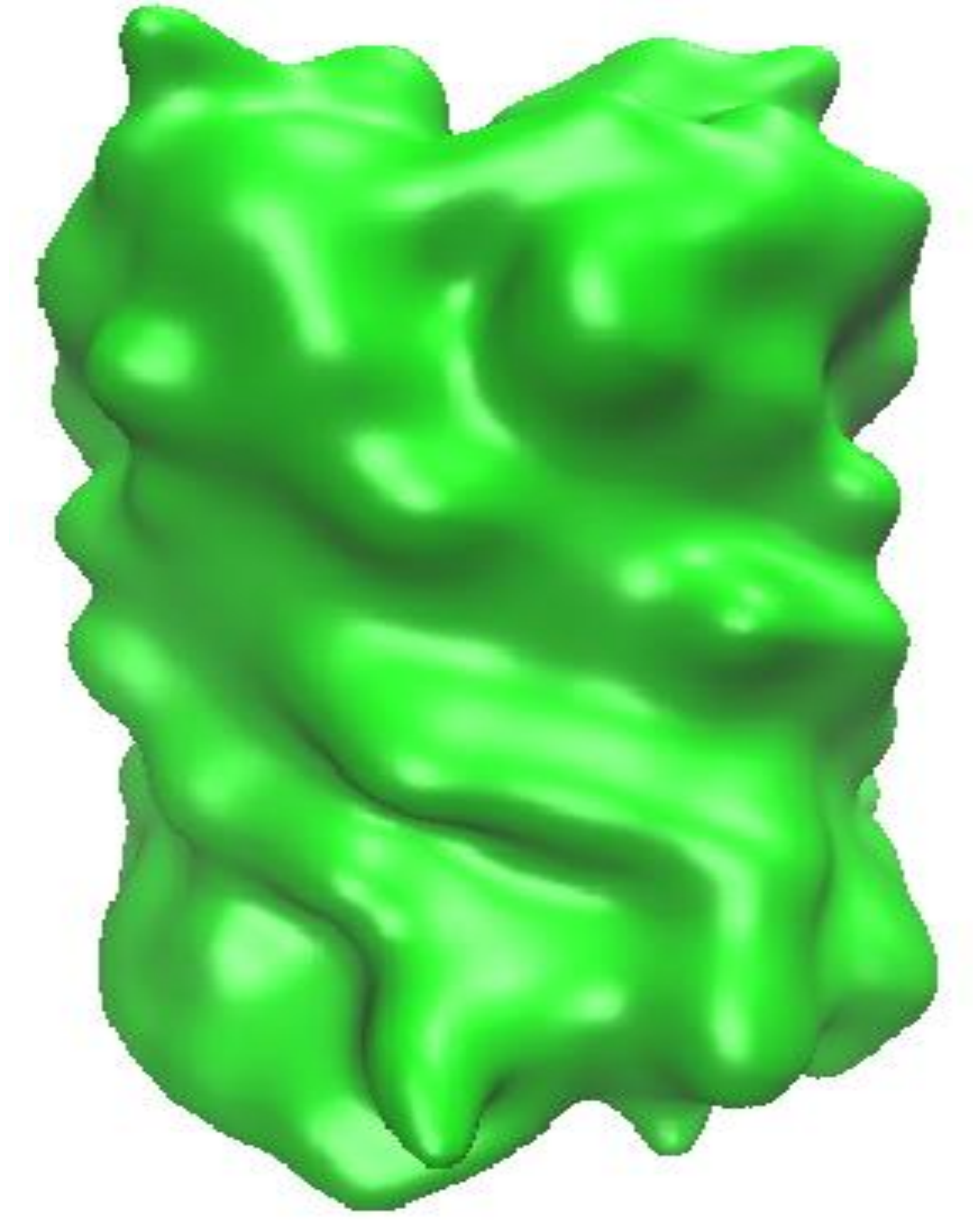}
\includegraphics[width=0.24\textwidth]{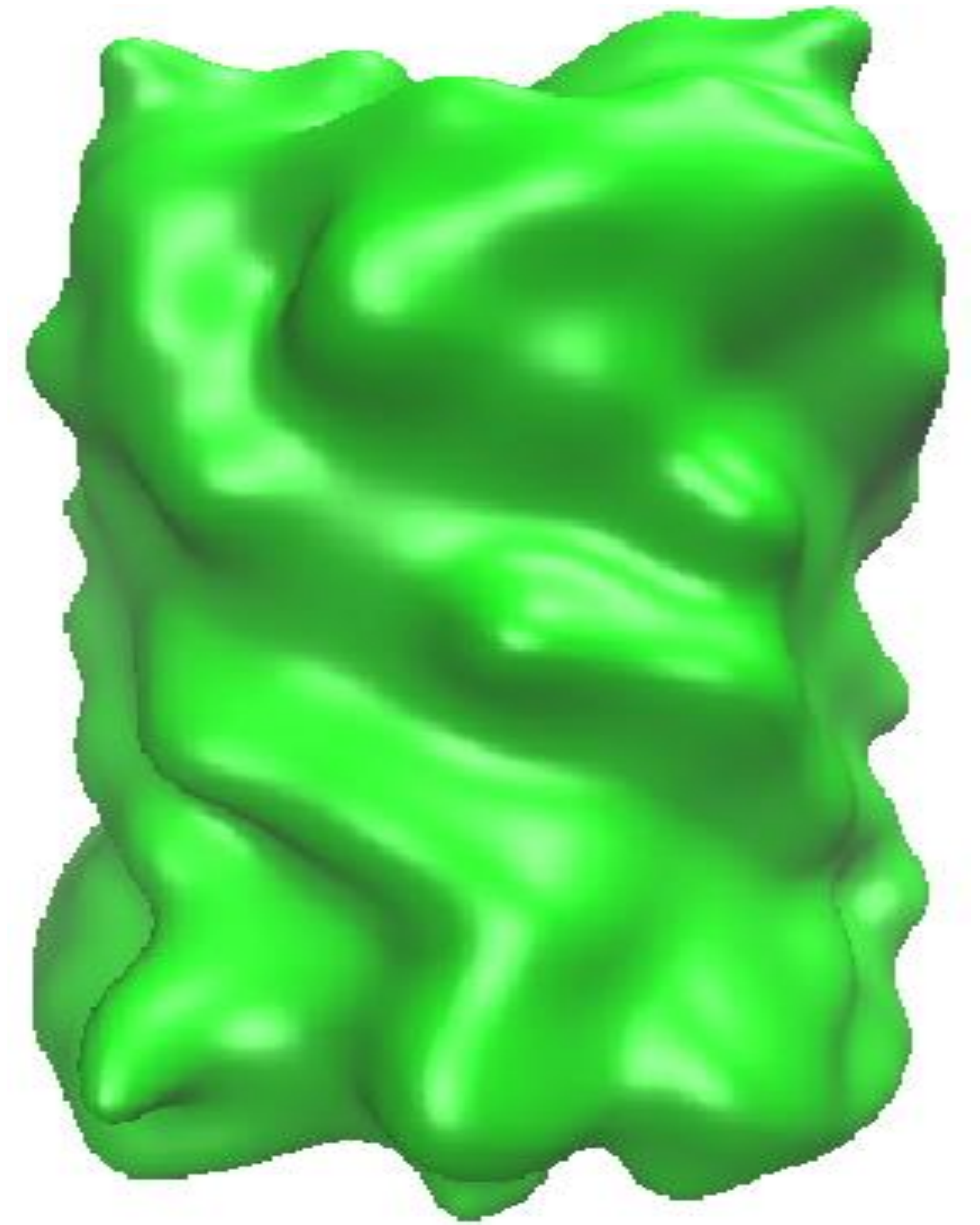}
\includegraphics[width=0.24\textwidth]{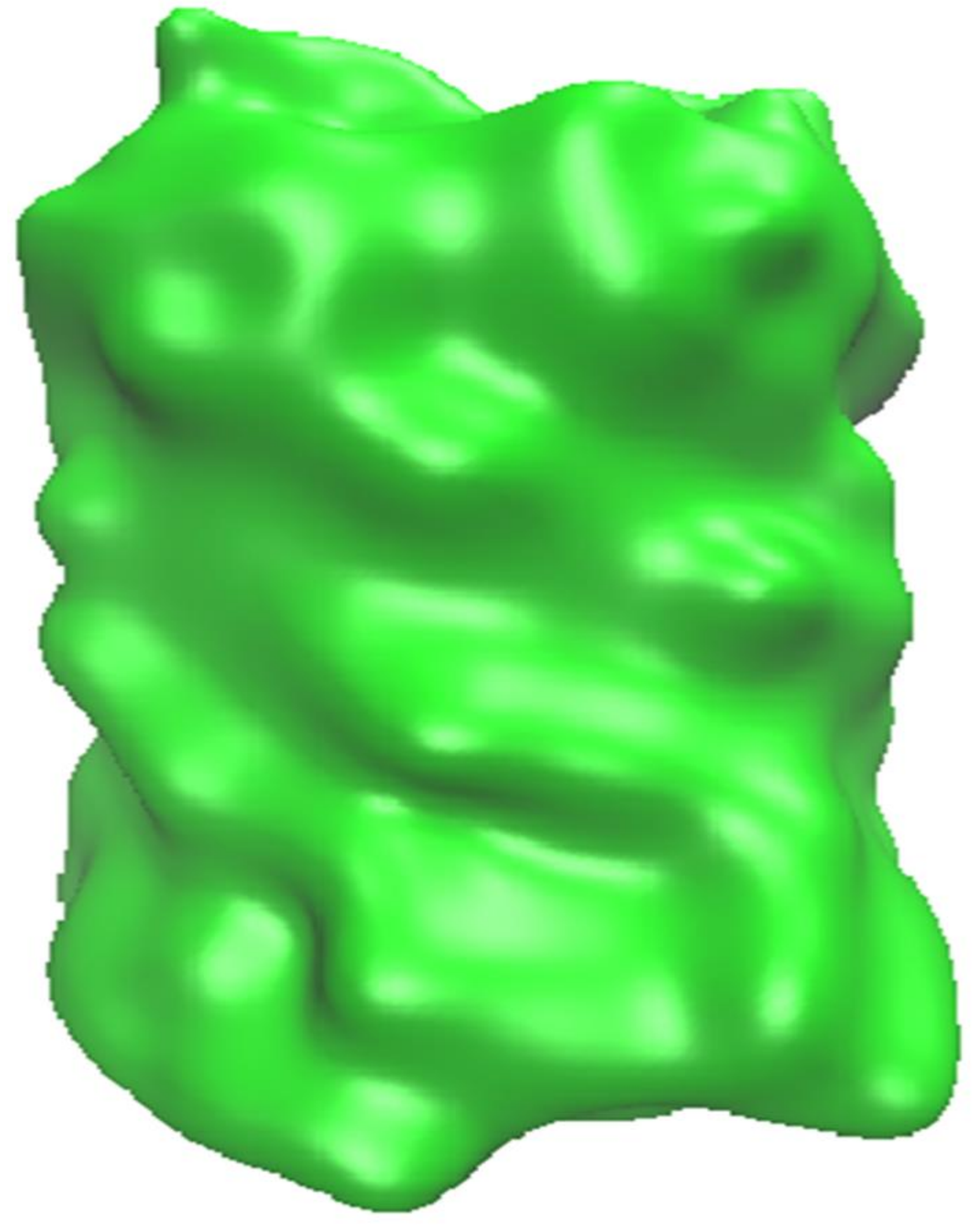}
\end{tabular}
\end{center}
\caption{Geometric evolution of protein  2GR8 under the Laplace-Beltrami flow. Charts from left to right and from top to bottom are frames 1 to 6, respectively.
}
\label{evo_pdb}
\end{figure}

We first consider the geometric  evolution of protein  2GR8 under the Laplace-Beltrami flow and then compute its  homology evolution.  Figure \ref{evo_pdb} depicts some frames generated from the time evolution process of the Laplace-Beltrami flow. The first two frames exhibit much atomic detail. As time progress, the atomic features disappear while beta strands are clearly demonstrated in frames 3-6. In fact, beta strand features diminish at the last two frames and the global cylindrical feature dominates.  Therefore, the   Laplace-Beltrami flow generates a multiscale representation of the protein as illustrated in our earlier work \cite{Wei:2005, KLXia:2014a}.

\begin{table}[!ht]
\centering
\caption{The time evolution of the topological invariants of protein 2GR8 under the Laplace-Beltrami flow.}
\begin{tabular}{lllll}
\cline{1-5}
Frame &Time  &$\beta_0$  &$\beta_1$  &$\beta_2$ \\
\hline
1 &0.10  &1  &263 &12  \\
2 &0.50  &1  &1   &21  \\
3 &1.00  &1  &1   &9  \\
4 &1.50  &1  &1   &2  \\
5 &1.70  &1  &1   &1  \\
6 &1.80  &1  &0   &0  \\
\hline
\end{tabular}
\label{beta_evo_topo}
\end{table}

\begin{figure}
\begin{center}
\begin{tabular}{ccc}
\includegraphics[width=0.3\textwidth]{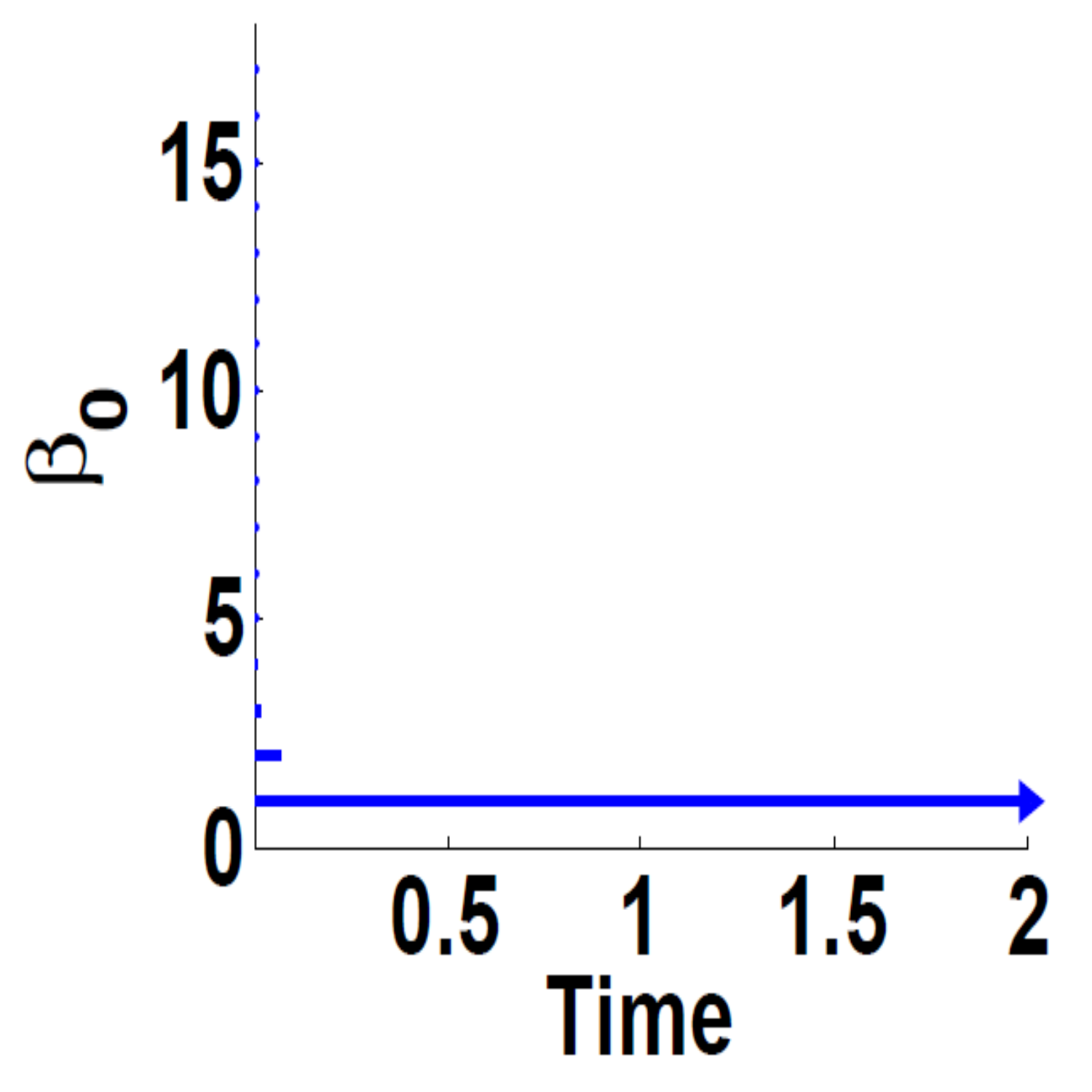}
\includegraphics[width=0.3\textwidth]{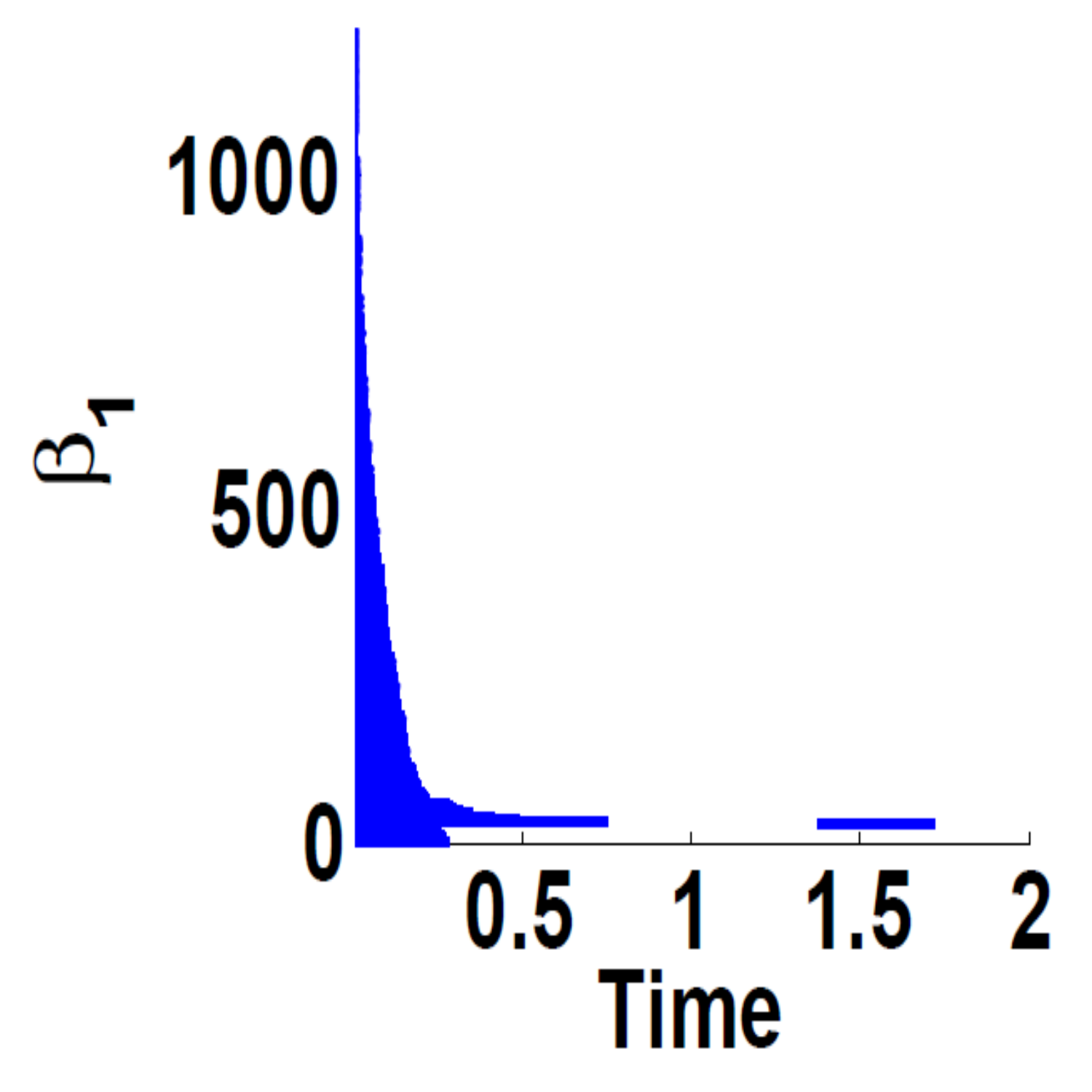}
\includegraphics[width=0.3\textwidth]{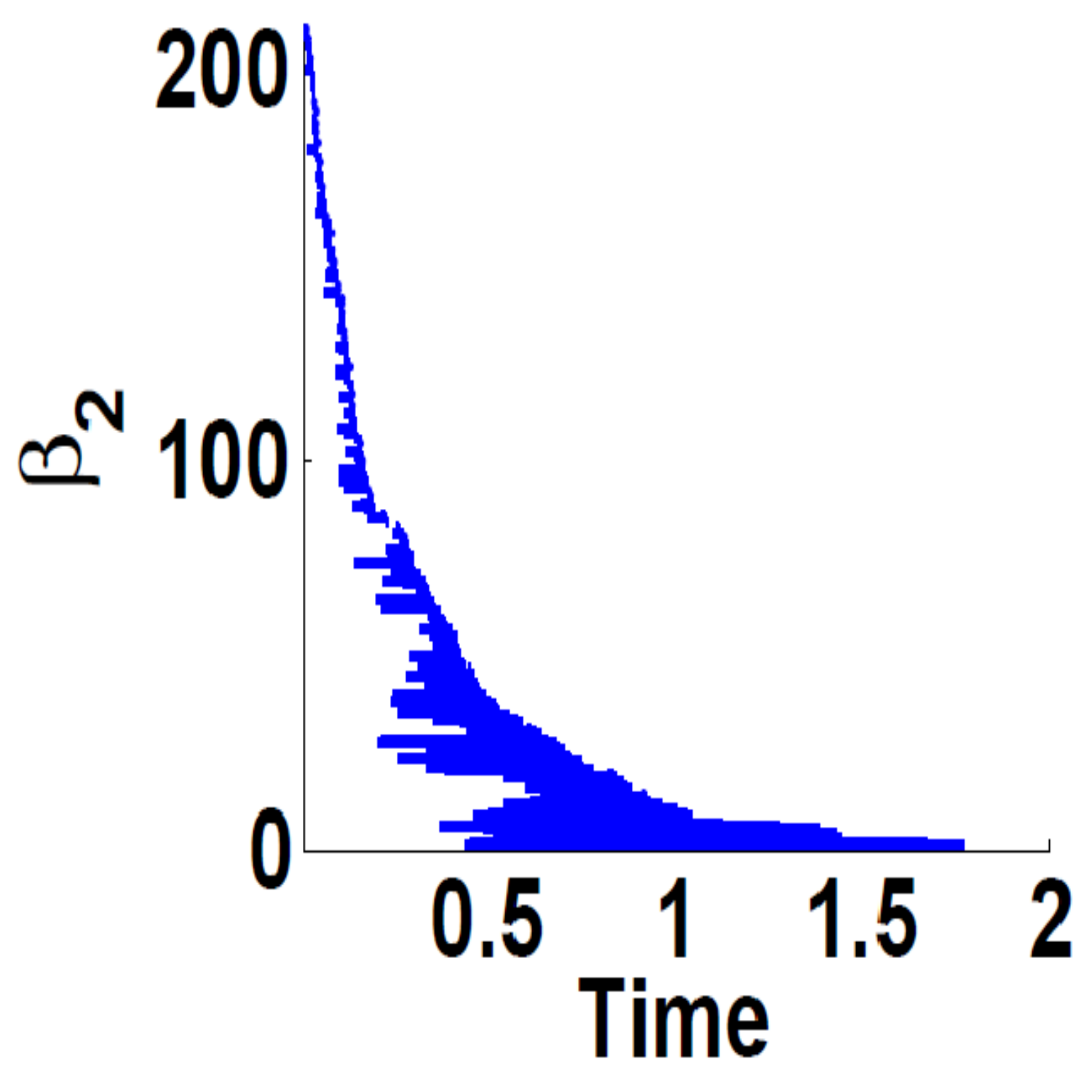}
\end{tabular}
\end{center}
\caption{The time evolution of the topological invariants of protein 2GR8 under the Laplace-Beltrami flow.
}
\label{pdb_bett}
\end{figure}

Table \ref{beta_evo_topo} gives the corresponding time evolution of topological invariants of the six frames  for protein  2GR8. One sees a large number of $\beta_1$ rings in the first frame. However, there is just one ring, in the second frame. The number of cavities reaches the highest values in the second frame  (among six frames) and gradually reduces to zero.    From Table \ref{beta_evo_topo},   we note that there is a ring in Frames 2-5.  However,  we cannot determine whether it is the same ring or not from the classical homology theory. There may be a different ring generated at each of Frame 2-5. Persistent homology is designed to reserve this issue. The persistence of the topological invariants during the time  evolution process is illustrated in Fig. \ref{pdb_bett}. It is confirmed that the ring  initially exists and is not generated in intermediate steps of the evolution. However, this ring is not a global one because it lasts for a relatively short period during the time evolution.

\subsubsection{A beta barrel}

\begin{figure}
\begin{center}
\begin{tabular}{cc}
\includegraphics[width=0.34\textwidth]{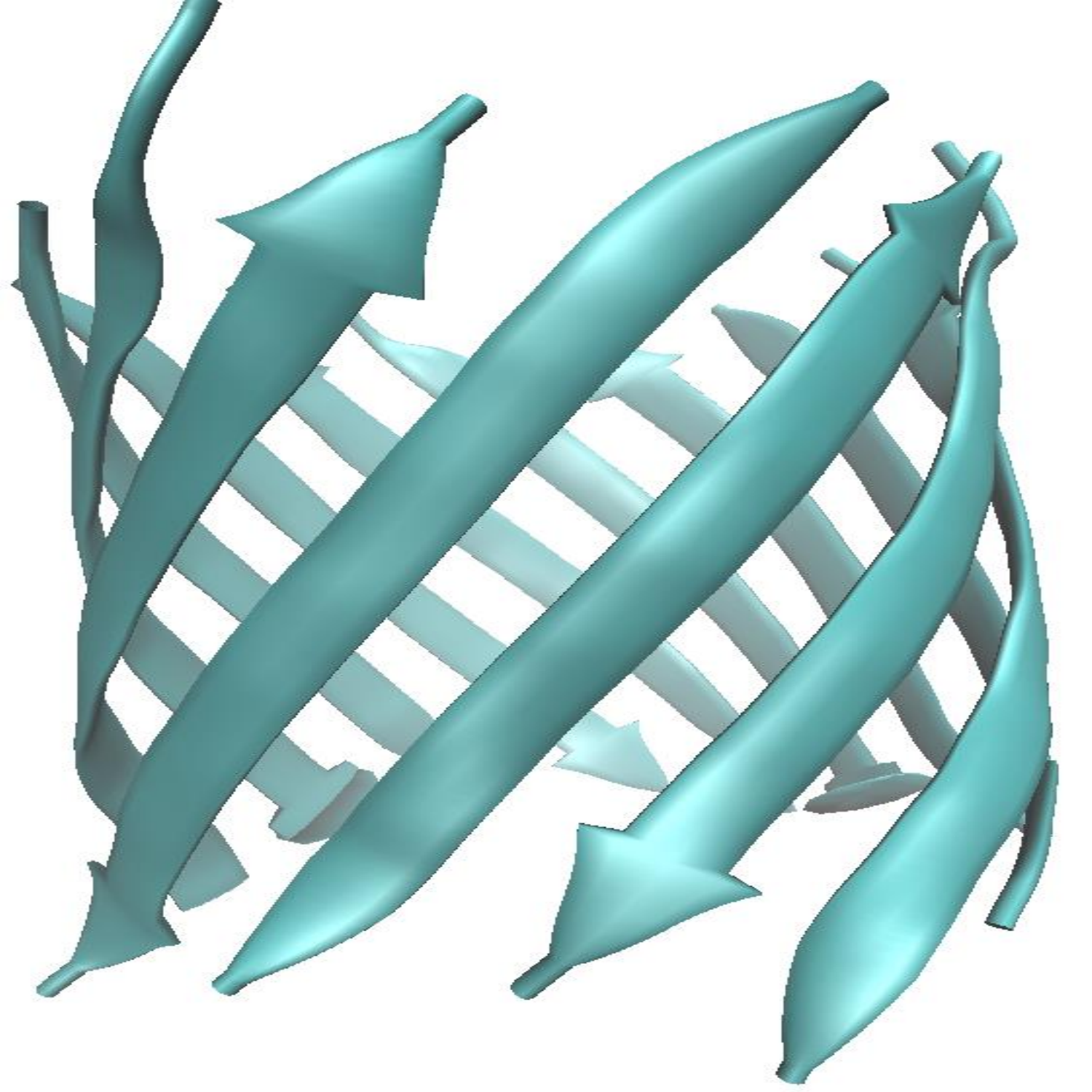}
\includegraphics[width=0.4\textwidth]{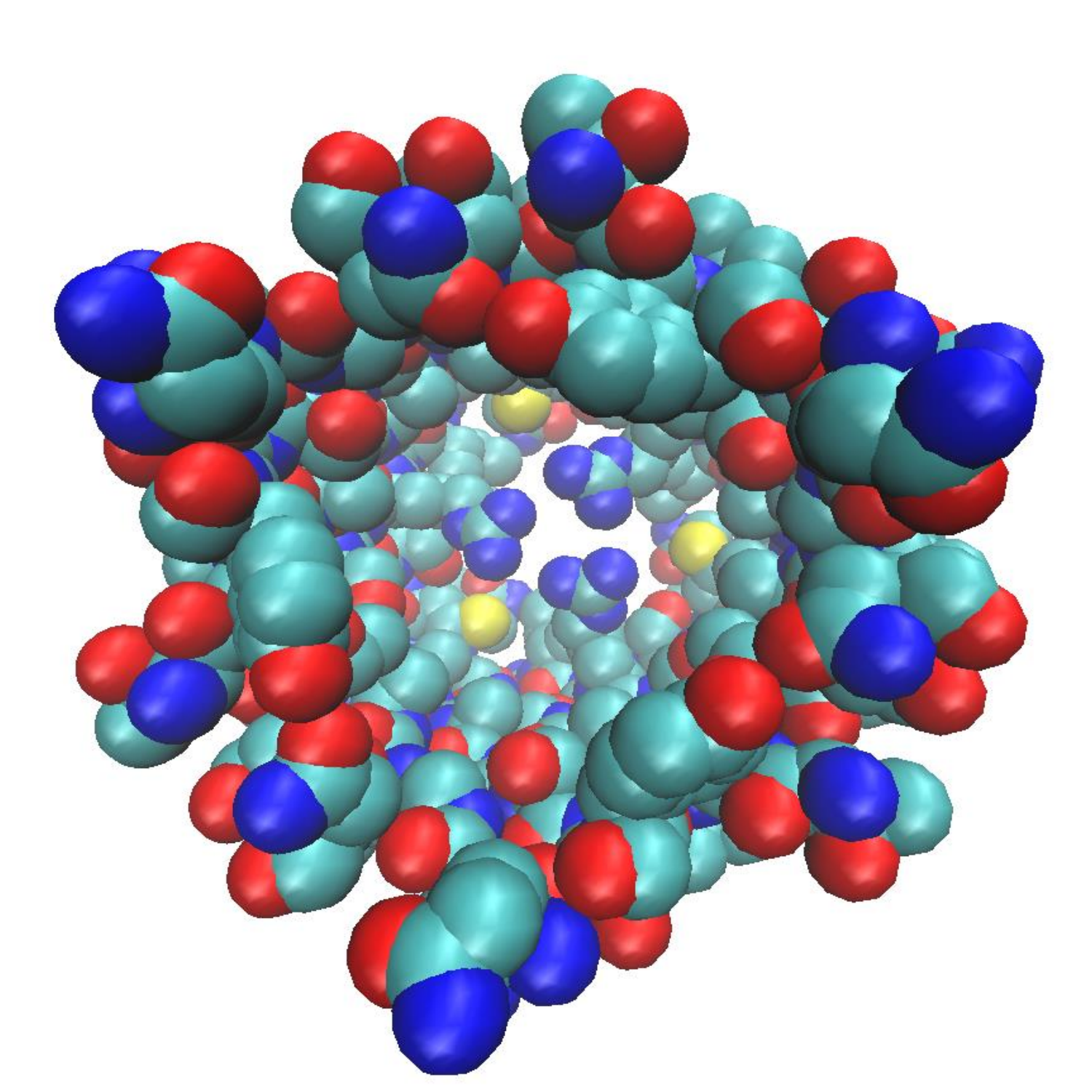}\\
\end{tabular}
\end{center}
\caption{The initial structure of  a beta barrel. Left chart: Secondary structure representation; Right chart: atomic representation. Colors indicate different types of  atoms.
}
\label{beta_sheet}

\end{figure}

\begin{figure}
\begin{center}
\begin{tabular}{ccc}
\includegraphics[width=0.25\textwidth]{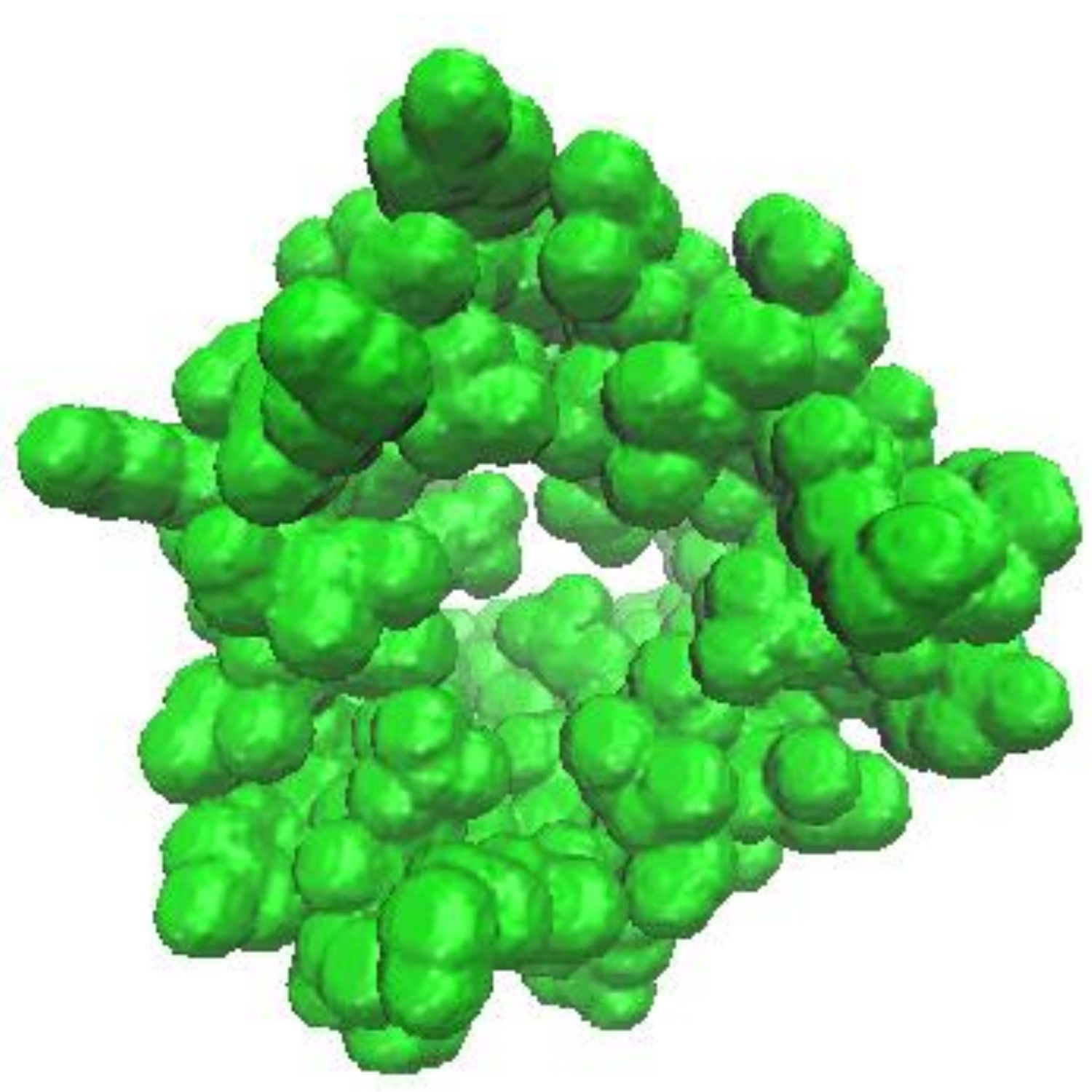}
\includegraphics[width=0.25\textwidth]{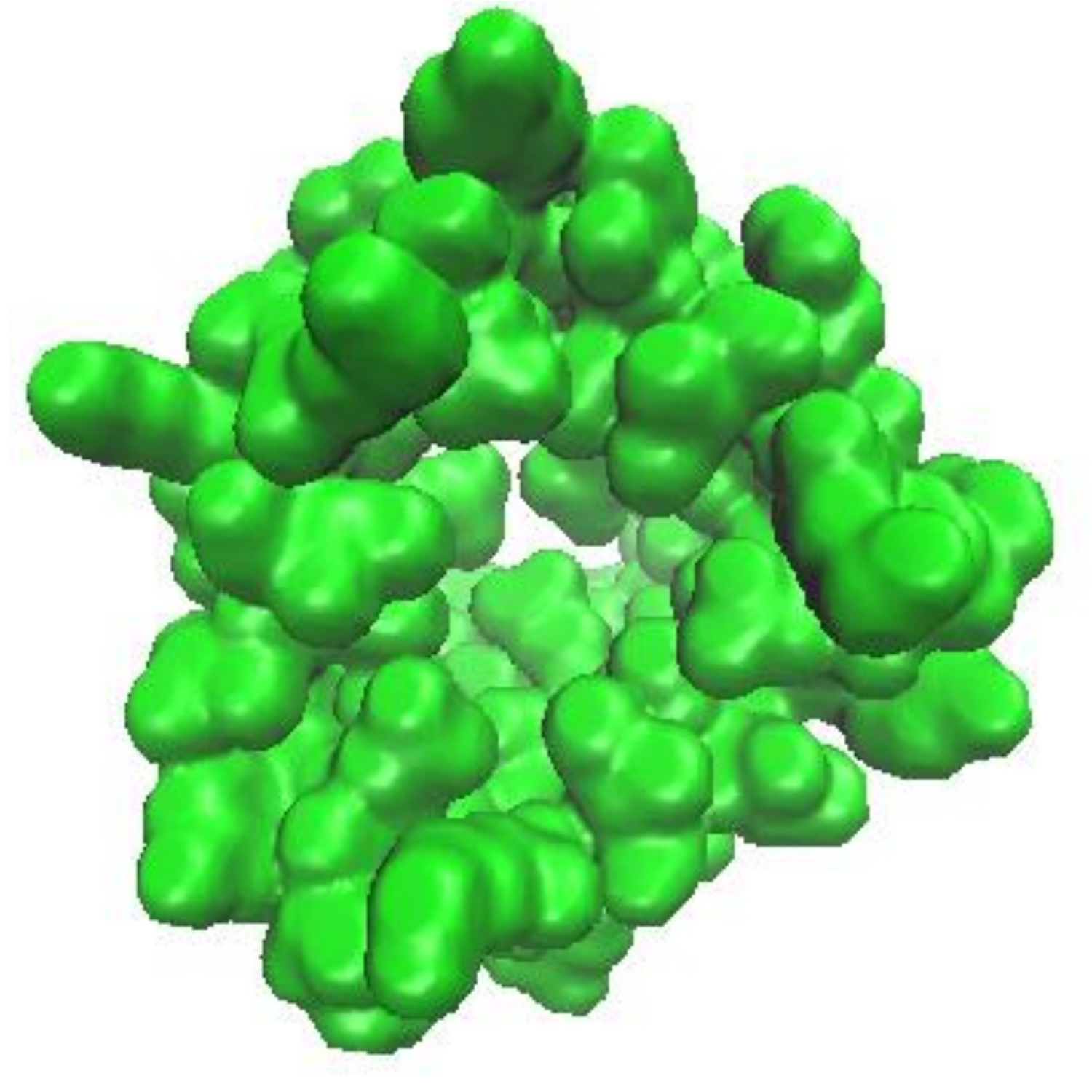}
\includegraphics[width=0.25\textwidth]{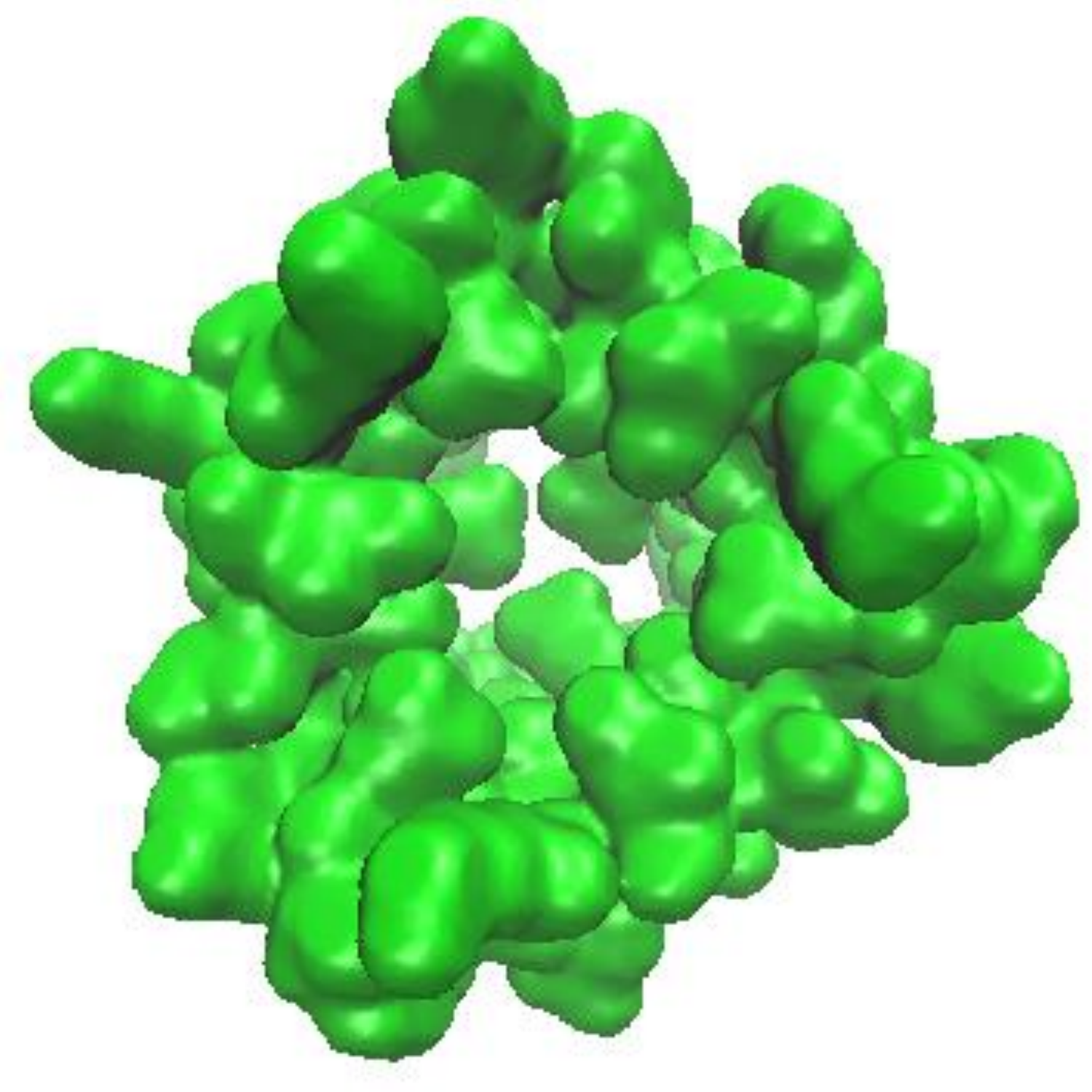}\\
\includegraphics[width=0.25\textwidth]{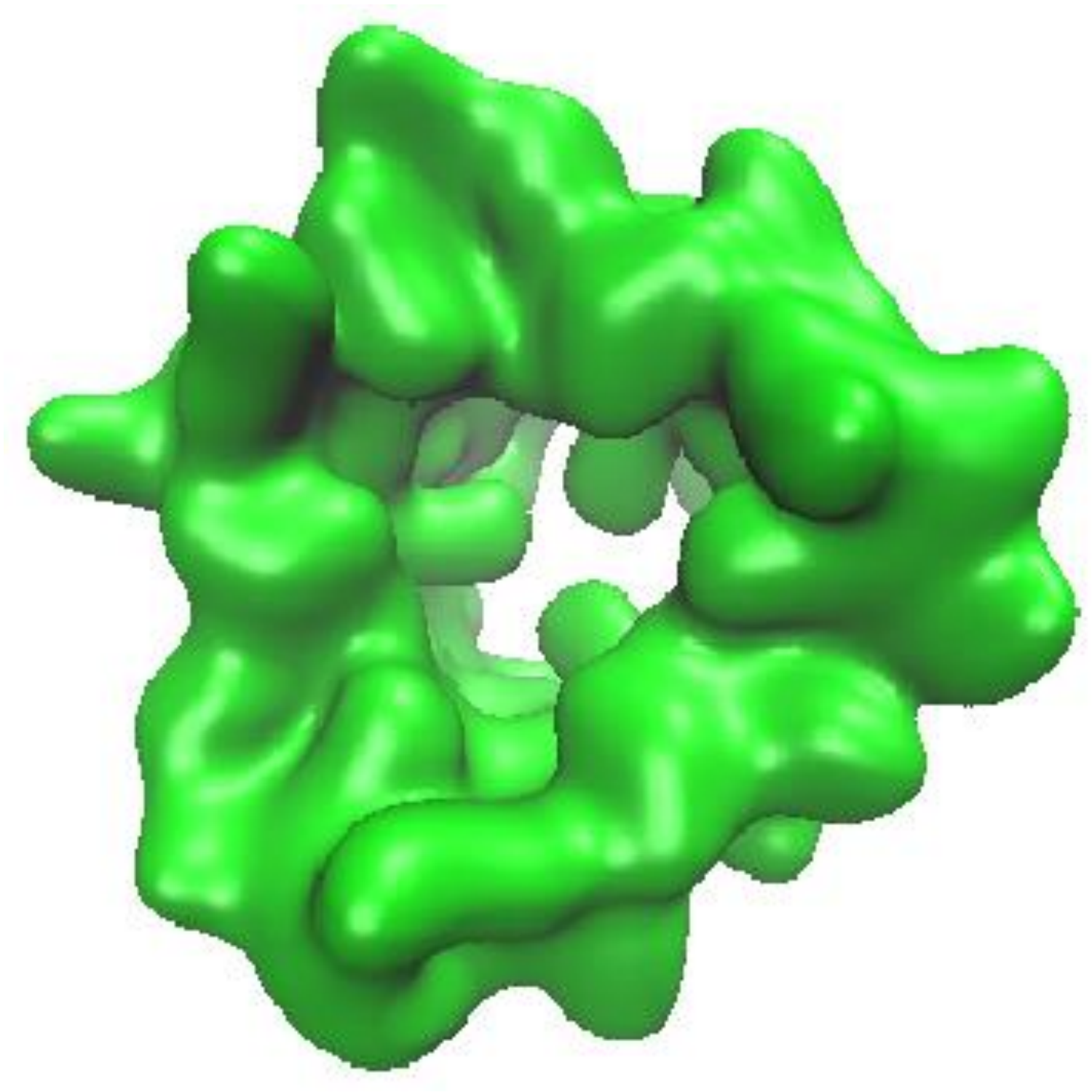}
\includegraphics[width=0.25\textwidth]{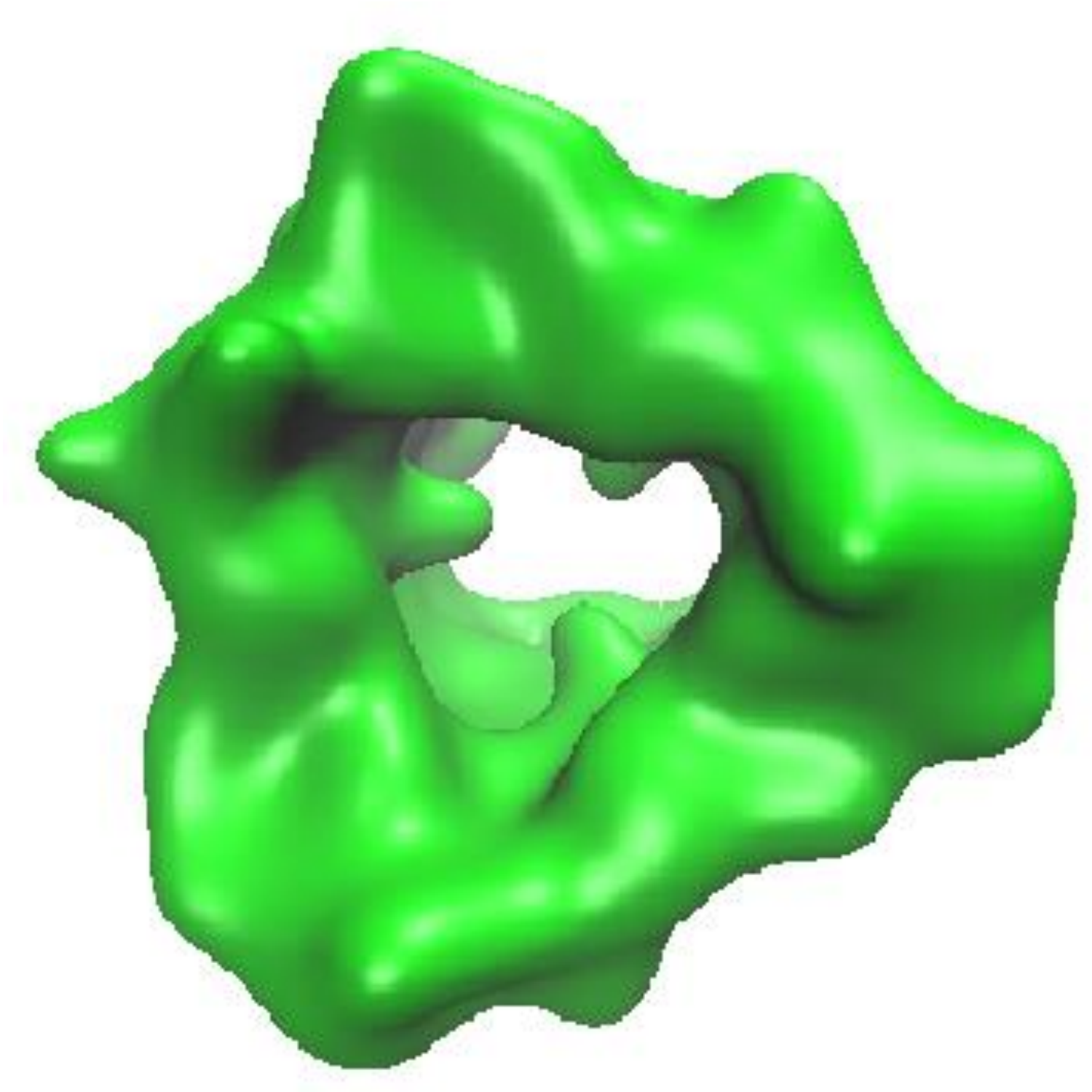}
\includegraphics[width=0.25\textwidth]{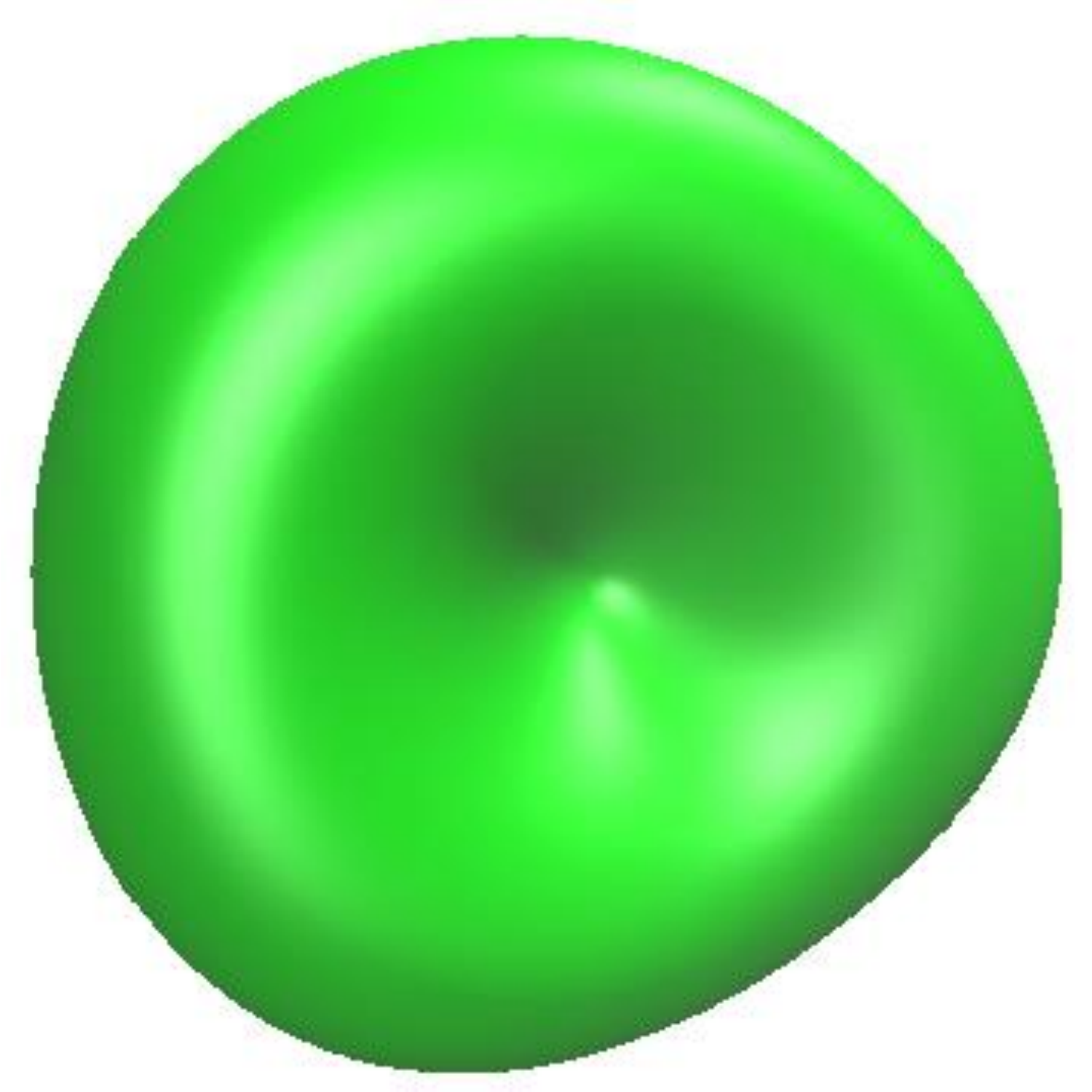}
\end{tabular}
\end{center}
\caption{The geometric evolution of  a beta barrel under the Laplace-Beltrami flow. Charts  from left to right  and from top to bottom are frames 1 to 6, respectively.
}
\label{evo_sh}
\end{figure}

We next create a pure beta barrel by removing three alpha  helices from  protein  2GR8, which enables us to observe the beta barrel ring geometry and topology  clearly. The initial structure of the beta barrel is shown in Fig. \ref{beta_sheet}.  The time evolution of the beta barrel is illustrated in Fig. \ref{evo_sh}.  Again, one sees atomic details in the first  few frames and global features in late frames.   Obviously, there is a large ring structure in the beta barrel.


Table \ref{Sheet_evo_topo} lists the corresponding topological invariants of six frames for the beta barrel. Although the number of $\beta_2$ varies dramatically, that of $\beta_1$ does not change over a long time period, indicating the global ring structure of the beta barrel.

\begin{table}[!ht]
\centering
\caption{The evolution the topological invariants of the beta barrel under the geometric flow.}
\label{Sheet_evo_topo}
\begin{tabular}{lllll}
\cline{1-5}
Frame &Time  &$\beta_0$  &$\beta_1$  &$\beta_2$ \\
\hline
1 &0.01  &1  &137 &0  \\
2 &0.10  &1  &62 &4  \\
3 &0.15  &1  &23 &2  \\
4 &1.00  &1  &4 &0  \\
5 &2.00  &1  &1  &0  \\
6 &29.0  &1  &0  &0  \\
\hline
\end{tabular}
\end{table}

\begin{figure}
\begin{center}
\begin{tabular}{ccc}
\includegraphics[width=0.3\textwidth]{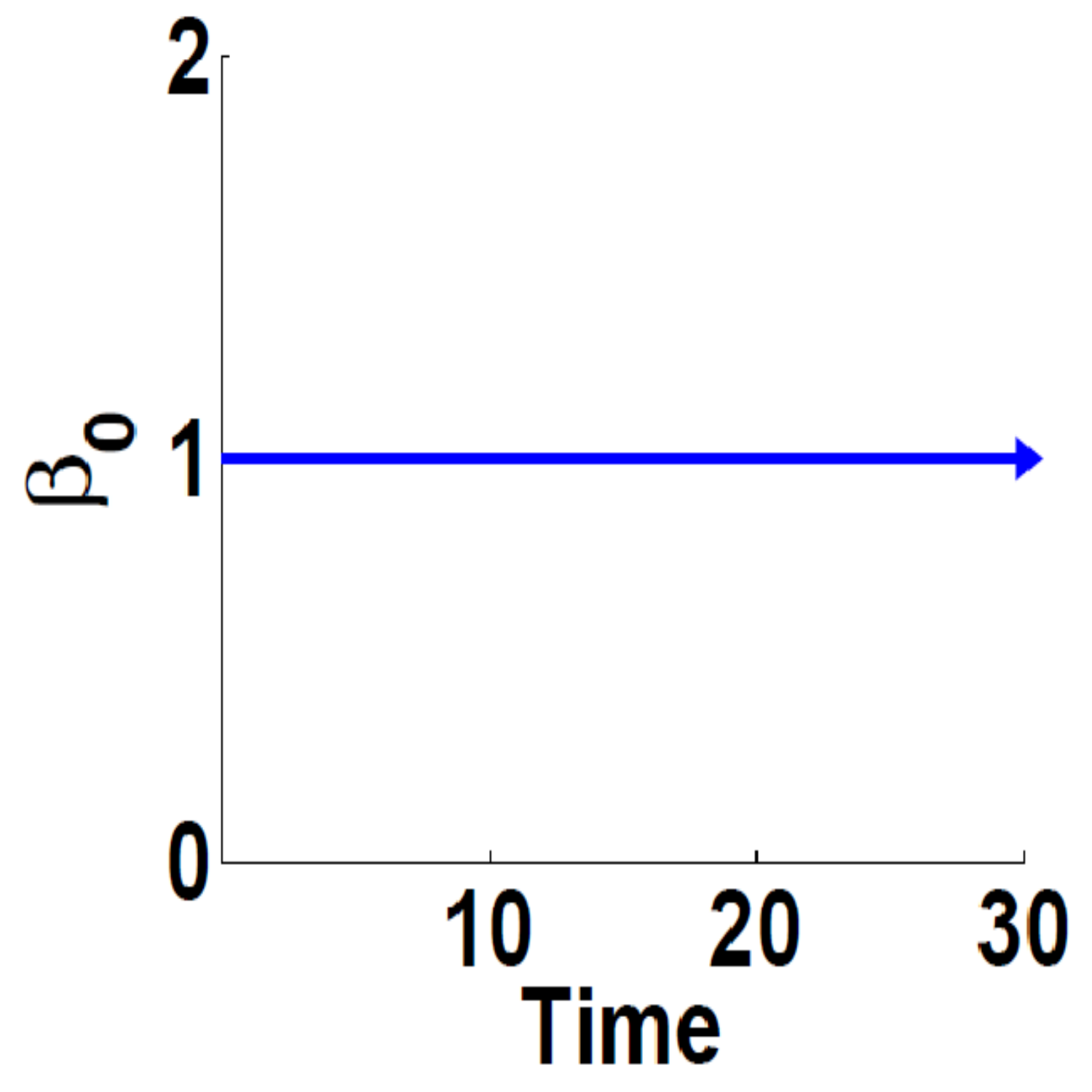}
\includegraphics[width=0.3\textwidth]{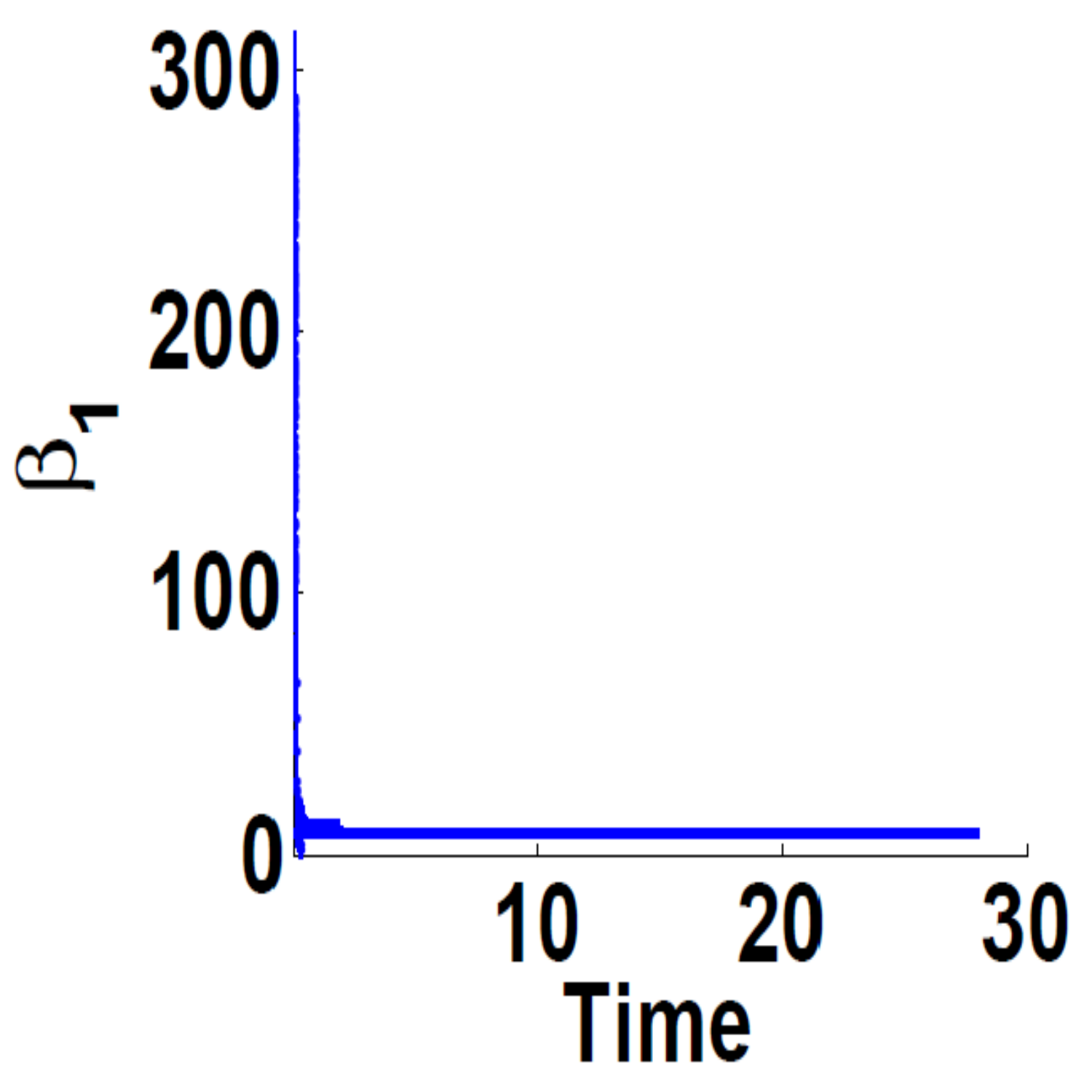}
\includegraphics[width=0.3\textwidth]{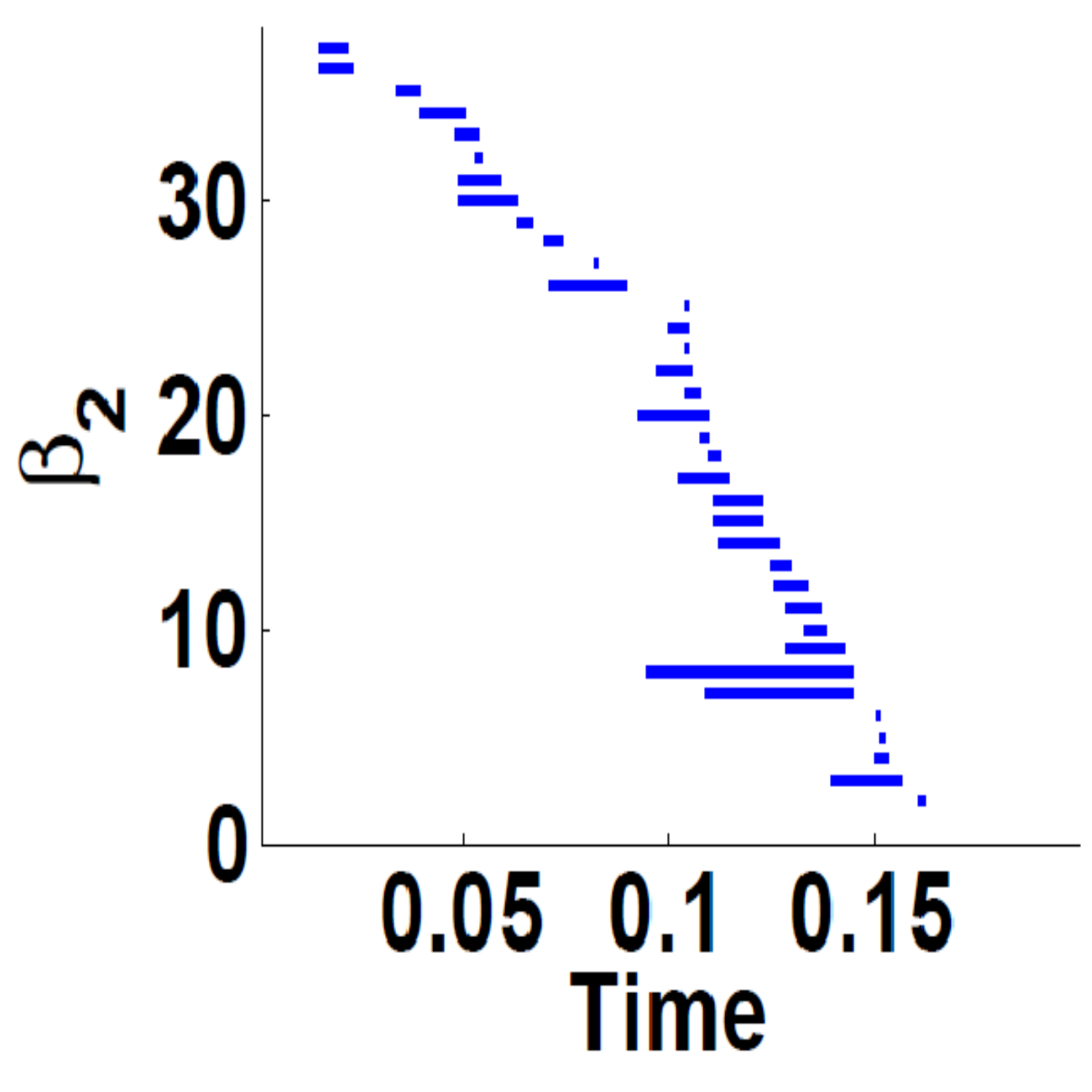}
\end{tabular}
\end{center}
\caption{The evolution of the topological invariants of the beta barrel under the Laplace-Beltrami flow.
}
\label{sh_betti_03}
\end{figure}

The persistence of the topological invariants over time evolution process for the  beta barrel is illustrated in Fig. \ref{sh_betti_03}. The $\beta_1$ panel has a long-lasting bar. A comparison with the time scale in the $\beta_1$ panel of Fig. \ref{pdb_bett} confirms that the present long-lasting bar corresponds to the intrinsic global structure of the beta barrel.

The above results demonstrate that the proposed Laplace-Beltrami flow based  persistent homology  is an efficient tool for analyzing the topological  structures of protein molecules.

\subsection{Fullerene total curvature energy prediction}\label{FullereneTotalCurvature}

\begin{figure}
\begin{center}
\begin{tabular}{ccc}
\includegraphics[width=0.4\textwidth]{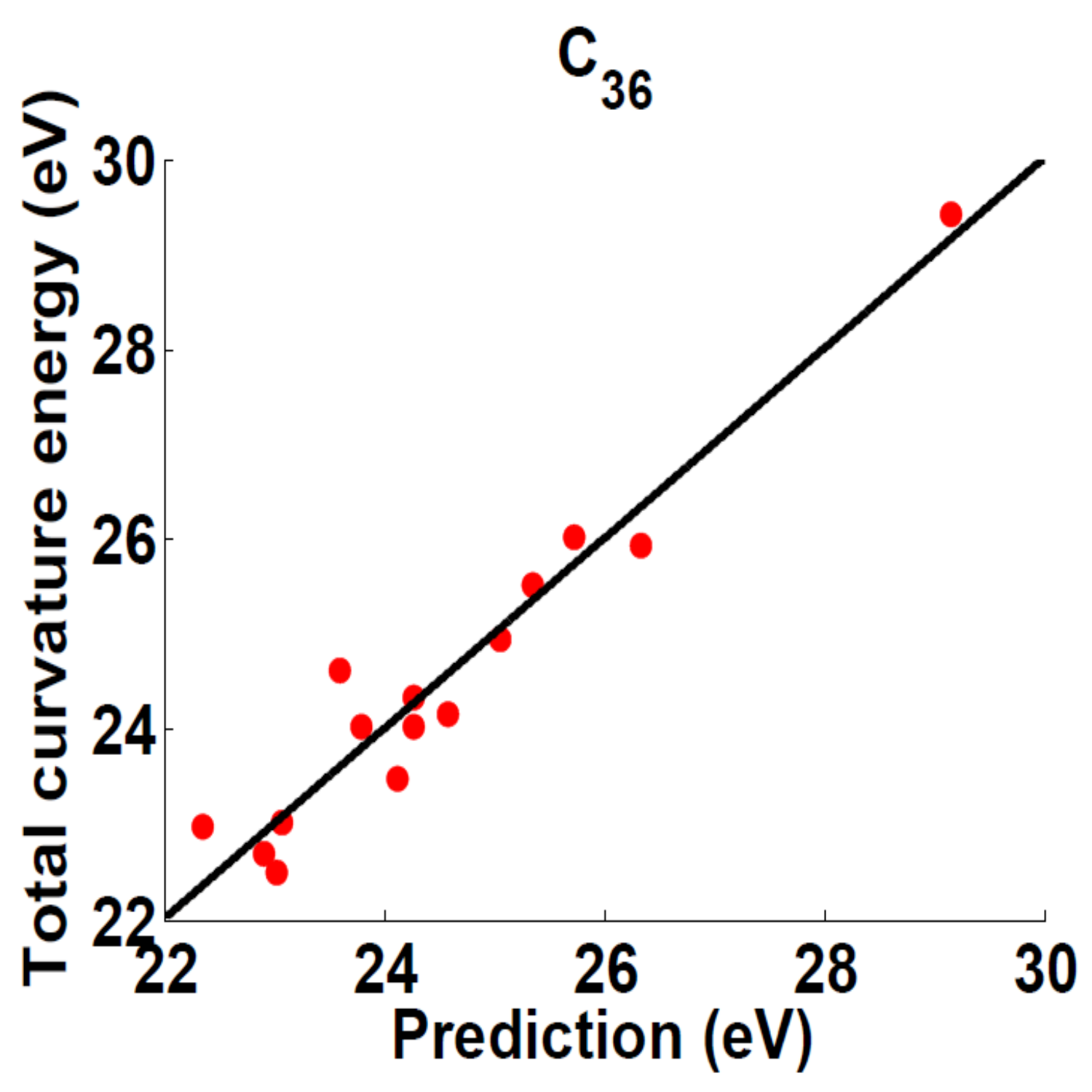}
\includegraphics[width=0.4\textwidth]{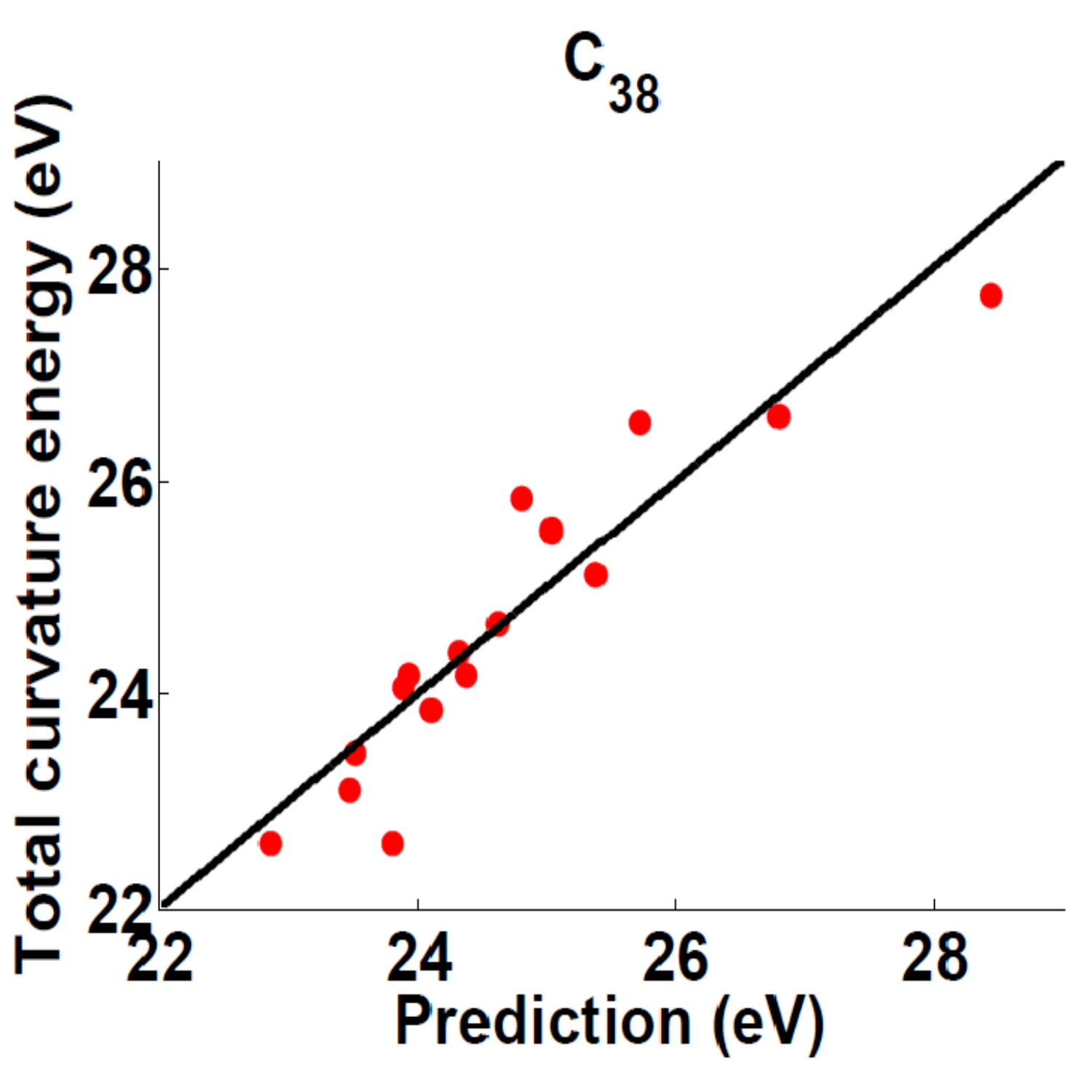}\\
\includegraphics[width=0.4\textwidth]{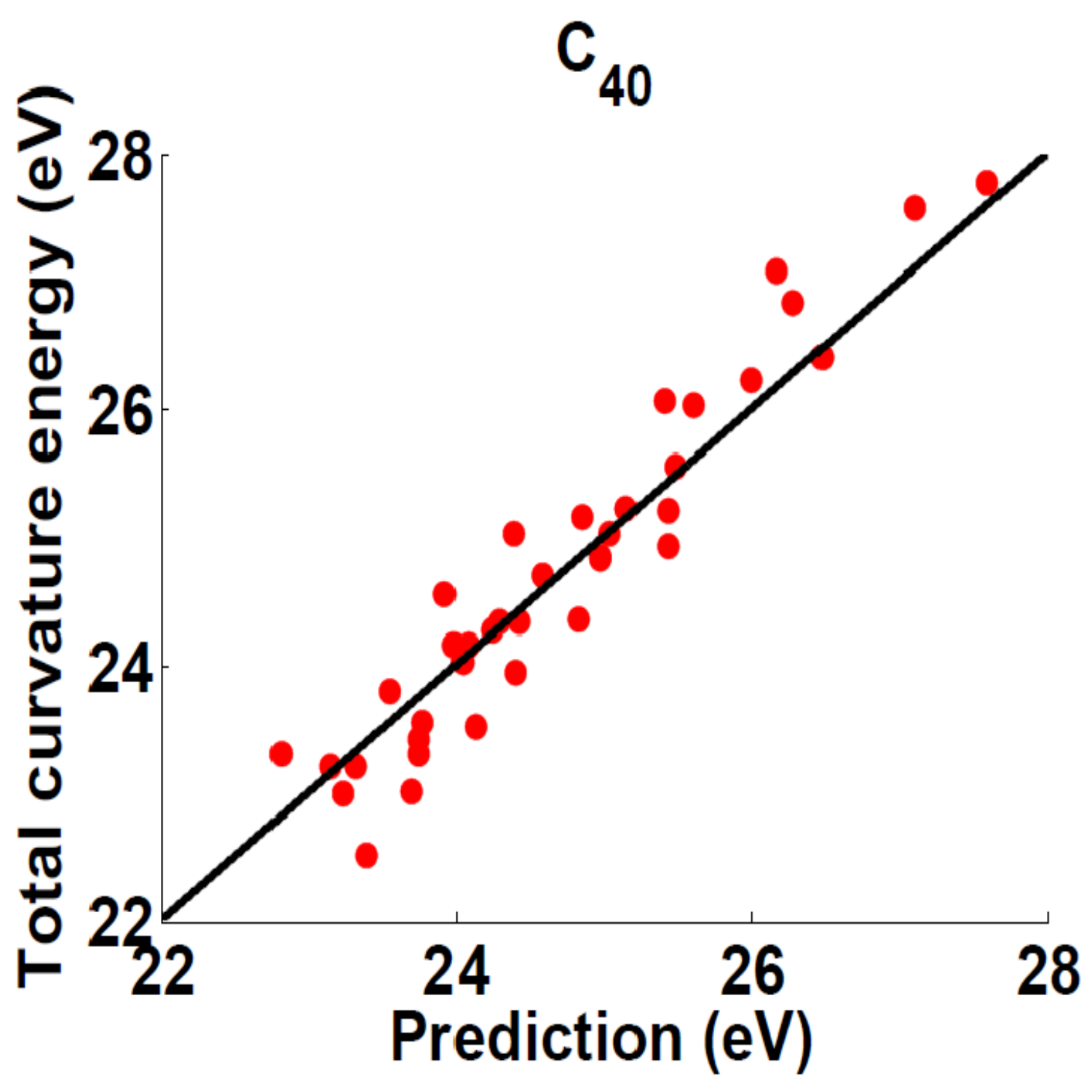}
\includegraphics[width=0.4\textwidth]{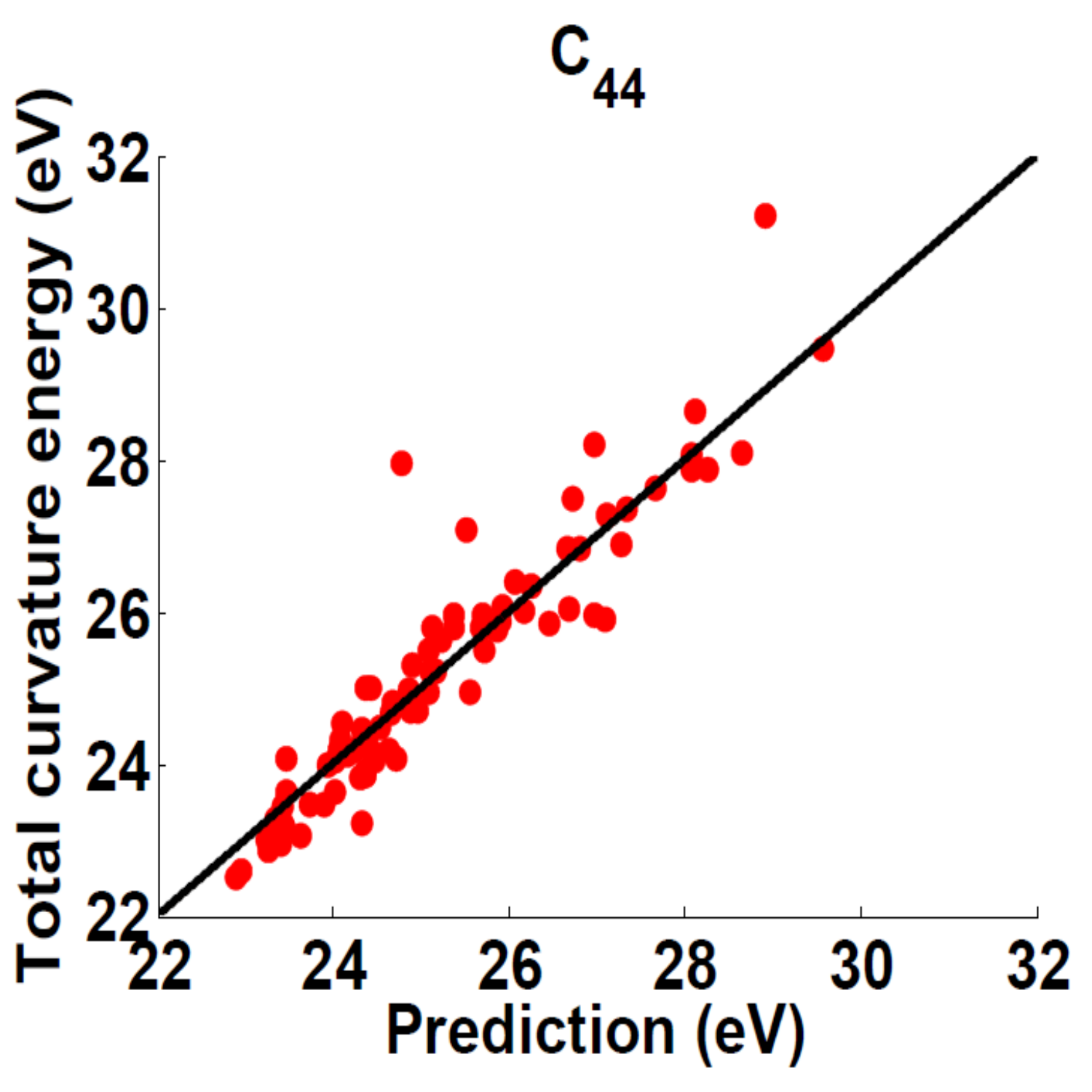}\\
\includegraphics[width=0.4\textwidth]{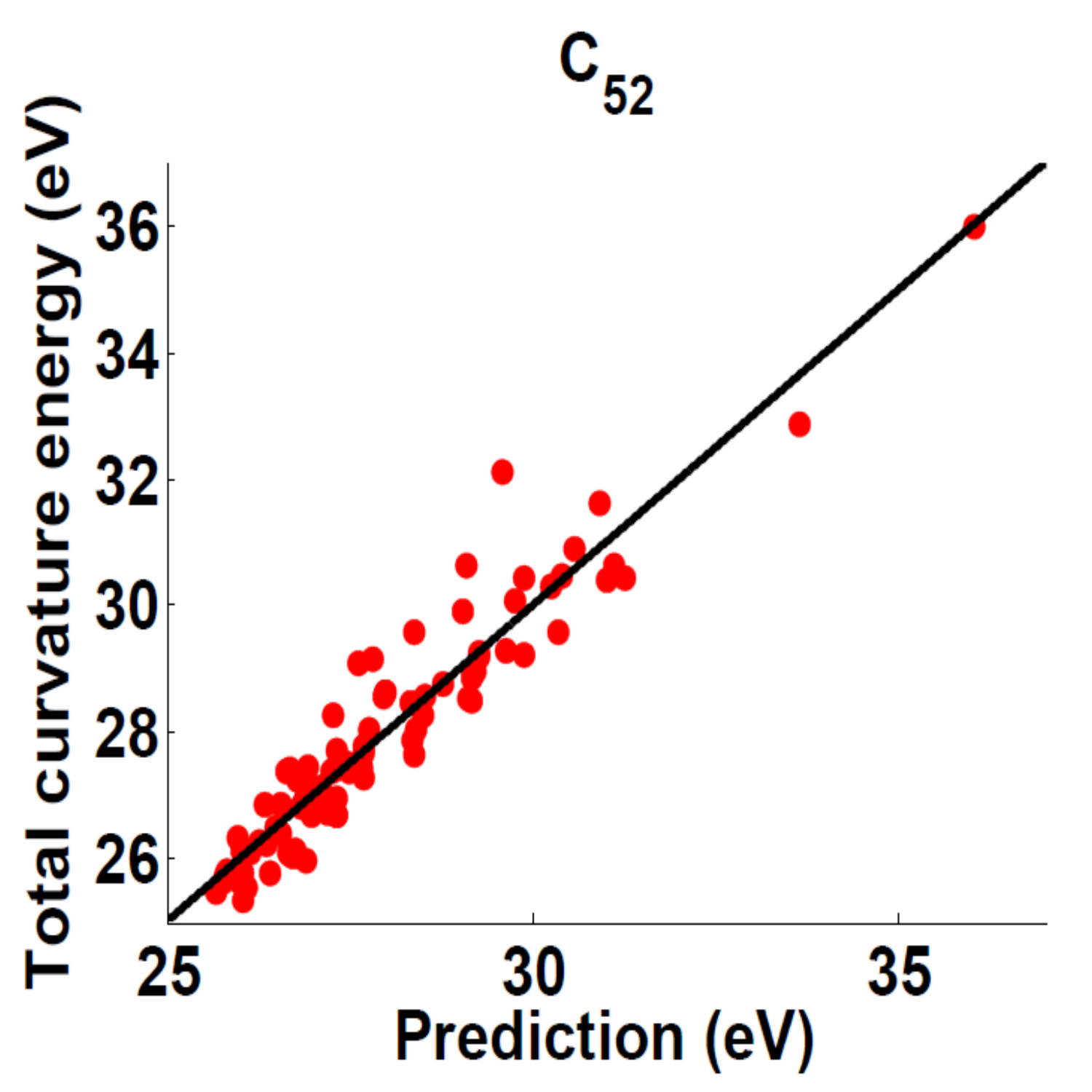}
\includegraphics[width=0.4\textwidth]{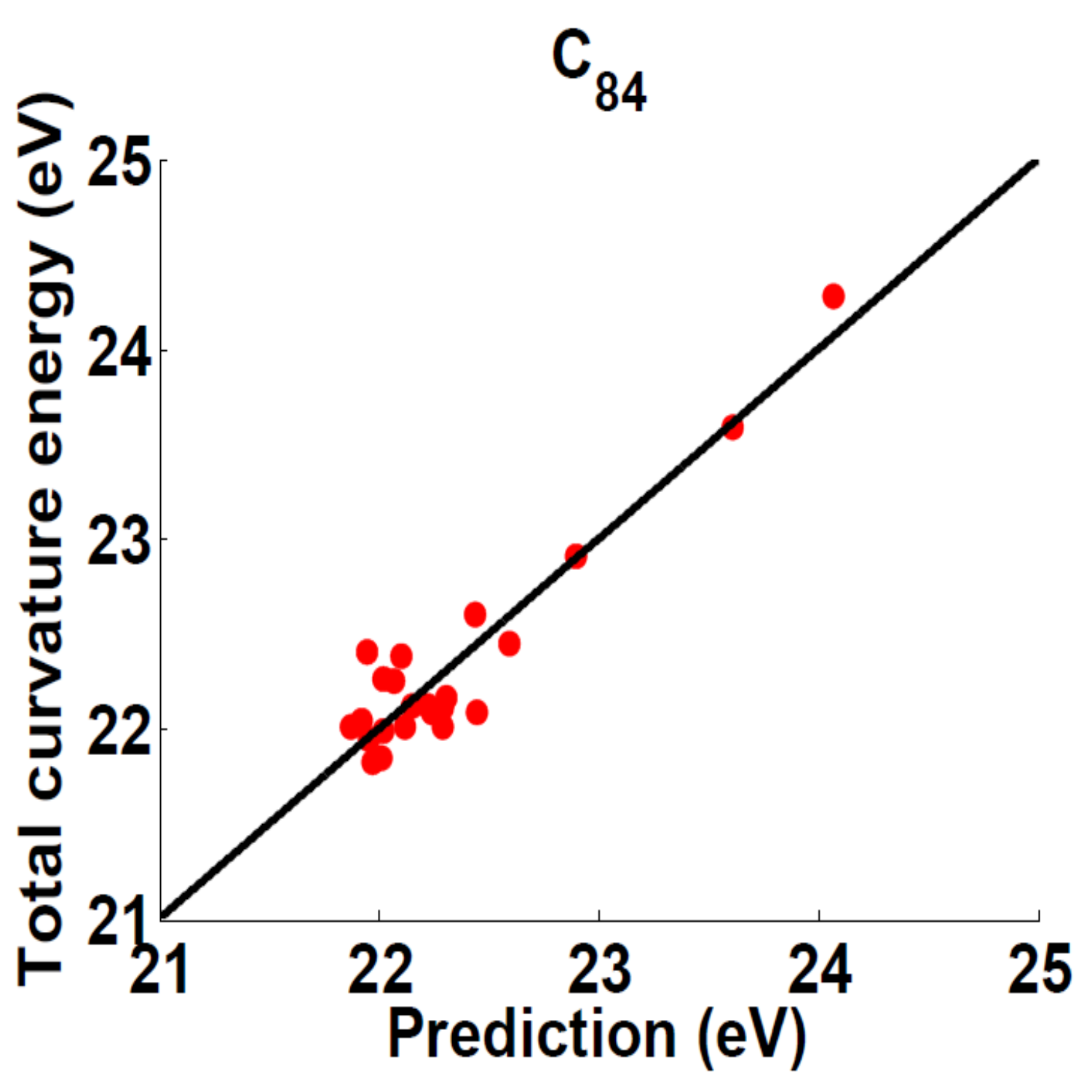}
\end{tabular}
\end{center}
\caption{The comparison of fullerene isomer  total curvature energies  and persistent homology theory predictions.
}
\label{pro_en}
\end{figure}

Having demonstrated the utility  of  the proposed Laplace-Beltrami flow based  persistent homology method for protein characterization, we are interested in the further application of this topological tool for quantitative analysis of carbon fullerene molecules. In particular, we explore the application of the present persistent homology method to the prediction of  the total curvature energies of the carbon fullerene isomers.  Fullerene  molecules admit a large number of isomers, especially when the number of atoms is large. Different isomers with the same chemical formula have different geometric structures which leads to the variations in their  total curvature energies. The stability of each given fullerene isomer is determined by its total curvature energy. In general, the higher  energy isomer is less stable.

\begin{figure}
\begin{center}
\begin{tabular}{ccc}
\includegraphics[width=0.4\textwidth]{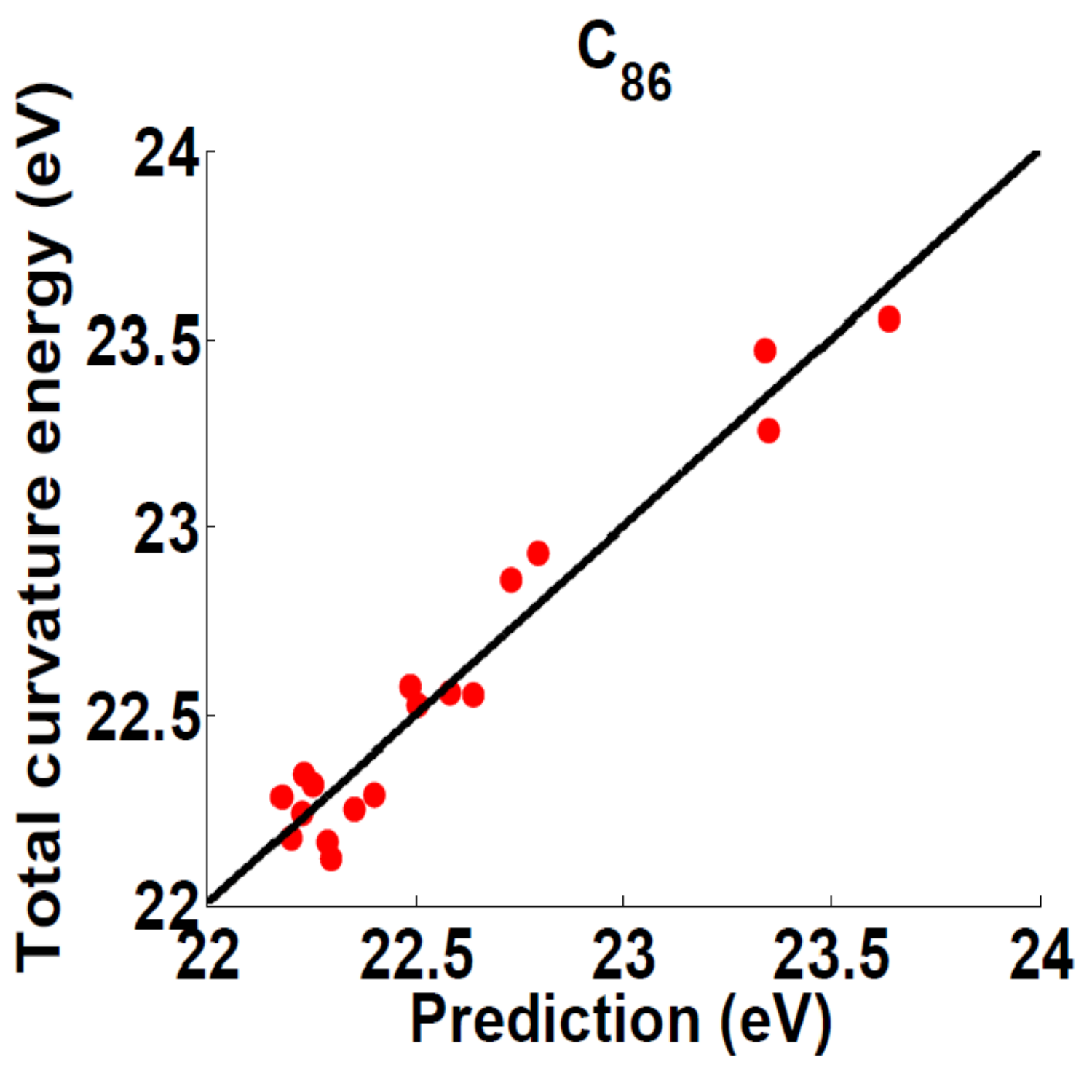}
\includegraphics[width=0.4\textwidth]{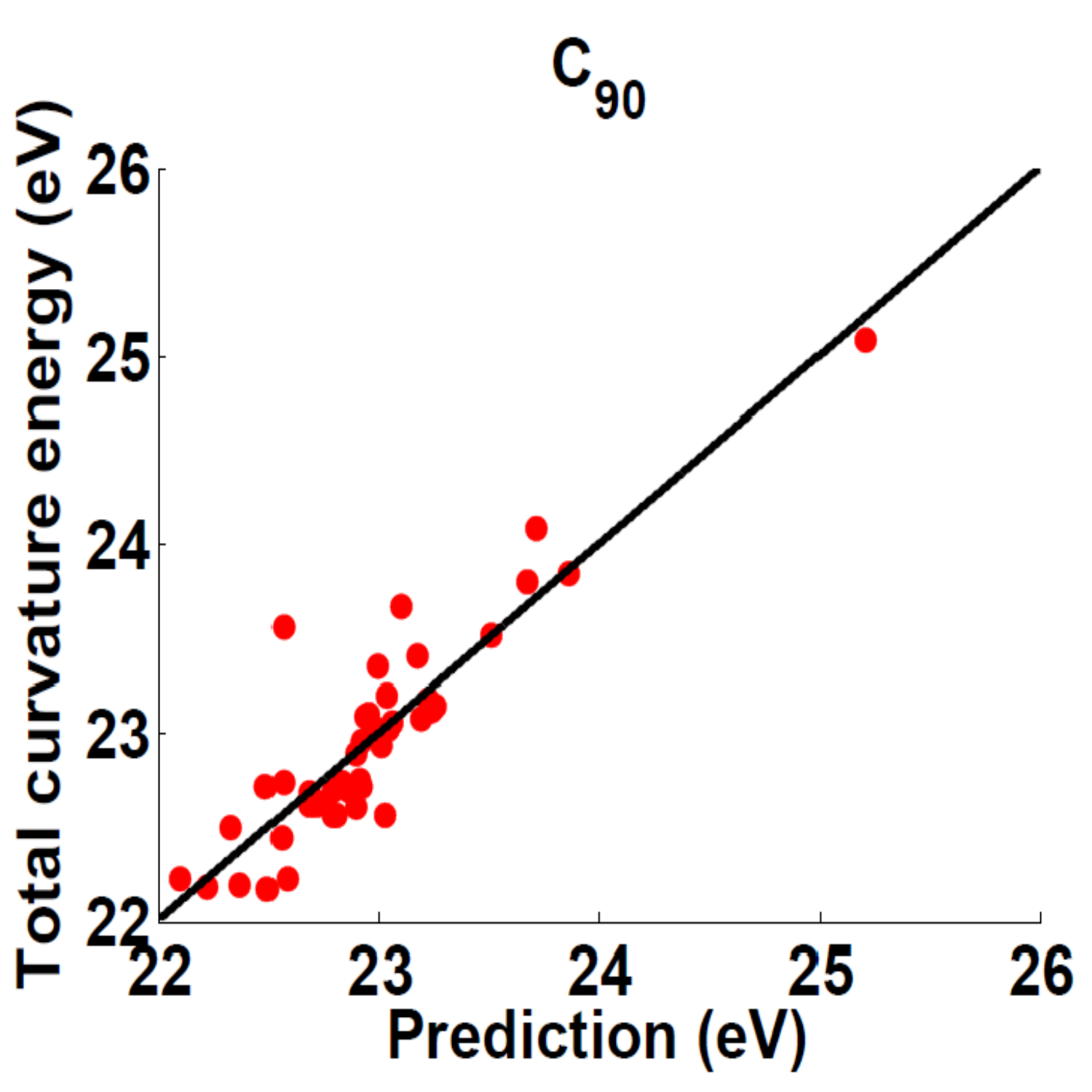}\\
\includegraphics[width=0.4\textwidth]{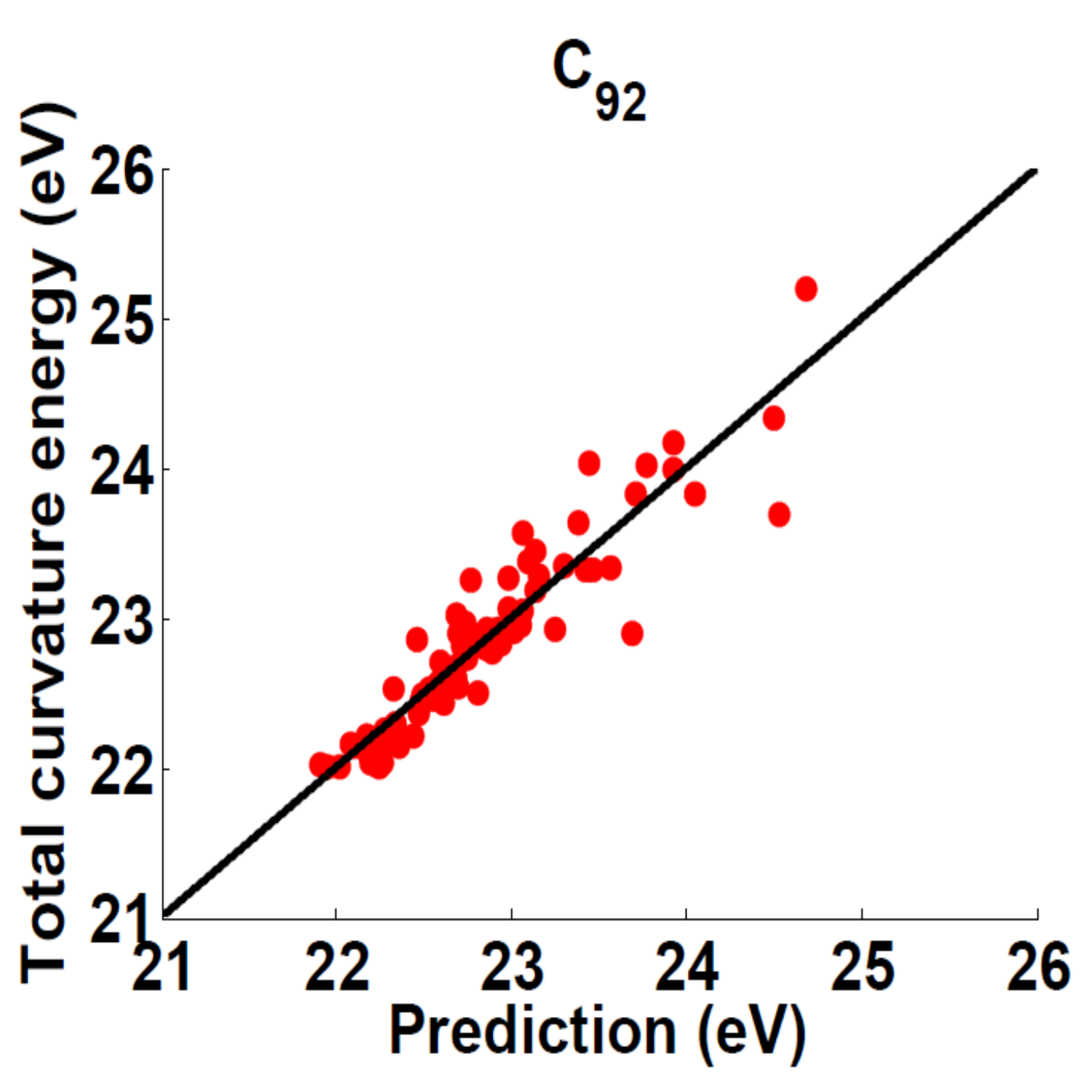}
\includegraphics[width=0.4\textwidth]{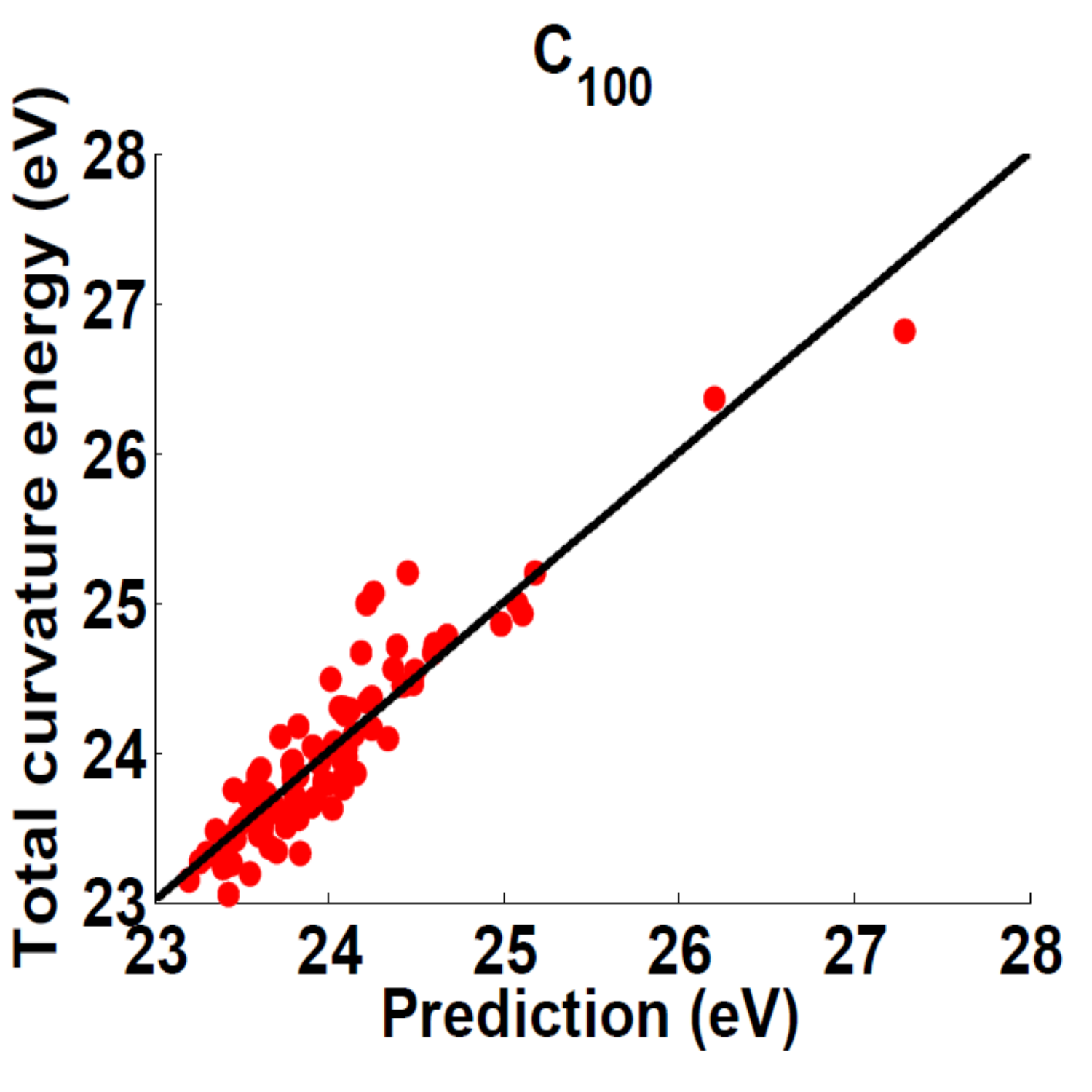}
\end{tabular}
\end{center}
\caption{The comparison of fullerene isomer  total curvature energies  and persistent homology theory predictions    (cont'd).
}
\label{pro_en2}
\end{figure}

\begin{table}[!ht]
\centering
\caption{The correlation coefficients and standard deviations of the predicted values with respect to  total curvature energy data.}
\begin{tabular}{lll}
\cline{1-3}
Fullerene molecule &Correlation coefficient  & Standard deviation \\
\hline
C$_{36}$  &0.9668  &0.4345  \\
C$_{38}$  &0.9280  &0.5263  \\
C$_{40}$  &0.9665  &0.4394  \\
C$_{44}$  &0.9485  &0.6211  \\
C$_{52}$  &0.9477  &0.5721  \\
C$_{84}$  &0.9389  &0.1932  \\
C$_{86}$  &0.9737  &0.0998  \\
C$_{90}$  &0.8956  &0.2469  \\
C$_{92}$  &0.9326  &0.2253  \\
C$_{100}$ &0.9253  &0.2364  \\
\hline
\end{tabular}
\label{corr700}
\end{table}

We assume that different isomers of a fullerene molecule  have the same surface area. This assumption is reasonable because all isomers share the same set of atoms and bonds.      However, these isomers may have different enclosed volumes as some isomers are more spherical than others. Those isomers that deviate from  the spherical shape must have  high curvature energies. The more deviation from the sphericity in the structure, the higher curvature energy an isomer has. Additionally, by the iso-perimetric inequality we know that for a class of isomers of  a given surface area, the volume is maximized when the isomer  is a perfect sphere. For fullerene isomers, more deviation from the sphericity in the structure, the earlier in the time evolution the $\beta_2$ bar dies, which leads to a shorter $\beta_2$ bar length. Therefore, we can establish a  relationship between the persistence of  $\beta_2$ invariant and the total curvature energy of a fullerene isomer.

In this work, the persistence or bar length of Betti 2, which essentially measures the size of the central cavity, is employed to predict the total curvature energy of  carbon fullerene isomers.   The Laplace-Beltrami flow is discretized with time stepping size $0.001$ and grid spacing size $0.25$. To quantitatively verify our prediction, the least squares method is employed to fit our predictions of the total curvature energies with the values provided in the web mentioned above. The accuracy of our prediction is evaluated by the  correlation coefficient (cc)
\begin{equation}
\label{corr}
{\rm cc}=\frac{\sum_{i=1}^N(L_i-\bar{L})(E_i-\bar{E})}{[\sum_{i=1}^N(L_i-\bar{L})^2(E_i-\bar{E})^2]^{1/2}},
\end{equation}
where $L_i$ represents the   bar length of $\beta_2$ generated by the Laplace-Beltrami flow based persistent homology method for the $i$th fullerene isomer of a given carbon fullerene family, $\bar{L}$ is the average of the  bar length of  $\beta_2$    over all the isomers of the fullerene, $E_i$ is the total curvature energy of  the $i$th fullerene isomer, $\bar{E}$ is the average of the total curvature energy over all the isomers of the fullerene. Note that we only count the $\beta_2$ bar that is due to the central cavity. 

We consider a total of ten different fullerene families with more than 500 fullerene isomers in this study, where the data are chosen from the following rules:
\begin{itemize}
\item For a specific carbon fullerene family, if it has less than or equal to  100 isomers, all the data are utilized.
\item For a given carbon fullerene family, if there are more than 100 isomers, the first 100 isomer molecules listed in the web are utilized.
\end{itemize}

\begin{table}[!ht]
\centering
\caption{Total curvature energies of C$_{36}$ isomers  vs lengths of the $\beta_2$ bar ($L(\beta_2)$) obtained  with different time stepping sizes.}
\label{c36t}
\begin{tabular}{lllll}
\hline
\cline{1-5}
Total curvature energy (eV)  &$L(\beta_2)$  ($\Delta t=0.0001$)   &$L(\beta_2)$ ($\Delta t=0.0002$)    &$L(\beta_2)$ ($\Delta t=0.0004$)   &$L(\beta_2)$  ($\Delta t=0.0008$) \\
\hline
22.493 &6.435         &6.436       &6.436      &6.440        \\
22.688 &6.464         &6.464       &6.464      &6.464        \\
22.965 &6.601         &6.602       &6.600      &6.608        \\
23.027 &6.422         &6.422       &6.424      &6.432        \\
23.469 &6.159         &6.560       &6.164      &6.160        \\
24.025 &6.240         &6.240       &6.240      &6.248        \\
24.031 &6.122         &6.124       &6.124      &6.128        \\
24.152 &6.044         &6.044       &6.048      &6.056        \\
24.335 &6.122         &6.124       &6.124      &6.128        \\
24.620 &6.292         &6.292       &6.296      &6.304        \\
24.938 &5.928         &5.930       &5.928      &5.936        \\
25.514 &5.852         &5.852       &5.856      &5.856        \\
25.937 &5.608         &5.606       &5.608      &5.608        \\
26.013 &5.760         &5.760       &5.760      &5.760        \\
29.424 &4.901         &4.902       &4.904      &4.904        \\
\hline
Correlation coefficient         &0.9668         &0.9643      &0.9669      &0.9659       \\
Standard deviation   &0.4345         &0.4499      &0.4339      &0.4401       \\
\hline
\end{tabular}
\end{table}
\clearpage

The predicted results and the corresponding total curvature energies are illustrated in Figs. \ref{pro_en} and \ref{pro_en2}.  Table  \ref{corr700}  gives the correlation coefficients and  standard deviations of the predicted total curvature energies based on the proposed persistent homology theory and the total curvature energy data. Our results for  ten different fullerene molecules show  good predictions of our differential geometry based persistent homology  model.

\begin{table}[!ht]
\centering
\caption{Total curvature energies of C$_{36}$ isomers  vs lengths of the $\beta_2$ bar $(L(\beta_2))$ obtained  with different spatial spacing sizes}
\label{c36s}
\begin{tabular}{lllll}
\hline
\cline{1-5}
Total curvature energy (eV)  &$L(\beta_2)$ ($h=0.15$)  &$L(\beta_2)$ ($h=0.20$)    & $L(\beta_2)$ ($h=0.25$) & $L(\beta_2)$ ($h=0.30$) \\
\hline
22.493 &6.837         &6.704       &6.435      &6.178        \\
22.688 &6.845         &6.729       &6.464      &6.410        \\
22.965 &6.771         &6.621       &6.601      &6.149        \\
23.027 &6.728         &6.541       &6.422      &6.283        \\
23.469 &6.638         &6.441       &6.159      &6.002        \\
24.025 &6.564         &6.440       &6.240      &5.986        \\
24.031 &6.363         &6.039       &6.122      &5.751        \\
24.152 &6.359         &6.345       &6.044      &5.762        \\
24.335 &6.363         &6.309       &6.122      &5.751        \\
24.620 &6.621         &6.459       &6.292      &5.926        \\
24.938 &6.291         &6.067       &5.928      &5.723        \\
25.514 &6.260         &6.048       &5.852      &5.740        \\
25.937 &6.042         &5.936       &5.608      &5.515        \\
26.013 &6.087         &5.822       &5.760      &5.626        \\
29.424 &5.385         &5.186       &4.901      &4.918        \\
\hline
Correlation coefficient             &0.9715         &0.9804      &0.9668      &0.9501       \\
Standard deviation       &0.4030         &0.3349      &0.4345      &0.5300       \\
\hline
\end{tabular}
\end{table}
\clearpage

To test the reliability and robustness of our method in the isomer total curvature energy prediction,  we  have carried out our analysis  with different  grid spacing sizes
and  time  stepping sizes for 15 C$_{36}$ isomers. Table \ref{c36t} lists the lengths  of $\beta_2$ bars obtained  with different time  stepping sizes ($\Delta t$) and the total curvature energies  of  C$_{36}$ isomers. A  uniform spatial spacing size of $h=0.25$ is used in this test. Similarly,  Table \ref{c36s} gives the lengths  of $\beta_2$ bars computed  with different grid  spacing sizes $h$ and the total curvature energies of   C$_{36}$ isomers. A  given time stepping size of $\Delta t=0.0001$  is adopted in this validation.  Again, we see a good  consistence among our results.

Based on the above spatial and temporal convergent analysis,  it is clear that our results are robust and reliable. Therefore,  the persistence of Betti 2 has a strong correlation with the total curvature energies of fullerene isomers. These  results demonstrate that the persistence of Betti 2 is indeed reversely proportional to the total curvature energies of fullerene isomers. Additionally,  the proposed Laplace-Beltrami flow based  persistent homology approach performs extremely well in quantitative prediction  of topology-function relationship for fullerene isomers.

\section{Conclusion}

It is well known that topology typically does not distinguish a doughnut and a mug, which implies there is too much reduction in the geometric information. Indeed, topology is seldom used for quantitative description and modeling.  In contrast, geometry gives rise to very detailed models for the physical world. At nano scale and/or atomic scale,  geometry based  models often involve too many degrees of freedom such that their simulations become intractable for many real world problems.  Persistent homology is a new branch of algebraic topology that has recently become quite popular for topological simplifications in scientific and engineering applications. Its essential idea is to embed topological invariants in a minimal amount of geometric variation, i.e., a filtration parameter. As a result, persistent homology bridges the traditional topology  and geometry.

In the past, most successful applications of persistent homology have been limited to characterization, identification and analysis (CIA) in the literature. Indeed, persistent homology has been rarely employed for quantitative prediction.  In our  recent work \cite{KLXia:2014c}, we have introduced molecular topological fingerprints, which treat all barcodes in an equal footing for data CIA. We have also proposed topology-function relationships, which utilize persistent homology  as an efficient tool for the physical modeling and quantitative prediction of biomoelcular systems.

In this work,  a general procedure is introduced to construct objective-oriented persistent homology approaches for the detection, extraction and/or enhancement of desirable topological traits in data. Our essential idea is to define an objective functional to optimize desirable properties. The optimization leads to a set of operators  whose actions enforce the   objective functional and  give rise to a multiscale representation  of the original data. When such a multiscale representation is utilized for filtration, the resulting objective-oriented  persistent homology automatically detect, extract and/or amplify the corresponding topological persistence of the data. As a proof of principle, we use the differential geometry theory of surfaces to construct a surface energy functional. The optimization of this functional leads to the Laplace-Beltrami operator, which is able to provide a geometry-embedded filtration of the data of interest. The resulting persistent homology enhances the corresponding geometric structure in topological persistence. The proposed method is intensively validated using benchmark tests and structures with known topological properties.

The application of the  proposed geometric flow based topological method is considered to   both the CIA and quantitative modeling of  proteins and carbon fullerene molecules. We first employ the present method for the analysis of a beta barrel protein. The structure of the beta barrel has a large ring. Topologically, it is interesting to observe a long-lived Betti-1 bar during the time evolution of the Laplace-Beltrami flow.

Another application of the proposed method is the   total curvature energy prediction of fullerene isomers. We propose a model to correlate isomer total curvature energy and its structural sphericity. The latter is measured by the length of the Betti 2 bar of the isomer central cavity.  Essentially, a more distorted isomer has a higher total curvature energy and a shorter period of persistence of the central cavity Betti 2 bar. In our quantitative energy prediction, we have utilized a total of ten sets of fullerene isomers. Our results indicate that both  the proposed  Laplace-Beltrami flow  based persistent homology method and the present quantitative model work extremely well. All the correlation coefficients are very high.

The present differential geometry   based persistent homology opens a new approach for the  topological simplification of big data. We expect that other  objective functionals can be designed  and corresponding    objective-oriented persistent homology methods can be developed for specific purposes in data sciences. This approach will also lead to the construction of new objective-oriented partial differential equations (PDEs), geometric PDEs and topological PDEs in the future.

\section*{Acknowledgments}

This work was supported in part by NSF grants   IIS-1302285 and DMS-1160352,   NIH grant R01GM-090208 and the  MSU Center for Mathematical Molecular Biosciences Initiative. The authors thank Gunnar Carlsson, Konstantin Mischaikow and   Kelin Xia for useful discussions.

\newpage
\appendix

As stated above, all fullerene data are downloaded from a web page: \href{http://www.nanotube.msu.edu/fullerene/fullerene-isomers.html}{fullerene-isomers}. However, it is well known that a web page may not exist after certain time. We therefore  present fullerene isomers and their total curvature energies used in the present work in the following tables. The corresponding structure data are available up on request.

\begin{table}[!ht]
\label{36}
\centering
\caption{Fullerene C36 isomers and total curvature energies}
\begin{tabular}{llllllll}
\cline{1-8}
Name &Energy (eV)   &Name  &Energy (eV)      &Name     &Energy (eV)       &Name     &Energy (eV)  \\
\hline
C36(C2)1 &25.937     &C36(D2)2     &29.424      &C36(C1)3   &24.938     &C36(Cs)4              &25.524    \\
C36(D2)5 &26.013     &C36(D2d)6    &24.335      &C36(C1)7   &24.031     &C36(Cs)8              &24.025    \\
C36(C2v)9 &22.965     &C36(C2)10    &24.152      &C36(C2)11   &23.469     &C36(C2)12              &23.027    \\
C36(D3h)13 &24.620     &C36(D2d)14    &22.493      &C36(D6h)15   &22.688     &              &    \\
\hline
\end{tabular}
\end{table}

\begin{table}[!ht]
\label{38}
\centering
\caption{Fullerene C38 isomers and total curvature energies}
\begin{tabular}{llllllll}
\cline{1-8}
Name &Energy (eV)   &Name  &Energy (eV)      &Name     &Energy (eV)       &Name     &Energy (eV)  \\
\hline
C38(C2)1 &26.613     &C38(D3h)2     &27.745      &C38(C1)3   &25.120     &C38(C1)4          &26.564    \\
C38(C1)5 &24.288     &C38(C2)6    &25.564      &C38(C1)7   &25.520     &C38(C1)8              &24.184    \\
C38(D3)9 &25.843     &C38(C2)10    &23.853      &C38(C1)11   &24.185     &C38(C2v)12              &24.665    \\
C38(C2)13 &23.440     &C38(C1)14    &23.111      &C38(C2v)15   &24.069     &C38(C3v)16              &22.610    \\
C38(C2)17 &22.603     &             &            &             &           &                    &       \\
\hline
\end{tabular}
\end{table}

\begin{table}[!ht]
\label{40}
\centering
\caption{Fullerene C40 isomers and total curvature energies}
\begin{tabular}{llllllll}
\cline{1-8}
Name &Energy (eV)   &Name  &Energy (eV)      &Name     &Energy (eV)       &Name     &Energy (eV)  \\
\hline
C40(D5d)1 &30.194     &C40(C2)2     &27.771      &C40(D2)3   &29.686     &C40(C1)4          &26.838    \\
C40(Cs)5 &26.233      &C40(C1)6    &27.587        &C40(C1)7   &27.587     &C40(C2v)8         &26.421    \\
C40(C2)9 &24.856      &C40(C1)10    &24.933       &C40(C2)11   &27.092     &C40(C1)12         &25.038    \\
C40(Cs)13 &24.830     &C40(Cs)14    &24.165       &C40(Cs)15   &24.343     &C40(C2)16         &25.035    \\
C40(C1)17 &24.549     &C40(C2)18    &26.062       &C40(C2)19   &25.165     &C40(C3v)20        &24.271     \\
C40(C2)21 &24.356     &C40(C1)22    &24.031       &C40(C2)23   &25.232     &C40(Cs)24        &23.522     \\
C40(C2)25 &24.377     &C40(C1)26    &23.301       &C40(C2)27   &23.805     &C40(Cs)28        &24.700     \\
C40(C2)29 &23.416     &C40(C3)30    &24.163       &C40(Cs)31   &23.205     &C40(D2)32        &25.212     \\
C40(D2h)33 &26.042    &C40(C1)34    &23.946       &C40(C2)35   &23.560     &C40(C2)36        &22.994     \\
C40(C2v)37 &23.015     &C40(D2)38    &22.522       &C40(D5d)39   &23.206     &C40(Td)40        &23.300     \\
\hline
\end{tabular}
\end{table}

\begin{table}[!ht]
\label{44}
\centering
\caption{Fullerene C44 isomers and total curvature energies}
\begin{tabular}{llllllll}
\cline{1-8}
Name &Energy (eV)   &Name  &Energy (eV)      &Name     &Energy (eV)       &Name     &Energy (eV)  \\
\hline
C44(C2)1 &29.456     &C44(D2)2     &32.322       &C44(D3d)3   &31.727      &C44(C2)4          &28.076    \\
C44(C2)5 &27.881      &C44(C2)6    &28.085        &C44(C1)7   &27.359      &C44(C1)8         &26.901    \\
C44(C1)9 &28.635      &C44(C1)10    &27.949       &C44(Cs)11   &27.882     &C44(C2)12         &27.618    \\
C44(C2v)13 &26.848     &C44(C2)14    &27.270       &C44(C1)15   &26.036     &C44(C1)16         &26.020    \\
C44(C1)17 &27.492     &C44(C1)18    &25.806       &C44(C1)19   &25.968     &C44(C2)20        &26.256     \\

C44(C1)21 &26.389     &C44(C1)22    &24.685       &C44(C1)23   &25.012     &C44(D2)24        &25.872     \\
C44(C1)25 &25.218     &C44(C1)26    &25.950       &C44(C1)27   &25.505     &C44(Cs)28        &24.845     \\
C44(C1)29 &24.164     &C44(C1)30    &24.487       &C44(C1)31   &25.802     &C44(C2)32        &24.446     \\
C44(Cs)33 &25.314    &C44(C2)34    &27.070       &C44(D3)35   &31.214     &C44(C2)36        &24.669     \\
C44(D3h)37 &25.905    &C44(D3d)38    &25.964       &C44(C2v)39   &24.944    &C44(C1)40        &24.708     \\

C44(C1)41 &25.793     &C44(C1)42     &24.935       &C44(C1)43   &25.754      &C44(C2)44          &25.488    \\
C44(C2)45 &25.834      &C44(C2)46    &26.079        &C44(C1)47   &24.285      &C44(C1)48         &25.210    \\
C44(C2)49 &24.380      &C44(C1)50    &24.960       &C44(C1)51   &24.174     &C44(C1)52         &23.454    \\
C44(Cs)53 &25.790     &C44(Cs)54    &24.117       &C44(C2v)55   &23.983     &C44(C1)56         &24.320    \\
C44(C1)57 &23.831     &C44(C1)58    &24.991       &C44(C1)59   &23.427     &C44(C1)60        &24.054     \\

C44(C2)61 &24.844     &C44(C1)62    &24.170       &C44(C1)63   &24.061     &C44(C1)64        &24.537     \\
C44(C1)65 &24.804     &C44(C2)66    &26.402       &C44(C1)67   &23.276     &C44(C2)68        &23.218     \\
C44(C1)69 &22.958     &C44(Cs)70    &23.619       &C44(Cs)71   &24.168     &C44(D3h)72        &22.846     \\
C44(T)73 &24.076    &C44(C2)74    &23.621       &C44(D2)75   &22.582     &C44(C2)76        &23.848     \\
C44(C1)77 &22.900    &C44(C1)78    &23.065       &C44(C2)79   &23.463    &C44(D3)80        &23.159     \\

C44(C2)81 &24.389     &C44(S4)82     &23.258       &C44(D2)83   &23.903      &C44(Cs)84          &24.040    \\
C44(D2)85 &25.617      &C44(D3d)86    &28.214        &C44(C2)87   &23.220      &C44(C1)88         &23.049    \\
C44(D2)89 &22.513      &           &           &         &         &             &       \\
\hline
\end{tabular}
\end{table}

\begin{table}[!ht]
\label{52}
\centering
\caption{Fullerene C52 isomers and total curvature energies}
\begin{tabular}{llllllll}
\cline{1-8}
Name &Energy (eV)   &Name  &Energy (eV)      &Name     &Energy (eV)       &Name     &Energy (eV)  \\
\hline
C52(C2)1 &32.862     &C52(D2)2     &35.990       &C52(Cs)3   &30.452      &C52(C1)4          &30.393    \\
C52(C2)5 &30.646      &C52(Cs)6    &29.299        &C52(C1)7   &29.177      &C52(C1)8         &28.759    \\
C52(C1)9 &30.445      &C52(C1)10    &29.177       &C52(C1)11   &28.855     &C52(C1)12         &29.210    \\
C52(C1)13 &30.490     &C52(C1)14    &30.907       &C52(C2)15   &31.612     &C52(C1)16         &29.602    \\
C52(C1)17 &31.948     &C52(C1)18    &29.269       &C52(C1)19   &28.916     &C52(C1)20        &29.258     \\

C52(C2)21 &29.903     &C52(C2)22    &28.553       &C52(C1)23   &27.864     &C52(C1)24        &27.635     \\
C52(C1)25 &28.527     &C52(C1)26    &28.045       &C52(C1)27   &28.271     &C52(C1)28        &28.514     \\
C52(C2)29 &28.955     &C52(C1)30    &28.527       &C52(C2)31   &30.073     &C52(C1)32        &30.317     \\
C52(C1)33 &28.577    &C52(C1)34    &27.383       &C52(C1)35   &26.944     &C52(C1)36        &29.248     \\
C52(C1)37 &27.645    &C52(Cs)38    &29.592       &C52(C1)39   &27.276    &C52(Cs)40        &26.724     \\

C52(C1)41 &27.721     &C52(C1)42     &27.676       &C52(C2)43   &26.717      &C52(Cs)44          &27.455    \\
C52(C1)45 &30.622      &C52(C1)46    &27.524        &C52(C1)47   &28.032      &C52(C1)48         &27.408    \\
C52(C1)49 &27.529      &C52(C2)50    &28.618       &C52(C1)51   &27.005     &C52(C1)52         &27.127    \\
C52(C1)53 &26.929     &C52(C1)54    &26.754       &C52(C1)55   &26.705     &C52(C1)56         &26.846    \\
C52(C1)57 &26.919     &C52(D2)58    &32.111       &C52(C1)59   &28.477     &C52(C1)60        &23.375     \\

C52(C1)61 &29.172     &C52(C1)62    &25.762       &C52(C1)63   &26.041     &C52(C1)64        &27.762     \\
C52(C1)65 &26.077     &C52(Cs)66    &25.729       &C52(Cs)67   &27.443     &C52(C1)68        &26.383     \\
C52(C1)69 &27.389     &C52(C1)70    &25.753       &C52(C1)71   &27.405     &C52(C1)72        &26.846     \\
C52(C1)73 &26.857    &C52(C1)74    &25.468       &C52(C1)75   &26.336     &C52(C1)76        &27.267     \\
C52(C1)77 &29.103    &C52(C1)78    &25.652       &C52(C1)79   &27.349    &C52(C1)80        &25.765     \\

C52(C1)81 &26.857     &C52(C2)82    &26.813       &C52(C2)83   &28.263     &C52(C2)84        &25.361     \\
C52(C1)85 &27.436     &C52(C1)86    &25.508       &C52(C1)87   &27.288     &C52(C2)88        &27.449     \\
C52(C1)89 &25.532     &C52(C1)90    &26.098       &C52(C1)91   &26.693     &C52(Cs)92        &25.805     \\
C52(C1)93 &26.103    &C52(D2d)94    &26.864       &C52(Cs)95   &25.937     &C52(C1)96        &26.124     \\
C52(C1)97 &26.130    &C52(C1)98    &25.646       &C52(C2)99   &27.367    &C52(C1)100        &26.255     \\
\hline
\end{tabular}
\end{table}

\begin{table}[!ht]
\label{84}
\centering
\caption{Fullerene C84 isomers and total curvature energies}
\begin{tabular}{llllllll}
\cline{1-8}
Name &Energy (eV)   &Name  &Energy (eV)      &Name     &Energy (eV)       &Name     &Energy (eV)  \\
\hline
C84(D2)1 &24.281     &C84(C2)2     &23.593      &C84(Cs)3   &22.389     &C84(D2d)4          &22.607    \\
C84(D2)5 &22.910     &C84(C2v)6    &22.408      &C84(C2v)7   &22.270     &C84(C2)8              &22.167    \\
C84(C2)9 &22.124     &C84(Cs)10    &22.043      &C84(C2)11   &22.088     &C84(C1)12              &22.011    \\
C84(C2)13 &22.109     &C84(Cs)14    &22.250      &C84(Cs)15   &22.012     &C84(Cs)16              &22.019    \\
C84(C2v)17 &22.124     &C84(C2v)18    &22.159      &C84(D3d)19   &22.090     &C84(Td)20              &22.453    \\
C84(D2)21 &21.950     &C84(D2)22    &21.854      &C84(D2d)23   &21.829     &C84(D6h)24              &21.990    \\
\hline
\end{tabular}
\end{table}

\begin{table}[!ht]
\label{86}
\centering
\caption{Fullerene C86 isomers and total curvature energies}
\begin{tabular}{llllllll}
\cline{1-8}
Name &Energy (eV)   &Name  &Energy (eV)      &Name     &Energy (eV)       &Name     &Energy (eV)  \\
\hline
C86(C1)1 &23.258     &C86(C2)2     &23.553      &C86(C2)3   &23.473     &C86(C2)4          &22.862    \\
C86(c1)5 &22.576     &C86(C2)6    &22.933      &C86(C1)7   &22.528     &C86(Cs)8           &22.562    \\
C86(C2v)9 &22.556     &C86(C2v)10    &22.285      &C86(C1)11   &22.242     &C86(C1)12              &22.256    \\
C86(C1)13 &22.169     &C86(C2)14    &22.292      &C86(Cs)15   &22.178     &C86(Cs)16              &22.348    \\
C86(C2)17 &22.211     &C86(C3)18    &22.320      &C86(D3)19   &22.123     &             &      \\
\hline
\end{tabular}
\end{table}

\begin{table}[!ht]
\label{90}
\centering
\caption{Fullerene C90 isomers and total curvature energies}
\begin{tabular}{llllllll}
\cline{1-8}
Name &Energy (eV)   &Name  &Energy (eV)      &Name     &Energy (eV)       &Name     &Energy (eV)  \\
\hline
C90(D5h)1 &25.081     &C90(C2v)2     &24.092      &C90(C1)3   &23.808     &C90(C2)4          &23.846    \\
C90(Cs)5 &23.521     &C90(C2)6    &23.080      &C90(C1)7   &22.894     &C90(C2)8              &23.407    \\
C90(C1)9 &22.716     &C90(Cs)10    &22.612      &C90(C1)11   &23.094     &C90(C2)12              &23.120    \\
C90(C2v)13 &23.672     &C90(C1)14    &23.195      &C90(C1)15   &23.180     &C90(C2v)16              &23.355    \\
C90(Cs)17 &23.147     &C90(C2)18    &23.023      &C90(C2)19   &22.687     &C90(C1)20              &22.625    \\
C90(C1)21 &22.561     &C90(C1)22    &22.695      &C90(C2)23   &22.569     &C90(C1)24              &23.020    \\
C90(C2v)25 &23.096     &C90(C1)26     &22.621      &C90(C1)27   &22.700     &C90(C2)28          &22.715    \\
C90(C1)29 &22.960     &C90(C1)30    &22.565      &C90(C2)31   &22.989     &C90(C1)32              &22.559    \\
C90(Cs)33 &23.060     &C90(Cs)34    &22.737      &C90(Cs)35   &22.497     &C90(C2v)36              &22.939    \\
C90(C2)37 &22.748     &C90(C1)38    &22.614      &C90(C2v)39   &22.742     &C90(C2)40              &22.227    \\
C90(C2)41 &22.190     &C90(C2)42    &22.222      &C90(C2)43   &22.177     &C90(C2)44              &22.443    \\
C90(C2)45 &22.174     &C90(C2v)46    &22.225      &   &     &              &      \\
\hline
\end{tabular}
\end{table}

\begin{table}[!ht]
\label{92}
\centering
\caption{Fullerene C92 isomers and total curvature energies}
\begin{tabular}{llllllll}
\cline{1-8}
Name &Energy (eV)   &Name  &Energy (eV)      &Name     &Energy (eV)       &Name     &Energy (eV)  \\
\hline
C92(D2)1 &25.193     &C92(C1)2     &23.825       &C92(C2)3   &24.330      &C92(C2)4          &23.995    \\
C92(Cs)5 &23.695      &C92(Cs)6    &23.638        &C92(C2)7   &23.316      &C92(C1)8         &23.051    \\
C92(C2)9 &23.351      &C92(C1)10    &22.769       &C92(C1)11   &22.837     &C92(C1)12         &22.888    \\
C92(C1)13 &23.188     &C92(Cs)14    &22.969       &C92(Cs)15   &22.675     &C92(Cs)16         &22.913    \\
C92(C2)17 &23.831     &C92(C1)18    &23.325       &C92(C2)19   &23.336     &C92(C1)20        &23.448     \\

C92(C2)21 &24.031     &C92(C2v)22    &24.207       &C92(C2)23   &23.276     &C92(Cs)24        &22.850     \\
C92(C2)25 &23.895     &C92(C2)26    &23.023       &C92(C2)27   &23.263     &C92(D3)28        &23.566     \\
C92(D2h)29 &24.174     &C92(C1)30    &22.941       &C92(C2)31   &22.878     &C92(C1)32        &22.733     \\
C92(C1)33 &22.922    &C92(C2)34    &22.706       &C92(C2v)35   &23.380     &C92(C2)36        &22.903     \\
C92(C1)37 &23.056    &C92(C1)38    &22.529       &C92(C1)39   &22.930    &C92(C1)40        &22.894     \\

C92(C1)41 &25.793     &C92(C1)42     &24.935       &C92(C1)43   &25.754      &C92(C2)92          &25.488    \\
C92(C2)45 &25.834      &C92(C2)46    &26.079        &C92(C1)47   &24.285      &C92(C1)48         &25.210    \\
C92(C2)49 &24.380      &C92(C1)50    &24.960       &C92(C1)51   &24.174     &C92(C1)52         &23.454    \\
C92(Cs)53 &25.790     &C92(Cs)54    &24.117       &C92(C2v)55   &23.983     &C92(C1)56         &24.320    \\
C92(C1)57 &23.831     &C92(C1)58    &24.991       &C92(C1)59   &23.427     &C92(C1)60        &24.054     \\

C92(C2)61 &24.892     &C92(C1)62    &24.170       &C92(C1)63   &24.061     &C92(C1)64        &24.537     \\
C92(C1)65 &24.804     &C92(C2)66    &26.402       &C92(C1)67   &23.276     &C92(C2)68        &23.218     \\
C92(C1)69 &22.958     &C92(Cs)70    &23.619       &C92(Cs)71   &24.168     &C92(D3h)72        &22.846     \\
C92(T)73 &24.076    &C92(C2)74    &23.621       &C92(D2)75   &22.582     &C92(C2)76        &23.848     \\
C92(C1)77 &22.900    &C92(C1)78    &23.065       &C92(C2)79   &23.463    &C92(D3)80        &23.159     \\

C92(C2)81 &24.389     &C92(S4)82     &23.258       &C92(D2)83   &23.903      &C92(Cs)84          &24.040    \\
C92(D2)85 &25.617      &C92(D3d)86    &28.214        &        &          &            &       \\
\hline
\end{tabular}
\end{table}

\begin{table}[!ht]
\label{100}
\centering
\caption{Fullerene C100 isomers and total curvature energies}
\begin{tabular}{llllllll}
\cline{1-8}
Name &Energy (eV)   &Name  &Energy (eV)      &Name     &Energy (eV)       &Name     &Energy (eV)  \\
\hline
C100(C2)1 &32.862     &C100(D2)2     &35.990       &C100(Cs)3   &30.4100      &C100(C1)4          &30.393    \\
C100(C2)5 &30.646      &C100(Cs)6    &29.299        &C100(C1)7   &29.177      &C100(C1)8         &28.759    \\
C100(C1)9 &30.445      &C100(C1)10    &29.177       &C100(C1)11   &28.855     &C100(C1)12         &29.210    \\
C100(C1)13 &30.490     &C100(C1)14    &30.907       &C100(C2)15   &31.612     &C100(C1)16         &29.602    \\
C100(C1)17 &31.948     &C100(C1)18    &29.269       &C100(C1)19   &28.916     &C100(C1)20        &29.258     \\

C100(C2)21 &29.903     &C100(C2)22    &28.553       &C100(C1)23   &27.864     &C100(C1)24        &27.635     \\
C100(C1)25 &28.1007     &C100(C1)26    &28.045       &C100(C1)27   &28.271     &C100(C1)28        &28.514     \\
C100(C2)29 &28.955     &C100(C1)30    &28.1007       &C100(C2)31   &30.073     &C100(C1)32        &30.317     \\
C100(C1)33 &28.577    &C100(C1)34    &27.383       &C100(C1)35   &26.944     &C100(C1)36        &29.248     \\
C100(C1)37 &27.645    &C100(Cs)38    &29.592       &C100(C1)39   &27.276    &C100(Cs)40        &26.724     \\

C100(C1)41 &27.721     &C100(C1)42     &27.676       &C100(C2)43   &26.717      &C100(Cs)44          &27.455    \\
C100(C1)45 &30.622      &C100(C1)46    &27.1004        &C100(C1)47   &28.032      &C100(C1)48         &27.408    \\
C100(C1)49 &27.1009      &C100(C2)50    &28.618       &C100(C1)51   &27.005     &C100(C1)52         &27.127    \\
C100(C1)53 &26.929     &C100(C1)54    &26.754       &C100(C1)55   &26.705     &C100(C1)56         &26.846    \\
C100(C1)57 &26.919     &C100(D2)58    &32.111       &C100(C1)59   &28.477     &C100(C1)60        &23.375     \\

C100(C1)61 &29.172     &C100(C1)62    &25.762       &C100(C1)63   &26.041     &C100(C1)64        &27.762     \\
C100(C1)65 &26.077     &C100(Cs)66    &25.729       &C100(Cs)67   &27.443     &C100(C1)68        &26.383     \\
C100(C1)69 &27.389     &C100(C1)70    &25.753       &C100(C1)71   &27.405     &C100(C1)72        &26.846     \\
C100(C1)73 &26.857    &C100(C1)74    &25.468       &C100(C1)75   &26.336     &C100(C1)76        &27.267     \\
C100(C1)77 &29.103    &C100(C1)78    &25.6100       &C100(C1)79   &27.349    &C100(C1)80        &25.765     \\

C100(C1)81 &26.857     &C100(C2)82    &26.813       &C100(C2)83   &28.263     &C100(C2)84        &25.361     \\
C100(C1)85 &27.436     &C100(C1)86    &25.508       &C100(C1)87   &27.288     &C100(C2)88        &27.449     \\
C100(C1)89 &25.532     &C100(C1)90    &26.098       &C100(C1)91   &26.693     &C100(Cs)92        &25.805     \\
C100(C1)93 &26.103    &C100(D2d)94    &26.864       &C100(Cs)95   &25.937     &C100(C1)96        &26.124     \\
C100(C1)97 &26.130    &C100(C1)98    &25.646       &C100(C2)99   &27.367    &C100(C1)100        &26.255     \\
\hline
\end{tabular}

\end{table}


\vspace{0.6cm}
\begin{thebibliography}{10}

\bibitem{Bates:2009}
P.~W. Bates, Z.~Chen, Y.~H. Sun, G.~W. Wei, and S.~Zhao.
\newblock Geometric and potential driving formation and evolution of
  biomolecular surfaces.
\newblock {\em J. Math. Biol.}, 59:193--231, 2009.

\bibitem{Bates:2006}
P.~W. Bates, G.~W. Wei, and S.~Zhao.
\newblock The minimal molecular surface.
\newblock {\em arXiv:q-bio/0610038v1}, [q-bio.BM], 2006.

\bibitem{Bates:2006f}
P.~W. Bates, G.~W. Wei, and S.~Zhao.
\newblock The minimal molecular surface.
\newblock {\em Midwest Quantitative Biology Conference}, Mission Point Resort,
  Mackinac Island, MI:September 29 -- October 1, 2006.

\bibitem{Bates:2008}
P.~W. Bates, G.~W. Wei, and Shan Zhao.
\newblock Minimal molecular surfaces and their applications.
\newblock {\em Journal of Computational Chemistry}, 29(3):380--91, 2008.

\bibitem{Bendich:2010}
Paul Bendich, Herbert Edelsbrunner, and Michael Kerber.
\newblock Computing robustness and persistence for images.
\newblock {\em IEEE Transactions on Visualization and Computer Graphics},
  16:1251--1260, 2010.

\bibitem{Blomgren:1998}
P.~Blomgren and T.F. Chan.
\newblock Color {TV}: total variation methods for restoration of vector-valued
  images.
\newblock {\em Image Processing, IEEE Transactions on}, 7(3):304--309, 1998.

\bibitem{Bubenik:2007}
Peter Bubenik and Peter~T. Kim.
\newblock A statistical approach to persistent homology.
\newblock {\em Homology, Homotopy and Applications}, 19:337--362, 2007.

\bibitem{Carlsson:2009}
G.~Carlsson.
\newblock Topology and data.
\newblock {\em Am. Math. Soc}, 46(2):255--308, 2009.

\bibitem{Carlsson:2008}
G.~Carlsson, T.~Ishkhanov, V.~Silva, and A.~Zomorodian.
\newblock On the local behavior of spaces of natural images.
\newblock {\em International Journal of Computer Vision}, 76(1):1--12, 2008.

\bibitem{Carstensen:1997}
V.~Carstensen, R.~Kimmel, and G.~Sapiro.
\newblock Geodesic active contours.
\newblock {\em International Journal of Computer Vision}, 22:61--79, 1997.

\bibitem{Cecil:2005}
Thomas Cecil.
\newblock A numerical method for computing minimal surfaces in arbitrary
  dimension.
\newblock {\em J. Comput. Phys.}, 206(2):650--660, 2005.

\bibitem{ChangHW:2013}
H.~W. Chang, S.~Bacallado, V.~S. Pande, and G.~E. Carlsson.
\newblock Persistent topology and metastable state in conformational dynamics.
\newblock {\em PLos ONE}, 8(4):e58699, 2013.

\bibitem{DuanChen:2012a}
Duan Chen, Zhan Chen, and G.~W. Wei.
\newblock Quantum dynamics in continuum for proton transport {II: Variational}
  solvent-solute interface.
\newblock {\em International Journal for Numerical Methods in Biomedical
  Engineering}, 28:25 -- 51, 2012.

\bibitem{DuanChen:2012b}
Duan Chen and G.~W. Wei.
\newblock Quantum dynamics in continuum for proton transport---{Generalized}
  correlation.
\newblock {\em J Chem. Phys.}, 136:134109, 2012.

\bibitem{ZhanChen:2010a}
Z.~Chen, N.~A. Baker, and G.~W. Wei.
\newblock Differential geometry based solvation models {I}: Eulerian
  formulation.
\newblock {\em J. Comput. Phys.}, 229:8231--8258, 2010.

\bibitem{ZhanChen:2010b}
Z.~Chen, N.~A. Baker, and G.~W. Wei.
\newblock Differential geometry based solvation models {II}: Lagrangian
  formulation.
\newblock {\em J. Math. Biol.}, 63:1139-- 1200, 2011.

\bibitem{ZhanChen:2011a}
Z.~Chen and G.~W. Wei.
\newblock Differential geometry based solvation models {III}: Quantum
  formulation.
\newblock {\em J. Chem. Phys.}, 135:194108, 2011.

\bibitem{ZhanChen:2012}
Z.~Chen, Shan Zhao, J.~Chun, D.~G. Thomas, N.~A. Baker, P.~B. Bates, and G.~W.
  Wei.
\newblock Variational approach for nonpolar solvation analysis.
\newblock {\em Journal of Chemical Physics}, 137(084101), 2012.

\bibitem{Cheng:2007e}
L.~T. Cheng, Joachim Dzubiella, Andrew~J. McCammon, and B.~Li.
\newblock Application of the level-set method to the implicit solvation of
  nonpolar molecules.
\newblock {\em Journal of Chemical Physics}, 127(8), 2007.

\bibitem{Chopp:1993}
David~L. Chopp.
\newblock Computing minimal surfaces via level set curvature flow.
\newblock {\em J. Comput. Phys.}, 106(1):77--91, 1993.

\bibitem{Dabaghian:2012}
Y.~Dabaghian, F.~Memoli, L.~Frank, and G.~Carlsson.
\newblock A topological paradigm for hippocampal spatial map formation using
  persistent homology.
\newblock {\em PLoS Comput Biol}, 8(8):e1002581, 08 2012.

\bibitem{Dey:2008}
T.~K. Dey, K.~Y. Li, J.~Sun, and C.~S. David.
\newblock Computing geometry aware handle and tunnel loops in 3d models.
\newblock {\em ACM Trans. Graph.}, 27, 2008.

\bibitem{Dey:2013}
Tamal~K. Dey and Y.~S. Wang.
\newblock Reeb graphs: Approximation and persistence.
\newblock {\em Discrete and Computational Geometry}, 49(1):46--73, 2013.

\bibitem{DiFabio:2011}
Barbara Di~Fabio and Claudia Landi.
\newblock A mayer-vietoris formula for persistent homology with an application
  to shape recognition in the presence of occlusions.
\newblock {\em Foundations of Computational Mathematics}, 11:499--527, 2011.

\bibitem{Edelsbrunner:2002}
H.~Edelsbrunner, D.~Letscher, and A.~Zomorodian.
\newblock Topological persistence and simplification.
\newblock {\em Discrete Comput. Geom.}, 28:511--533, 2002.

\bibitem{Edelsbrunner:2010}
Herbert Edelsbrunner and John Harer.
\newblock {\em Computational topology: an introduction}.
\newblock American Mathematical Soc., 2010.

\bibitem{Feng:2004a}
X.~Feng and A.~Prohl.
\newblock Analysis of a fully discrete finite element method for the phase
  field model and approximation of its sharp interface limits.
\newblock {\em Mathematics of Computation}, 73:541--567, 2004.

\bibitem{XFeng:2013b}
X.~Feng, K.~L. Xia, Y.~Y. Tong, and G.~W. Wei.
\newblock Multiscale geometric modeling of macromolecules {II:} lagrangian
  representation.
\newblock {\em Journal of Computational Chemistry}, 34:2100--2120, 2013.

\bibitem{Frosini:1999}
Patrizio Frosini and Claudia Landi.
\newblock Size theory as a topological tool for computer vision.
\newblock {\em Pattern Recognition and Image Analysis}, 9(4):596--603, 1999.

\bibitem{Frosini:2013}
Patrizio Frosini and Claudia Landi.
\newblock Persistent betti numbers for a noise tolerant shape-based approach to
  image retrieval.
\newblock {\em Pattern Recognition Letters}, 34:863--872, 2013.

\bibitem{Gameiro:2013}
M.~Gameiro, Y.~Hiraoka, S.~Izumi, M.~Kramar, K.~Mischaikow, and V.~Nanda.
\newblock Topological measurement of protein compressibility via persistence
  diagrams.
\newblock {\em preprint}, 2013.

\bibitem{Ghrist:2008}
R.~Ghrist.
\newblock Barcodes: {The} persistent topology of data.
\newblock {\em Bull. Amer. Math. Soc.}, 45:61--75, 2008.

\bibitem{Gomes:2001}
J.~Gomes and O.~D. Faugeras.
\newblock Using the vector distance functions to evolve manifolds of arbitrary
  codimension.
\newblock {\em Lecture Notes in Computer Science}, 2106:1--13, 2001.

\bibitem{Greer:2004}
J.~B. Greer and A.~L. Bertozzi.
\newblock H-1 solutions of a class of fourth order nonlinear equations for
  image processing.
\newblock {\em Discrete and Continuous Dynamical Systems}, 10:349--366, 2004.

\bibitem{Greer:2004b}
J.~B. Greer and A.~L. Bertozzi.
\newblock Traveling wave solutions of fourth order pdes for image processingl.
\newblock {\em SIAM Journal on Mathematics Analysis}, 36:38--68, 2004.

\bibitem{Guan:2014}
J.~Guan, Z.~Q. Jin, Z.~Zhu, and D.~Tom\'{a}nek.
\newblock Local curvature and stability of two-dimensional systems.
\newblock {\em preprint}, 2014.

\bibitem{Harker:2010}
Shaun Harker, Konstantin Mischaikow, Marian Mrozek, Vidit N, Hubert Wagner, and
  Mateusz Juda.
\newblock The efficiency of a homology algorithm based on discrete morse theory
  and coreductions.
\newblock {\em Proceeding of the 3rd International Workshop on Computational
  Topology in Image Context, Image A}, pages 41--47, 2010.

\bibitem{Harker:2013}
Shaun Harker, Konstantin Mischaikow, Marian Mrozek, and Vidit Nanda.
\newblock {Discrete Morse theoretic algorithms for computing homology of
  complexes and maps}.
\newblock {\em Found. Comput. Math.}, pages doi:10.1007/s10208--013--9145--0,
  2013.

\bibitem{Hatcher:2001}
Allen Hatcher.
\newblock {\em Algebraic Topology}.
\newblock Cambridge University Press, 2001.

\bibitem{Horak:2009}
D.~Horak, S~Maletic, and M.~Rajkovic.
\newblock Persistent homology of complex networks.
\newblock {\em Journal of Statistical Mechanics: Theory and Experiment},
  2009(03):P03034, 2009.

\bibitem{ZMJin:2010}
Z.~M. Jin and X.~P. Yang.
\newblock Strong solutions for the generalized perona-malik equation for image
  restoration.
\newblock {\em Nonlinear Analysis-Theory Methods and Applications},
  73:1077--1084, 2010.

\bibitem{Kaczynski:2004}
T.~Kaczynski, K.~Mischaikow, and M.~Mrozek.
\newblock {\em Computational homology}.
\newblock Springer-Verlag, 2004.

\bibitem{Kasson:2007}
P.~M. Kasson, A.~Zomorodian, S.~Park, N.~Singhal, L.~J. Guibas, and V.~S.
  Pande.
\newblock Persistent voids a new structural metric for membrane fusion.
\newblock {\em Bioinformatics}, 23:1753--1759, 2007.

\bibitem{Krishnamoorthy:2007}
Bala Krishnamoorthy, Scott Provan, and Alexander Tropsha.
\newblock A topological characterization of protein structure.
\newblock In {\em Data Mining in Biomedicine, Springer Optimization and Its
  Applications}, pages 431--455, 2007.

\bibitem{LeeH:2012}
H~Lee, H.~Kang, M.~K. Chung, B.~Kim, and D.~S. Lee.
\newblock Persistent brain network homology from the perspective of dendrogram.
\newblock {\em Medical Imaging, IEEE Transactions on}, 31(12):2267--2277, Dec
  2012.

\bibitem{Li:1996a}
Y.~Li and F.~Santosa.
\newblock A computational algorithm for minimizing total variation in image
  restoration.
\newblock {\em IEEE Transactions on Image Processing}, 5(6):987--95, 1996.

\bibitem{XuLiu:2012}
Xu~Liu, Zheng Xie, and Dongyun Yi.
\newblock A fast algorithm for constructing topological structure in large
  data.
\newblock {\em Homology, Homotopy and Applications}, 14:221--238, 2012.

\bibitem{Mikula:2004}
Karol Mikula and Daniel Sevcovic.
\newblock A direct method for solving an anisotropic mean curvature flow of
  plane curves with an external force.
\newblock {\em Mathematical Methods in the Applied Sciences},
  27(13):1545--1565, 2004.

\bibitem{Mischaikow:1999}
K.~Mischaikow, M~Mrozek, J.~Reiss, and A.~Szymczak.
\newblock Construction of symbolic dynamics from experimental time series.
\newblock {\em Physical Review Letters}, 82:1144--1147, 1999.

\bibitem{Mischaikow:2013}
K.~Mischaikow and V.~Nanda.
\newblock Morse theory for filtrations and efficient computation of persistent
  homology.
\newblock {\em Discrete and Computational Geometry}, 50(2):330--353, 2013.

\bibitem{Mumford:1989}
David Mumford and Jayant Shah.
\newblock Optimal approximations by piecewise smooth functions and associated
  variational problems.
\newblock {\em Communications on Pure and Applied Mathematics}, 42(5):577--685,
  1989.

\bibitem{Perseus}
Vidit Nanda.
\newblock Perseus: the persistent homology software.
\newblock Software available at \url{http://www.sas.upenn.edu/~vnanda/perseus}.

\bibitem{Niyogi:2011}
P.~Niyogi, S.~Smale, and S.~Weinberger.
\newblock A topological view of unsupervised learning from noisy data.
\newblock {\em SIAM Journal on Computing}, 40:646--663, 2011.

\bibitem{Opron:2014}
K.~Opron, K.~L. Xia, and G.~W. Wei.
\newblock Fast and anisotropic flexibility-rigidity index for protein
  flexibility and fluctuation analysis.
\newblock {\em Journal of Chemical Physics}, 140:234105, 2014.

\bibitem{SOsher:1988}
S.~Osher and J.A. Sethian.
\newblock {Fronts propagating with curvature-dependent speed: algorithms based
  on Hamilton-Jacobi formulations}.
\newblock {\em Journal of computational physics}, 79(1):12--49, 1988.

\bibitem{Osher:2001}
Stanley Osher and Ronald~P. Fedkiw.
\newblock Level set methods: An overview and some recent results.
\newblock {\em J. Comput. Phys.}, 169(2):463--502, 2001.

\bibitem{Osher:1990}
Stanley Osher and Leonid~I. Rudin.
\newblock Feature-oriented image enhancement using shock filters.
\newblock {\em SIAM Journal on Numerical Analysis}, 27(4):919--940, 1990.

\bibitem{Pachauri:2011}
D.~Pachauri, C.~Hinrichs, M.K. Chung, S.C. Johnson, and V.~Singh.
\newblock Topology-based kernels with application to inference problems in
  alzheimer's disease.
\newblock {\em Medical Imaging, IEEE Transactions on}, 30(10):1760--1770, Oct
  2011.

\bibitem{JKPark:2013a}
J.~K. Park and G.~W. Wei.
\newblock A molecular level prototype for mechanoelectrical transducers in
  mammalian hair cells.
\newblock {\em Journal of Computational Neuroscience}, 35:231--241, 2014.

\bibitem{Rieck:2012}
Bastian Rieck, Hubert Mara, and Heike Leitte.
\newblock Multivariate data analysis using persistence-based filtering and
  topological signatures.
\newblock {\em IEEE Transactions on Visualization and Computer Graphics},
  18:2382--2391, 2012.

\bibitem{Robins:1999}
Vanessa Robins.
\newblock Towards computing homology from finite approximations.
\newblock In {\em Topology Proceedings}, volume~24, pages 503--532, 1999.

\bibitem{Rudin:1992}
Leonid~I. Rudin, Stanley Osher, and Emad Fatemi.
\newblock Nonlinear total variation based noise removal algorithms.
\newblock In {\em Proceedings of the eleventh annual international conference
  of the Center for Nonlinear Studies on Experimental mathematics :
  computational issues in nonlinear science}, pages 259--268, Amsterdam, The
  Netherlands, The Netherlands, 1992. Elsevier North-Holland, Inc.

\bibitem{Sapiro:1996}
G.~Sapiro and D.~L. Ringach.
\newblock Anisotropic diffusion of multivalued images with applications to
  color filtering.
\newblock {\em Image Processing, IEEE Transactions on}, 5(11):1582--1586, 1996.

\bibitem{Sarti:2002}
A.~Sarti, R.~Malladi, and J.~A. Sethian.
\newblock Subjective surfaces: A geometric model for boundary completion.
\newblock {\em International Journal of Computer Vision}, 46(3):201--221, 2002.

\bibitem{Sethian:2001}
J.~A. Sethian.
\newblock Evolution, implementation, and application of level set and fast
  marching methods for advancing fronts.
\newblock {\em J. Comput. Phys.}, 169(2):503--555, 2001.

\bibitem{Silva:2005}
V.~D. Silva and R~Ghrist.
\newblock Blind swarms for coverage in 2-d.
\newblock In {\em In Proceedings of Robotics: Science and Systems}, page~01,
  2005.

\bibitem{Singh:2008}
G.~Singh, F.~Memoli, T.~Ishkhanov, G.~Sapiro, G.~Carlsson, and D.~L. Ringach.
\newblock Topological analysis of population activity in visual cortex.
\newblock {\em Journal of Vision}, 8(8), 2008.

\bibitem{Smereka:2003}
Peter Smereka.
\newblock Semi-implicit level set methods for curvature and surface diffusion
  motion.
\newblock {\em Journal of Scientific Computing}, 19(1):439--456, 2003.

\bibitem{Sochen:1998}
N.~Sochen, R.~Kimmel, and R.~Malladi.
\newblock A general framework for low level vision.
\newblock {\em Image Processing, IEEE Transactions on}, 7(3):310--318, 1998.

\bibitem{Strombom:2007}
D.~Strombom.
\newblock {Persistent Homology in the cubical setting}.
\newblock {\em Master's Thesis, Lulea University of Technology}, 2007.

\bibitem{Javaplex}
Andrew Tausz, Mikael Vejdemo-Johansson, and Henry Adams.
\newblock Javaplex: A research software package for persistent (co)homology.
\newblock Software available at \url{http://code.google.com/p/javaplex}, 2011.

\bibitem{Wagner:2012}
C.~Wagner, C.~Chen, and E.~Vucini.
\newblock {Efficient computation of persistent homology for cubical data}.
\newblock {\em Topological Methods in Data Analysis and Visualization II}, page
  Springer Heidelberg Dordrecht London New York, 2012.

\bibitem{BeiWang:2011}
Bei Wang, Brian Summa, Valerio Pascucci, and M.~Vejdemo-Johansson.
\newblock Branching and circular features in high dimensional data.
\newblock {\em IEEE Transactions on Visualization and Computer Graphics},
  17:1902--1911, 2011.

\bibitem{YWang:2011c}
Y.~Wang, G.~W. Wei, and Si-Yang Yang.
\newblock Partial differential equation transform -- {Variational formulation
  and Fourier analysis}.
\newblock {\em International Journal for Numerical Methods in Biomedical
  Engineering}, 27:1996--2020, 2011.

\bibitem{YWang:2012b}
Y.~Wang, G.~W. Wei, and Si-Yang Yang.
\newblock {Mode decomposition evolution equations}.
\newblock {\em Journal of Scientific Computing}, 50:495--518, 2012.

\bibitem{YWang:2012d}
Y.~Wang, G.~W. Wei, and Si-Yang Yang.
\newblock Selective extraction of entangled textures via adaptive pde
  transform.
\newblock {\em International Journal in Biomedical Imaging}, 2012:Article ID
  958142, 2012.

\bibitem{Wei:1999}
G.~W. Wei.
\newblock Generalized { Perona-Malik} equation for image restoration.
\newblock {\em IEEE Signal Processing Lett.}, 6:165--167, 1999.

\bibitem{Wei:2009}
G.~W. Wei.
\newblock Differential geometry based multiscale models.
\newblock {\em Bulletin of Mathematical Biology}, 72:1562 -- 1622, 2010.

\bibitem{Wei:2002a}
G.~W. Wei and Y.~Q. Jia.
\newblock Synchronization-based image edge detection.
\newblock {\em Europhysics Letters}, 59(6):814--819, 2002.

\bibitem{Wei:2005}
G.~W. Wei, Y.~H. Sun, Y.~C. Zhou, and M.~Feig.
\newblock Molecular multiresolution surfaces.
\newblock {\em arXiv:math-ph/0511001v1}, pages 1 -- 11, 2005.

\bibitem{Wei:2013}
Guo-Wei Wei.
\newblock Multiscale, multiphysics and multidomain models {I: Basic} theory.
\newblock {\em Journal of Theoretical and Computational Chemistry},
  12(8):1341006, 2013.

\bibitem{Wei:2012}
Guo-Wei Wei, Qiong Zheng, Zhan Chen, and Kelin Xia.
\newblock Variational multiscale models for charge transport.
\newblock {\em SIAM Review}, 54(4):699 -- 754, 2012.

\bibitem{Willmore:1997}
T.~J. Willmore.
\newblock {\em Riemannian Geometry}.
\newblock Oxford University Press, USA, 1997.

\bibitem{KLXia:2014a}
K.~L. Xia, X.~Feng, Y.~Y. Tong, and G.~W. Wei.
\newblock Multiscale geometric modeling of macromolecules i: Cartesian
  representation.
\newblock {\em Journal of Computational Physics}, 275:912--936, 2014.

\bibitem{KLXia:2014c}
K.~L. Xia and G.~W. Wei.
\newblock Persistent homology analysis of protein structure, flexibility and
  folding.
\newblock {\em International Journal for Numerical Methods in Biomedical
  Engineerings}, 30:814--844, 2014.

\bibitem{MXu:2007}
M.~Xu and S.~L. Zhou.
\newblock Existence and uniqueness of weak solutions for a fourth-order
  nonlinear parabolic equation.
\newblock {\em Journal of Mathematical Analysis and Applications},
  325(1):636--654, 2007.

\bibitem{YaoY:2009}
Y.~Yao, J.~Sun, X.~H. Huang, G.~R. Bowman, G.~Singh, M.~Lesnick, L.~J. Guibas,
  V.~S. Pande, and G.~Carlsson.
\newblock Topological methods for exploring low-density states in biomolecular
  folding pathways.
\newblock {\em The Journal of Chemical Physics}, 130:144115, 2009.

\bibitem{Yu:2008g}
Z.~Y. Yu and C.~Bajaj.
\newblock Computational approaches for automatic structural analysis of large
  biomolecular complexes.
\newblock {\em IEEE/ACM Trans Comput Biol Bioinform}, 5:568--582, 2008.

\bibitem{Zhang:2006c}
Y.~Zhang, G.~Xu, and C.~Bajaj.
\newblock Quality meshing of implicit solvation models of biomolecular
  structures.
\newblock {\em Computer Aided Geometric Design}, 23(6):510--30, 2006.

\bibitem{SZhao:2011a}
Shan Zhao.
\newblock Pseudo-time-coupled nonlinear models for biomolecular surface
  representation and solvation analysis.
\newblock {\em International Journal for Numerical Methods in Biomedical
  Engineering}, 27:1964--1981, 2011.

\bibitem{SZhao:2014a}
Shan Zhao.
\newblock Operator splitting adi schemes for pseudo-time coupled nonlinear
  solvation simulations.
\newblock {\em Journal of Computational Physics}, 257:1000 -- 1021, 2014.

\bibitem{QZheng:2012}
Q.~Zheng, S.~Y. Yang, and G.~W. Wei.
\newblock { Molecular surface generation using PDE transform}.
\newblock {\em International Journal for Numerical Methods in Biomedical
  Engineering}, 28:291--316, 2012.

\bibitem{QZheng:2011a}
Qiong Zheng, Duan Chen, and G.~W. Wei.
\newblock Second-order {Poisson-Nernst-Planck} solver for ion transport.
\newblock {\em Journal of Comput. Phys.}, 230:5239 -- 5262, 2011.

\bibitem{QZheng:2011b}
Qiong Zheng and G.~W. Wei.
\newblock {Poisson-Boltzmann-Nernst-Planck model}.
\newblock {\em Journal of Chemical Physics}, 134:194101, 2011.

\bibitem{Zomorodian:2005}
A.~Zomorodian and G.~Carlsson.
\newblock Computing persistent homology.
\newblock {\em Discrete Comput. Geom.}, 33:249--274, 2005.

\bibitem{Zomorodian:2008}
Afra Zomorodian and Gunnar Carlsson.
\newblock Localized homology.
\newblock {\em Computational Geometry - Theory and Applications},
  41(3):126--148, 2008.

\end{thebibliography}
\end{document}